\documentclass[a4paper]{article} 

\usepackage[margin=3.5 cm]{geometry}
\usepackage{amssymb,amsmath,amsfonts}
\usepackage{amsthm} 
\usepackage{esvect} 
\usepackage{bm} 
\usepackage{graphicx} 
\usepackage{comment}
\usepackage{amssymb,amsmath,amsfonts}
\usepackage{amsthm} 
\usepackage{esvect} 
\usepackage{bm} 
\usepackage{bbm}
\usepackage{graphicx} 
\usepackage{comment}
\usepackage{lmodern}
\usepackage{amsmath,amssymb}
\usepackage{todonotes}
\usepackage{caption}
\usepackage{url}
\usepackage{pdfsync}
\usepackage{adjustbox}
\usepackage{multirow}
\usepackage{bbm}
\usepackage{algorithm}
\usepackage{amsopn}
\usepackage{algorithm}
\usepackage{algorithmic}
\usepackage{cleveref}
\usepackage{tikz}
\tikzset{
  treenode/.style = {shape=rectangle, rounded corners,
                     draw, align=center,
                     top color=white, bottom color=blue!20},
  root/.style     = {treenode, font=\Large, bottom color=blue!50},
  env/.style      = {treenode, font=\ttfamily\normalsize},
  dummy/.style    = {circle,draw}
}

\newcommand{\e}{{\rm e}}

\newcommand{\E}{{\mathbb E}}

\newcommand{\R}{{\mathbb R}}
\newcommand{\N}{{\mathbb N}}

\newcommand{\f}{{\mathbf f}}
\newcommand{\V}{{\mathbf V}}

\newcommand{\W}{{\mathbf W}}
\newcommand{\g}{{\mathbf g}}
\newcommand{\x}{{\mathbf x}}
\newcommand{\cc}{{\mathbf c}}

\newcommand{\bigO}{\mathcal{O}}

\newtheorem{proposition}{Proposition}[section]
\newtheorem{lemma}[proposition]{Lemma}

\newtheorem{definition}[proposition]{Definition}

\newtheorem{remark}[proposition]{Remark}

\title{Weighted Monte Carlo with least squares and randomized extended Kaczmarz for option pricing\footnote{The authors would like to thank Daniel Kressner for helpful discussions on this paper.}}
\author{Damir Filipovi\'c\footnote{EPFL and Swiss Finance Institute, 1015 Lausanne, Switzerland. Email: \tt{damir.filipovic@epfl.ch}} \quad Kathrin Glau\footnote{Queen Mary University of London, Mile End Road, E1 4NS London, United Kingdom. Email: {\tt k.glau@qmul.ac.uk}. This project has received funding from the European Union's Horizon 2020 research and innovation program under grant 665667.} \quad Yuji Nakatsukasa\footnote{Mathematical Institute, University of Oxford, Oxford, OX2 6GG, UK, and National Institute of Informatics, Japan. Email: {\tt nakatsukasa@maths.ox.ac.uk}} \quad Francesco Statti \footnote{EPFL, 1015 Lausanne, Switzerland. Email: {\tt francesco.statti@epfl.ch}}}
\date{September 30, 2019}
\begin{document}
\maketitle
\begin{abstract}
We propose a methodology for computing single and multi-asset European option prices, and more generally expectations of scalar functions of (multivariate) random variables. This new approach combines the ability of Monte Carlo simulation to handle high-dimensional problems with the efficiency of function approximation. Specifically, we first generalize the recently developed method for multivariate integration in [arXiv:1806.05492] to integration with respect to probability measures. The method is based on the principle ``approximate and integrate'' in three steps i) sample the integrand at points in the integration domain, ii) approximate the integrand by solving a least-squares problem, iii) integrate the approximate function. 
In high-dimensional applications we face memory limitations due to large storage requirements in step ii). Combining weighted sampling and the randomized extended Kaczmarz algorithm we obtain a new efficient approach to solve large-scale least-squares problems. Our convergence and cost analysis along with numerical experiments show the effectiveness of the method in both low and high dimensions, and under the assumption of a limited number of available simulations.

\end{abstract}


\section{Introduction}
Recently, a new algorithm combining function approximation and integration with Monte Carlo simulation has been developed in \cite{nakatsukasa2018approximate}. This algorithm, called MCLS (Monte Carlo with Least Squares),\footnote{Note that the similarity between the names MCLS and Least-Squares Monte Carlo for American option pricing developed in \cite{Longstaff01valuingamerican} is only owed to the fact that both methods use Monte Carlo and least-squares.} draws on Monte Carlo's ability to deal with the curse of dimensionality and reduces the variance through function approximation. 

In this paper we extend MCLS to efficiently price European options in low and high dimensions.
That is we approximate integrals of the form 
\[
I_\mu=\int_{E} f(\x) d\mu(\x),
\]
where $(E, \mathcal{A},\mu)$ is a probability space. MCLS is based on the principle ``approximate and integrate'' and mainly consists of three steps: i) generate $N$ sample points $\{\x_i\}_{i=1}^N \in E$, from $\mu$; ii) approximate the integrand $f$ with a linear combination of a priori chosen basis functions $f(\x) \approx p(\x):= \sum_{j=0}^n c_j \phi_j(\x)$, where the coefficients $\mathbf{c}=(c_0,\ldots,c_n)^\top$ are computed by solving the least-squares problem $\min_{\mathbf{c}\in\mathbb{R}^{n+1}}\|\V\mathbf{c}-\f\|_2$ for the Vandermonde\footnote{Usually the notion ``Vandermonde matrix'' is used for the special case when $\phi_i(x)=x^i$ and the matrix we are dealing with therefore is a generalized Vandermonde matrix. In statistics, such a matrix is usually referred to as ``design matrix''.  For the rest of the paper we call it Vandermonde matrix as in \cite{nakatsukasa2018approximate}.} matrix $\V=(\phi_j(\x_i))_{i=1,\ldots,N;j=0,\ldots,n}$ and $\mathbf{f}=(f(x_1),\ldots,f(x_N))^\top$; iii) integrate $p$ to approximate the integral $\sum_{j=0}^n c_j \int_{E} \phi_j(\x)\mu(d\x) \approx \int_{E} f(\x)\mu(d\x)$. 
Key to exploit the advantage of function approximation for integration is to choose basis functions $\{\phi_j\}_{j=0}^n$ that can be easily, possibly exactly, integrated with respect to $\mu$, and that $p$ approximates $f$ well. It can be shown that for $n=0$ the MCLS estimator coincides with the standard MC estimator. For a given fixed $N$, MCLS can outperform MC for any reasonably chosen basis $\{\phi_j\}_{j=0}^n$ with $n>0$.

Our contribution is fourfold. First, we provide a more detailed convergence, error and cost analysis for the MCLS estimator of $I_\mu$. This extends \cite{nakatsukasa2018approximate} which only considered $E=[0,1]^d$ and $\mu(d\x)=d\x$. In particular, our cost analysis reveals that MCLS asymptotically becomes more accurate than MC at the same cost, see Proposition \ref{prop:asymptotic}. This is of practical use whenever high accuracy is required and a large number of simulations is available, as for instance in option pricing. For a typical task in portfolio risk management, instead, only a limited budget of simulations is available. This is because evaluating $\f$ is extremely costly on the IT infrastructure of standard financial institutions, such as insurance companies. Our error analysis suggests that MCLS provides a better approximation than MC for such limited budget situations, compare Proposition \ref{thm_CM} and the subsequent discussion.

Second, we note that a computational bottleneck in MCLS is the storage requirement arising when solving the least-squares problem. Indeed, when the number of simulations $N$ and of the number of basis functions $n+1$ are too large, it is not feasible to explicitly store the $N\times (n+1)$ Vandermonde matrix $\V$. This severely limits the scalability of MCLS.
In order to overcome this limitation we enhance MCLS by the randomized extended Kaczmarz (REK) algorithm \cite{zouzias2013randomized} for solving the least-squares problem. The benefit of REK is that no explicit storage of $\V$ is needed. However REK only applies efficiently to well conditioned least-squares problems. Here we profit from the reduction of the condition number of $\V$ thanks to a weighted sampling scheme due to \cite{cohen2017optimal}. Moreover, under this weighted sampling scheme the rows of $\V$ have equal Euclidean norm, which further speeds up the REK.

Third, we apply MCLS to efficiently price European options in low and high dimensions. Here, the method turns out to be especially favorable for the class of polynomial models \cite{filipovic2016polynomial, filipovic2017polynomial} which covers widely used asset models such as the Black Scholes, Heston and Cox Ingersoll Ross models.
In this framework conditional moments and thus expectations of polynomials in the underlying price process are given in closed form. This naturally suggests the choice of polynomials as basis functions $\{\phi_j\}_{j=0}^n$, for which step iii) of MCLS becomes very efficient.

Fourth, we approximate the high-dimensional integral 
$\int_{[0,1]^d} \sin \Big (\sum_{j=1}^d x_j \Big ) d\x$
for $d=10$ and $d=30$, which shows that the limitations of the approach in \cite{nakatsukasa2018approximate} due to the high dimensionality have been indeed overcome considerably thanks to REK. 

The rest of the paper is organized as follows. In Section \ref{sec-MCLS} we review the main ingredients of our methodology and we explain how to combine them. In particular, in Section \ref{MCLS-review} we review MCLS as in \cite{nakatsukasa2018approximate} and we extend it to arbitrary probability measures. Then, in Section \ref{sec-optimally} we present the weighted sampling strategy proposed in \cite{cohen2017optimal}. In Section \ref{sec-kaczmarz} we review the randomized extended Kaczmarz algorithm and combine it with MCLS. We provide a convergence and a cost analysis in Section \ref{sec-cost analysis}. In Section \ref{sec-MCLS option pricing} we apply MCLS to option pricing. Here, we present numerical results for different polynomial models and payoff profiles in both low and high dimensionality. In Section \ref{sec-high dimensional application} we analyze the performance of MCLS for a standard high-dimensional integration test problem. We conclude in Section \ref{sec-conclusion}.

\section{Core methods}\label{sec-MCLS}
For the reader's convenience we present the three main ingredients of our methodology. First, we extend MCLS to arbitrary probability measures. Then, we present its combination with weighted sampling strategies as in \cite{nakatsukasa2018approximate}. Finally, we recap the randomized extended Kaczmarz algorithm and we propose our combined strategy.
\subsection{MCLS}\label{MCLS-review}
In this section we introduce a methodology to compute the definite integral 
\[
I_\mu:=\int_{E} f(\x) d\mu(\x),
\]
for some probability space $(E, \mathcal{A},\mu)$ and
for a function $f : E \to \R$, which we assume to be square-integrable, i.e.\ in
\begin{equation*}
L^2_\mu = \{f:E \to\R \enskip | \enskip \| f \|^2_\mu = \int_E f(\x)^2 d\mu(\x) < \infty, f \text{ measurable} \},
\end{equation*}
which is a Hilbert space with the inner product
$\langle f, g \rangle_\mu= \int_E f(\x) g(\x) d\mu(\x)$. The method is an extension of the method proposed in \cite{nakatsukasa2018approximate} for integrals with respect to the Lebesgue measure. 

To start, we choose a set of $n$ basis functions $\{\phi_j\}_{j=1}^n$, with $\phi_0 \equiv 1$, which will be used to approximate the integrand $f$. The idea is to choose basis functions $\phi_j$ that can be easily integrated. For instance, polynomials can be a good choice. Then, the steps of MCLS are as follows. First, as in standard Monte Carlo methods, one generates $N$ sample points $\{\x_i\}_{i=1}^N \in E$, according to $\mu$.

Second, the integrand $f$ and the set of basis functions are evaluated at all simulated points $\{\x_i\}_{i=1}^N$ leading to the following least-squared problem:
\begin{equation}\label{LS}
\min_{\mathbf{c}\in\mathbb{R}^{n+1}}
\left\|
\underbrace{
  \begin{bmatrix}
1&\phi_1(\x_1)&\phi_2(\x_1)&\ldots &\phi_n(\x_1)\\
1&\phi_1(\x_2)&\phi_2(\x_2)&\ldots&\phi_n(\x_2)\\
\vdots&\vdots&\vdots\\
1&\phi_1(\x_N)&\phi_2(\x_N)&\ldots&\phi_n(\x_N)\\
  \end{bmatrix}}_{=: \V}
  \begin{bmatrix}
c_0\\c_1\\\vdots \\c_n    
  \end{bmatrix}
-  \underbrace{\begin{bmatrix}
    f(\x_1)\\f(\x_2)\\\vdots\\ f(\x_N)
  \end{bmatrix}}_{=:\f}
\right\|_2, 
\end{equation}
which we denote as $\min_{\mathbf{c}\in\mathbb{R}^{n+1}}\|\V\mathbf{c}-\f\|_2$. Note that \eqref{LS} can be seen as a discrete version of the projection problem $\min_\mathbf{c} \| f-\sum_{j=0}^n c_j\phi_j \|_\mu$.
Third, one solves \eqref{LS}, whose solution is known to be explicitly given by 
\[
\mathbf{\hat{c} }= (\V^T\V)^{-1}\V^T\f.
\]
At this point, the linear combination $p(\x) := \sum_{j=0}^n\hat{c}_j \phi_j(\x)$ is an approximation of $f$.  

Finally, the last step consists of computing the integral of the approximant $p$, and $I$ is approximated by 
\begin{equation*}
I \approx \hat I_{\mu,N}=\int_{E} p(\x)d\mu(\x)=\hat{c}_0+\sum_{j=1}^n\hat{c}_j\int_{E} \phi_j(\x)d\mu(\x).  
\end{equation*}
We summarize the procedure in Algorithm \ref{Algo1}.

We remark that there is an interesting connection between MCLS and the standard Monte Carlo method: If one takes $n=0$, i.e. one approximates $f$ with a constant function, the resulting approximation is the solution of the least-squares problem 
\begin{equation*}
\min_{\mathbf{c}\in\mathbb{R}^{n+1}}
\left\|
  \begin{bmatrix}
1\\
1\\
\vdots\\
1\\
  \end{bmatrix}
 c_0
-  \begin{bmatrix}
    f(\x_1)\\f(\x_2)\\\vdots\\ f(\x_N)
  \end{bmatrix}
\right\|_2,
\end{equation*}
which is exactly given by $\hat{c}_0:= \frac{1}{N} \sum_{i=1}^N f(\x_i)$, the standard Monte Carlo estimator. We recall that in the standard MC method, asymptotically for large $N$ the error scales like\footnote{See e.g. \cite{caflisch1998monte} for the first term and recall that the expectation of a random variable $X$ is the constant with minimal distance to $X$ in the $L^2$-norm, rescaling yields the identity \eqref{error standardmc}.}
\begin{equation}\label{error standardmc}
\frac{\big(\int_E (f(\x)-I_{\mu})^2 d\mu(\x)\big )^{\frac{1}{2}}}{\sqrt{N}}
=\frac{\min_{c \in \R} \|f-c\|_\mu}{\sqrt{N}}=:\frac{\sigma(f)}{\sqrt{N}}.
\end{equation}
The quantity $(\sigma(f))^2$ is usually referred to as \textsl{the variance of $f$}.
This relation between MC and MCLS leads to an asymptotic error analysis, which we detail in Section \ref{sec-convergence}.

This connection leads to an asymptotic error analysis, which we detail in Section \ref{sec-convergence}.
This connection can also be exploited in order to increase the speed of convergence by combining it with quasi-Monte Carlo. In \cite{nakatsukasa2018approximate} also other ways to speed up the procedure are proposed, for example by an  adaptive choice of the basis functions (MCLSA).

It is observed in \cite{nakatsukasa2018approximate} that the method performs well for dimensions $d$ up to $d=6$. For higher dimensions solving the least-squares problem \eqref{LS} becomes computationally expensive, this is mainly due to two effects:
\begin{itemize}
\item[(i)] The size of the matrix $V$, being $N \times (n+1)$, rapidly becomes very large, posing memory limitations.
\item[(ii)] The condition number of the Vandermonde matrix $V$ typically gets large.
\end{itemize}
In the following we address these issues by combining MCLS with weighted sampling strategies and with the randomized extended Kaczmarz algorithm for solving the least-squares problem.
\begin{algorithm}[t]
\caption{Generalized MCLS}
\begin{algorithmic}[1]\label{Algo1}
\REQUIRE{Function $f$, basis functions $\{\phi_j\}_{j=1}^n$, $\phi_0\equiv 1$, integer $N (>n)$, probability distribution $\mu(\x)$ over domain $E$.}
\ENSURE{Approximate integral $\hat I_{\mu,N}\approx \int_{E} f(\x)d\mu(\x)$}
\STATE Generate sample points $\{\x_i\}_{i=1}^N \in E$, according to $\mu$.
\STATE Evaluate $f(\x_i)$ and $\phi_j(\x_i)$, for $i=1,\ldots,N$ and $j=1,\ldots,n$.
\STATE Solve the least-squares problem \eqref{LS} for $\mathbf{c}=[c_0,c_1,\ldots,c_n]^T$.
\STATE Compute $\hat I_{\mu,N}=\hat{c}_0+\sum_{j=1}^n\hat{c}_j\int_{E} \phi_j(\x)d\mu(\x)$.
\end{algorithmic}
\end{algorithm}

\subsection{Well conditioned least-squares problem via weighted sampling}\label{sec-optimally}


It is crucial that the coefficient matrix $\V$ in~\eqref{LS} be well conditioned, from both a computational and (more importantly) a function approximation perspective. 
Computationally, an ill-conditioned $\V$ means the least-squares problem is harder to solve using e.g. the conjugate gradient method, and the randomized Kaczmarz method described Section~\ref{sec-kaczmarz}. From an approximation viewpoint, $\V$ having a large condition number\footnote{We denote by $\kappa_2(\V)$ the 2-norm condition number of the matrix $\V$.} $\kappa_2(\V)$ implies that the function approximation error (in the continuous setting) can be large: $\| f-\sum_{j=0}^n c_j \phi_j\|^2_{\mu}$ is bounded roughly by $\kappa_2(\V)\| f-\sum_{j=0}^n c_{j}^* \phi_j\|^2_\mu$ (see~\cite[\S~5.4]{nakatsukasa2018approximate}), where $\cc^*:=\text{argmin}_{\mathbf{c}\in\mathbb{R}^{n+1}} \| f-\sum_{j=0}^n c_j\phi_j \|_\mu$. 
Hence in practice we devise the MCLS setting (choice of $\phi$ and sample strategy) so that $\V$ is well conditioned with high probability. 

A first step to obtain a well-conditioned Vandermonde matrix $V$, is to choose the basis to be orthonormal with respect to the scalar product $\langle \cdot \rangle_\mu$, for instance by applying a Gram-Schmidt orthonormalization procedure. Next, we observe that the strong law of large numbers yields
\begin{equation*}
\frac{1}{N}(\V^T\V)_{i+1,j+1} = 
\frac{1}{N}\sum_{l=1}^N\phi_i(\x_l)\phi_j(\x_l)
\stackrel{p}{\rightarrow}
\int_{E}\phi_i(\x)\phi_j(\x)d\mu(\x)
=\delta_{ij}
\end{equation*} as $N\rightarrow \infty$.
Therefore, for a large number of samples $N$ we expect $\frac{1}{N}\V^T \V$ to be close to the identity matrix $\mathbf{Id}_{n+1} \in \R^{(n+1)\times(n+1)}$. This implies that $\kappa_2(\V)$ is close to $1$. In practice, however, the condition number often is large. This is because the number $N$ of sample points required to obtain a well-conditioned $\V$ might be very large. For example, if we consider the one-dimensional interval $E=[-1,1]$ with the uniform probability measure and an orthonormal basis of Legendre polynomials, one can show that at least $N=\bigO(n^2 \log(n))$ sample points are needed to obtain a well conditioned $\V$. 
This example and others are discussed in \cite{chkifa2015discrete, cohen2017optimal}.

To overcome this problem, Cohen and Migliorati~\cite{cohen2017optimal} 
introduce a \emph{weighted} sampling for least-squares fitting. 
Its use for MCLS was suggested in~\cite{nakatsukasa2018approximate}, which we summarize here. 
Define the nonnegative function 
$w$ via 
\begin{equation}
  \label{eq:optsample}
  \frac{1}{w(\x)}=\frac{\sum_{j=0}^n\phi_{j}(\x)^2}{n+1}, 
\end{equation}
which is a probability distribution since $\frac{1}{w}\geq 0$ on $E$ and $\int_E \frac{1}{w(\x)}d\mu(\x)=1$. 
We then take samples $\{\tilde\x_i\}_{i=1}^N$ according to $\frac{d\mu}{w}$. Intuitively this means that we sample more often in areas where $\sum_{i=0}^n\phi_{i}(\x)^2$ takes large values. 

The least-squares problem \eqref{LS} with the samples $\sim \frac{d\mu}{w}$
becomes
\begin{equation}  \label{eq:MCLSweight}
\min_{{\mathbf c}}\|\sqrt{\W}(\V{\mathbf c}-\f)\|_2,
\end{equation}
where $\sqrt{\W} = \mbox{diag}(\sqrt{w(\tilde\x_1)},\sqrt{w(\tilde\x_2)},\ldots, \sqrt{w(\tilde\x_N)})$, and $\V,\f$ are as before in~\eqref{LS} with $\x\leftarrow \tilde\x$.  
This is again a least-squares problem $\min_{\mathbf c}\|\widetilde\V{\mathbf c}-\tilde\f\|_2$, with coefficient matrix $\widetilde\V:=\sqrt{\W}\V$ and right-hand side $\tilde\f := \sqrt{\W}\f$, whose solution is
${\mathbf c}=(\widetilde\V^T\widetilde\V)^{-1}\widetilde\V^T\sqrt{\W}\f$.
With high probability, the matrix $\widetilde\V$ is well conditioned, provided that $N\gtrsim n\log n$, see Theorem 2.1 in \cite{cohen2017optimal}. 
\begin{remark}\label{rem-rows}
Note that the left-multiplication by $\sqrt{\W}$ forces all the rows of $\widetilde\V$ to have the same norm (here $\sqrt{n+1}$); a property that proves useful in Section \ref{sec-kaczmarz}. 
\end{remark}
A simple strategy to sample from $w$ is as follows: for each of the $N$ samples, choose a basis function $\phi_j$ from $\{\phi_j\}_{j=0}^n$ uniformly at random, and sample from a probability distribution proportional to $\phi_j^2$. We refer to \cite{2017arXiv170700026H} for more details. 

\subsection{Randomized extended Kaczmarz to solve the least-squares problem}\label{sec-kaczmarz}

A standard least-squares solver that uses the QR factorization~\cite[Ch.~5]{golubbook4th} costs $O(Nn^2)$ operations, which quickly becomes prohibitive (relative to standard MC) when $n\gg 1$ . 
As an alternative, the conjugate gradient method (CG) applied to the normal equation 
$(\V^T\V)\cc = \V^T\f$ is suggested in~\cite{nakatsukasa2018approximate}. 
For $\kappa_2(\V)=O(1)$
this reduces the computational cost to $O(Nn)$. 
However,
CG still requires the storage of the whole matrix $\V$, 
which is $O(Nn)$. Indeed in practice, building and storing the matrix $\V$ becomes a major bottleneck in MCLS. 

To overcome this issue, here we suggest a further alternative, the randomized extended Kaczmarz (REK) algorithm developed by Zouzias and Freris~\cite{zouzias2013randomized}. REK is a particular stochastic gradient method to solve least-squares problems. It builds upon Strohmer and Vershynin's pioneering work \cite{strohmer2009randomized} and
Needell's extension to inconsistent systems~\cite{needell2010randomized}, and 
converges to the minimum-norm solution 
by simultaneously performing projection and solution refinement at each iteration. 
The convergence is geometric in expectation, and as already observed in~\cite{strohmer2009randomized}, Kaczmarz methods can sometimes even outperform the conjugate gradient method in speed for well-conditioned systems.
 A block version of REK was introduced in~\cite{needell2015randomized}, which sometimes additionally improves the performance.
 
Here we focus on REK and consider its application to MCLS. 
A pseudocode of REK is given in Algorithm~\ref{Algo2}. 
MATLAB notation is used, in which $\V(:,j)$ denotes the $j$th column of $\V$ and $\V(i,:)$ the $i$th row. 
The ${\mathbf z}^{(k)}$ iterates are the projection steps, which converge to $\f^\perp$, the part of $\f$ that lies in the orthogonal complement of $\V$'s column space. REK works by simultaneously projecting out the $\f^\perp$ component while refining the least-squares solution.

\begin{algorithm}[t]
\caption{REK: Randomized extended Kaczmarz method}
\begin{algorithmic}[1]\label{Algo2}
\REQUIRE{
$\V\in\mathbb{R}^{N\times n}$ and $\f\in\mathbb{R}^{N}$. 
}
\ENSURE{Approximate solution $\cc $ for $\min_\cc \|\V\cc -\f\|_2$}
\STATE{Initialize $\cc ^{(0)}=0$ and ${\mathbf z}^{(0)}=\f$}
\FOR{$k=1,2,\ldots,M$}
\STATE{Pick $i=i_k\in\{1,\ldots,N\}$ with probability $\|\V(i,:)\|_2^2/\|\V\|_F^2$}
\STATE{Pick $j=j_k\in\{1,\ldots,n+1\}$ with probability $\|\V(:,j)\|_2^2/\|\V\|_F^2$}
\STATE{Set ${\mathbf z}^{(k+1)}={\mathbf z}^{(k)}-\frac{\V(:,j_k)^T{\mathbf z}^{(k)}}{\|\V(:,j_k)\|_2^2}\V(:,j_k)$}
\STATE{Set $\cc ^{(k+1)}=\cc ^{(k)}+\frac{f_{i_k}-z_{i_k}^{(k)}-\V(i_k,:)^T\cc ^{(k)}}{\|\V(i_k,:)\|_2^2}\V(i_k,:)$}
\ENDFOR
\STATE{$\cc =\cc ^{M}$}
\end{algorithmic}
\end{algorithm}
Let us comment on REK (Algorithm~\ref{Algo2}) and its implementation, particularly in the  MCLS context:
\begin{itemize}
\item Employing the weighted sampling strategy of Section~\ref{sec-optimally} significantly simplifies Algorithm \ref{Algo2}. Following Remark \ref{rem-rows}, the norm of the rows of $\widetilde{\V}$ are constant and equal to $\sqrt{n+1}$. This also implies that $\|\widetilde{\V}\|_F^2 = N(n+1)$. The index $i_k$ in line 3 is therefore simulated uniformly at random. 
This has a practical significance in MCLS as the probability distribution $(\|\V(i,:)\|_2^2/\|\V\|_F^2)_{i=1,\ldots,N}$ does not have to be computed before starting the REK iterates. 
This results in (a potentially enormous) computational reduction; an additional benefit of using the weighted sampling strategy, besides improving conditioning. 
\item The number of iterations $M$ is usually not chosen a priori but by checking convergence of $\cc ^{(k)}$ infrequently. The suggestion in~\cite{zouzias2013randomized} is to check every $8\min(N,n)$ iterations for the conditions 
  \begin{equation*}
\frac{\|\V \cc^{(k)}-(\f-{\mathbf z}^{(k)}) \|_2}{\|\V\|_F\|\cc^{(k)}\|_2}    \leq \varepsilon,\qquad \mbox{and}\qquad 
\frac{\|\V^T {\mathbf z}^{(k)}) \|_2}{\|\V\|_F\|\cc^{(k)}\|_2}    \leq \varepsilon
  \end{equation*}
for a prescribed tolerance $\varepsilon>0$.
\item A significant advantage of REK is that it renders unnecessary the storage of the whole matrix $\V$: only the $i_k$th row and the $j_k$th column are needed, taking $O(N)$ memory cost. In practice, one can even sample in an online fashion: early samples can be discarded once the REK update is completed. 
\end{itemize}
The convergence of REK is known to be geometric in the expected mean squared sense~\cite[Thm~4.1]{zouzias2013randomized}: after $M$ iterations, we have 
\begin{equation}  \label{eq:REKconvergence}
\mathbb{\widetilde{E}}  \|\cc ^{(M)}-\mathbf{\hat{c}}\|_2^2\leq \left(
1-\frac{(\sigma_{\min}(\V))^2}{\|\V\|_F^2}\right)^{\lfloor \frac{M}{2}\rfloor}(1+2\kappa_2^2(\V))\|\mathbf{\hat{c}}\|, 
\end{equation}
where $\mathbf{\hat{c}}$ is the solution for $\min_\cc \|\V\cc -\f\|_2$ and the expectation $\mathbb{\widetilde{E}}$ is taken over the random choices of the algorithm. When $\V$ is close to having orthonormal columns (as would hold with weighted sampling and/or $N\rightarrow\infty$ with orthonormal basis functions $\phi$), the convergence in~\eqref{eq:REKconvergence} becomes $O((1-\frac{1}{n})^{\frac{M}{2}})$.

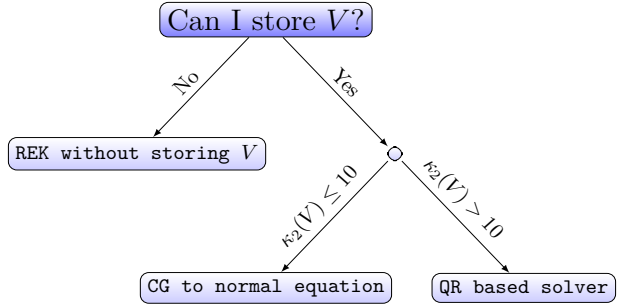
\begin{figure}
\begin{center}
\resizebox{8cm}{4cm}{\begin{tikzpicture}
  [
    grow                    = down,
    sibling distance        = 12em,
    level distance          = 6.5em,
    edge from parent/.style = {draw, -latex},
    sloped
  ]
  \node [root] {Can I store $V$?}
    child { node [env] {REK without storing $V$}
      edge from parent node [above] {\hspace{-0.2 cm}No} }
    child { node [env] {}
      child { node [env] {CG to normal equation}
        edge from parent node [above] {\hspace{-0.1cm}$\kappa_2(V) \leq 10$} }
      child { node [env] {QR based solver}
              edge from parent node [above, align=center]
                {\hspace{-0.3cm}$\kappa_2(V)>10$}
              }
              edge from parent node [above] {Yes} };
\end{tikzpicture}}
\end{center}
\caption{Choice of algorithm to solve the least-squares problem.}\label{scheme-algo}
\end{figure}

Our experiments suggest that conjugate gradients applied to the normal equation is faster than Kaczmarz, so we recommend CG whenever it is feasible. 
However, as mentioned above, an advantage of (extended) Kaczmarz is that there is no need to store the whole matrix to execute the iterations. For these reasons, we suggest to choose the solver for the LS problem \eqref{LS} according to the scheme shown in Figure~\ref{scheme-algo}.
Preliminary numerical experiments indicate that the threshold $10$ for $\kappa_2(\V)$ is a good choice.

\section{Convergence and cost analysis}\label{sec-cost analysis}

In this section we first present convergence results, on which basis we will derive a cost analysis.

\subsection{Convergence}\label{sec-convergence}

First, we obtain a convergence result and consequently asymptotic confidence intervals, applying the central limit theorem (CLT). The following statement and proof is a straightforward generalization of \cite[Theorem 5.1]{nakatsukasa2018approximate} for an arbitrary integrating probability measure $\mu$.

\begin{proposition}\label{thm_convW}
Fix $n$ and the $L^2_{\mu}$-basis functions $ \{\phi_j\}_{j=0}^n$ and let either $w=1$ or $w$ as in \eqref{eq:optsample}. 
Then with the weighted sampling $\frac{d\mu}{w}$, the corresponding MCLS estimator $\hat I_{\mu,N}$, as $N \to \infty$ we have
\begin{equation*}
\sqrt{N} (\hat I_{\mu,N} - I_\mu) \xrightarrow[]{d} \mathcal{N} (0, \min_\mathbf{c} \| \sqrt{w} (f-\sum_{j=0}^n c_j \phi_j)\|^2_\mu),
\end{equation*}
where $\xrightarrow[]{d}$ denotes convergence in distribution.
\begin{proof} The proof is provided in the Appendix.
\end{proof}
\end{proposition}
We observe that MCLS converges like  
$
\frac{\min_\mathbf{c} \| \sqrt{w}(f-\sum_{j=0}^n c_j\phi_j) \|_{\mu}}{\sqrt{N}}$ (when $\{\phi_j\}_{j=1}^n$ is fixed), highlighting the fact that the speed of convergence is still $1/\sqrt{N}$, but with variance reduced from $\min_c \| f-c \|_2$ (standard MC) to $\min_\mathbf{c} \| \sqrt{w} (f-\sum_{j=0}^n c_j \phi_j) \|_2$ (MCLS). In other words, the variance is reduced thanks to the approximation of the function $f$.

The above proposition shows that the MCLS estimator yields an approximate integral $\hat I_{\mu,N}$ that asymptotically (for $N \to \infty$ and $\{\phi_j\}_{j=1}^n$ fixed) satisfies\footnote{We use the notation ``$\approx$'' with the statement ``for $N \to \infty$'' to mean that the relation holds for sufficiently large $N$. E.g. \eqref{eq:error_asympt} means $\mathbb{E}[|\hat I_{\mu,N}-I_\mu|] = \frac{ \min_{\mathbf{c} \in \R^{n+1}} \| \sqrt{w} (f-\sum_{j=0}^n c_j \phi_j)\|_\mu}{\sqrt{N}}+o(\frac{1}{\sqrt{N}})$ for $N \to \infty$.}
\begin{equation}\label{eq:error_asympt}
\mathbb{E}[|\hat I_{\mu,N}-I_\mu|] \approx \frac{ \min_{\mathbf{c} \in \R^{n+1}} \| \sqrt{w} (f-\sum_{j=0}^n c_j \phi_j)\|_\mu}{\sqrt{N}},
\end{equation}
highlighting the fact that the asymptotic error is still $\bigO(1/\sqrt{N})$ (as in the standard MC), but with variance $(\sigma(f))^2$ reduced from $\min_{c \in \R} \| f-c \|_2^2$ (standard MC, see \eqref{error standardmc}) to $\min_{\mathbf{c}\in \R^{n+1}} \| \sqrt{w} (f-\sum_{j=0}^n c_j \phi_j) \|_\mu^2$ (MCLS). In other words, the variance is reduced thanks to the approximation of the function $f$ and the constant in front of the $\bigO(1/\sqrt{N})$ convergence in MCLS is equal to the function approximation error in the $L^2_\mu$ norm.

After solving the least-squares problem \eqref{eq:MCLSweight}, the variance $\min_{\textbf{c}\in \R^{n+1}} \| \sqrt{w} (f-\sum_{j=0}^n c_j \phi_j)\|_\mu^2$ can be estimated via\footnote{This approximation is commonly used in linear regression, see e.g. \cite{hastie2009elements}.}
\begin{equation}\label{sigmaLS}  
\widetilde\sigma_{LS}^2 := 
\frac{1}{N-n-1}\sum_{i=1}^N(w(\tilde \x_i))^2(f(\tilde\x_i)-p(\tilde\x_i))^2
=\frac{1}{N-n-1}\|\W({\V} \hat{\mathbf{c}}-{\f})\|^2_2, 
\end{equation}
where the samples $\tilde\x_i, i=1,\cdots,N$ are taken according to $\frac{d\mu}{w}$. This leads to approximate confidence intervals, for example the $95\%$ confidence interval is approximately given by 
\begin{equation}\label{Cinterval}
\Big [\hat I_{\mu,N} - 1.96\frac{\widetilde\sigma_{LS}}{\sqrt{N}}, \hat I_{\mu,N} - 1.96\frac{\widetilde\sigma_{LS}}{\sqrt{N}} \Big].
\end{equation}
As explained in \cite{nakatsukasa2018approximate}, the MCLS estimator is not unbiased, in the sense that $\E(\hat I_{\mu,N}) \neq I_{\mu}$. However, one can show along the same lines as in the proof of~\cite[Proposition 3.1]{nakatsukasa2018approximate} that with the MCLS estimator $\hat I_{\mu,N}$ with $n$ and $\{\phi_j\}_{j=0}^n$ fixed,
\begin{equation*}
|I_\mu-\E(\hat I_{\mu,N})|=O\bigg(\frac{1}{N}\bigg).
\end{equation*}
 This shows that the bias is of a smaller order than the error.

In the case of weighted sampling, we moreover have a finite sample error bound, which follows directly from \cite[Theorem 2.1 (iv)]{cohen2017optimal}. Note that as this is a non-asymptotic result, it is especially useful in practice. 
\begin{proposition}\label{thm_CM}
Assume that we adopt the weighted sampling $ \frac{d\mu}{w}$.
For any $r > 0$, if $n$ and $N$ are such that $n \leq \kappa \frac{N}{\log(N)}-1$ for $\kappa= \frac{1-\log(2)}{2 + 2r}$, then
\begin{equation}\label{CM-bound}
\mathbb{E}[\|f-\tilde{p}\|_\mu^2] \leq \Big (1+ \frac{4\kappa}{\log(N)} \Big)\min_{\mathbf c}\|f-\sum_{j=0}^nc_j\phi_j\|_\mu^2 +2 \|f\|_\mu^2 N^{-r},
\end{equation}
where $\tilde{p}$ is defined as
\begin{equation*}
\tilde{p}:=
\begin{cases}
p,& \text{if } \|\frac{1}{N}\V^T\V-\mathbf{I}\|_2\leq \frac{1}{2}\\
0, & \text{otherwise},
\end{cases}
\end{equation*}
with $p = \sum_{j=0}^n \hat{c}_j \phi_j$, for $\hat{\mathbf{c}}$ being the solution of \eqref{eq:MCLSweight}. Note that the simulation is done with respect to $\frac{d\mu}{w}$.
\end{proposition}
We note the slight difference between $p$ and $\tilde{p}$; this is introduced to deal with the tail case in which $\V$ becomes ill-conditioned (which happens with low probability). This is used for a theoretical purpose, but in practice, this modification is not necessary and we do not employ it in our experiments. 

Proposition~\ref{thm_CM} allows us to define a non-asymptotic,  proper bound for the expected error we commit when estimating the vector $\mathbf{c}^\ast$, solving the LS problem \eqref{eq:MCLSweight}. To see this, we first decompose the function $f$ into a sum of orthogonal terms
\begin{equation}\label{fdecomp}
f=\sum_{j=0}^n c^\ast_j \phi_j + g=: f_1+g,
\end{equation}
for some coefficients $c^\ast_j$, $j=0,\cdots,n$ and where $g$ satisfies $\int_E g(\x) \phi_j(\x) d\mu(\x)=0$ for all $j=0,\cdots, n$. Note that $\|g\|_\mu=\min_{\mathbf c}\|f-\sum_{j=0}^nc_j\phi_j\|_\mu$. Then,
\begin{align*}
\mathbb{E}[\|f- \sum_{j=0}^n \hat{c}_j \phi_j\|_\mu^2]&=
\mathbb{E}[\|f -f_1- \sum_{j=0}^n \hat{c}_j \phi_j+f_1\|_\mu^2] \\
&= \|f -f_1\|^2_\mu+ \mathbb{E}[\|\mathbf{c}^\ast - \hat{\mathbf{c}}\|^2_2]=\|g\|_\mu^2 + \mathbb{E}[\|\mathbf{c}^\ast - \hat{\mathbf{c}}\|^2_2].
\end{align*}
This, together with the bound \eqref{CM-bound} yields 
\begin{equation}\label{eq:boundvec}
\mathbb{E}[\|\mathbf{c}^\ast - \hat{\mathbf{c}}\|^2_2]\leq 
 \frac{4\kappa}{\log(N)}\min_{\mathbf{c}} \| f-\sum_{j=0}^n c_j \phi_j\|_\mu^2 +2 \|f\|_\mu^2 N^{-r}.
\end{equation}
When we are primarily interested in integration, we aim at an upper bound for the expected error of the first component of $\mathbf{c}$. The bound \eqref{eq:boundvec} clearly holds for the first component and this gives us a bound for $\mathbb{E}(|\hat I_{\mu,N}-I_\mu|^2)$. 
Intuitively, we expect that the error in the elements of $\mathbf{c}$ are not concentrated in any of the components. This suggests a heuristic bound 
\begin{equation}\label{heuristic-bound}
\mathbb{E}[|\hat I_{\mu,N}-I_\mu|^2] \lessapprox 
\frac{1}{n}\left( \frac{4\kappa}{\log(N)}\min_{\mathbf{c}} \| f-\sum_{j=0}^n c_j \phi_j\|_\mu^2 +2 \|f\|_\mu^2 N^{-r}\right).
\end{equation}
This argument has already been proposed in \cite{nakatsukasa2018approximate}. A rigorous argument still remains an open problem.
Observing that the first term of the right hand side is the dominant one (for $N \to \infty$) and assuming $n \approx \frac{N}{\log(N)}$, we can see that the heuristic bound \eqref{heuristic-bound} matches the asymptotic result derived in Proposition \ref{thm_convW}. 

\subsection{Cost Analysis}
The purpose of this section is to reveal the relationship between the error vs. cost (in flops).
The cost of MCLS is analyzed in \cite{nakatsukasa2018approximate} and in Table \ref{tab:compare} we report a cost and error comparison between MC and MCLS as given in Table 3.1 in \cite{nakatsukasa2018approximate}. Here, we highlight some cases for which MCLS outperforms MC in terms of accuracy or cost.

\begin{remark}\label{remark: cost MCLS}
Note that the cost of MCLS in Table \ref{tab:compare} is reported to be $C_fN+\bigO(Nn)$. As already mentioned at the beginning of Section \ref{sec-kaczmarz}, this reflects the cost of MCLS when applying the CG algorithm to solve the least-squares problem (whenever $\kappa_2(\V)=\bigO(1)$). In the case that we combine MCLS with the REK algorithm and $\kappa_2(\V)=\bigO(1)$, which happens with high probability whenever the weighted sampling strategy is used (see \cite[Theorem 2.1]{cohen2017optimal}), the cost is also $C_fN+\bigO(Nn)$, as shown in \cite[Lemma 9]{zouzias2013randomized} and in the subsequent discussion. The following cost analysis includes therefore the two options CG and REK.
\end{remark}

\begin{table}[t]
  \centering
  \begin{tabular}{lcc}
& Cost & Convergence \\ \hline  \rule{0pt}{1.5\normalbaselineskip}
MC     &  
$C_fN$ & $\displaystyle\frac{1}{\sqrt{N}}\min_{c}\|f-c\|_\mu$\\
MCLS & $C_fN+O(Nn)$ & $\displaystyle\frac{1}{\sqrt{N}}\min_{\mathbf{c}}\| \sqrt{w}(f-\sum_{j=0}^nc_j\phi_j)\|_\mu$\\
 \end{tabular}
  \caption{Comparison between MC and MCLS. $N$ is the number of sample points and $C_f$ denotes the cost for evaluating $f$ at a single point. 
  \label{tab:compare}}
\end{table}

First, consider the situation of a limited budget of sample points $N$ that can not be increased further, and the goal is to approximate the integral $I_\mu$ in the best possible way. This is a typical task in financial institutions. For instance, in portfolio risk management, simulation can be extremely expensive because a large number of risk factors and positions contribute to the company's portfolio. In this case even if MCLS is more expensive than MC (second column of Table \ref{tab:compare}), MCLS is preferable to MC as it yields a more accurate approximation (third column of Table \ref{tab:compare}).
 
Second, we show under mild conditions that MCLS also asymptotically becomes more accurate than MC at the same cost. This can be of practical relevance whenever the integral $I_\mu$ needs to be computed at a very high accuracy 
and one is able to spend a high computational cost. 
Let us fix some notation:
\begin{align*}
&e_n:= \min_{\mathbf{c}} \|\sqrt{w} (f-\sum_{j=0}^n c_j \phi_j)\|_\mu \enskip \text{for } n \geq 0,\\
&\text{Cost}_{MC}(N):=C_f N,\\
&\text{Cost}_{MCLS}(N',n):=C_f N' + C_M N'n \enskip \text{for some } C_M>0,\\
& \text{error}_{MC}(N):= \frac{e_0}{\sqrt{N}},\\
& \text{error}_{MCLS}(N',n):= \frac{e_n}{\sqrt{N'}},
\end{align*}
where the last two definitions reflect the asymptotic error behaviour for large $N$ and $N'$ (for a fixed $n$), depicted in Table \ref{tab:compare}. We are now in the position to present the result.
\begin{proposition}\label{prop:asymptotic}
Assume that $e_n =o \big(\frac{1}{\sqrt{n}}\big)$. Then there exists $\tilde{n} \in \N$ such that for any fixed $n>\tilde{n}$, $\text{error}_{MCLS}<\text{error}_{MC}$ as $\text{Cost}_{MCLS}=\text{Cost}_{MC} \to \infty$.
\begin{proof}
We first determine the value of $N=N(N',n)$ such that $\text{Cost}_{MCLS}=\text{Cost}_{MC}$:
\begin{align*}
\text{Cost}_{MCLS}=\text{Cost}_{MC} \iff N=N' \big ( 1+\frac{C_M}{C_f} n\big ).
\end{align*}
Consider now the error ratio under the constraint $\text{Cost}_{MCLS}=\text{Cost}_{MC}$, given by
\begin{equation*}
ER:= \frac{\text{error}_{MC}}{\text{error}_{MCLS}}=\frac{e_0}{e_n \sqrt{1+\frac{C_M}{C_f}n}},
\end{equation*}
yielding
\begin{equation*}
ER>1 \iff e_n \sqrt{1+\frac{C_M}{C_f}n} < e_0.
\end{equation*}
The assumption $e_n = o\big(\frac{1}{\sqrt{n}}\big)$ implies that there exists some $\tilde{n}$ such that $ER>1$ for all $n>\tilde{n}$. Now, fixing an arbitrary $n>\tilde{n}$ and letting $N'$ and consequently $N$ going to infinity yields the result. 
\end{proof}
\end{proposition}
\begin{remark}
Note that the quantity $ER$ in the proof of Proposition \ref{prop:asymptotic} only reflects the error ratio asymptotically for $N,N'\to\infty$. Therefore we restrict the statement of the result to the asymptotic case where $\text{Cost}_{MCLS}=\text{Cost}_{MC} \to \infty$.
\end{remark}

\begin{figure}[ht]
\centering
  \begin{minipage}[t]{0.49\hsize}
\includegraphics[width=1.0\textwidth]{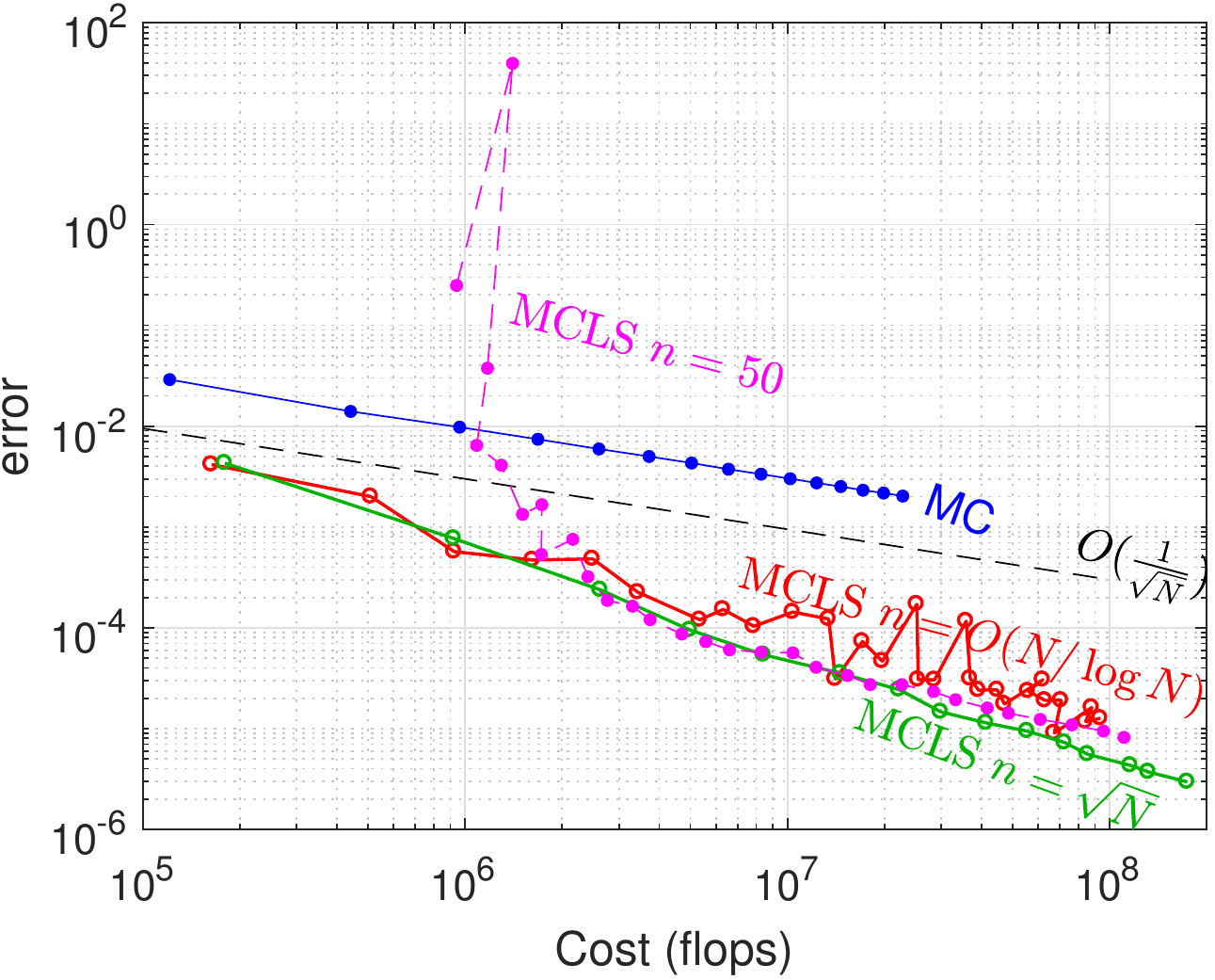}
  \end{minipage}
  \begin{minipage}[t]{0.5\hsize}
\includegraphics[width=1.0\textwidth]{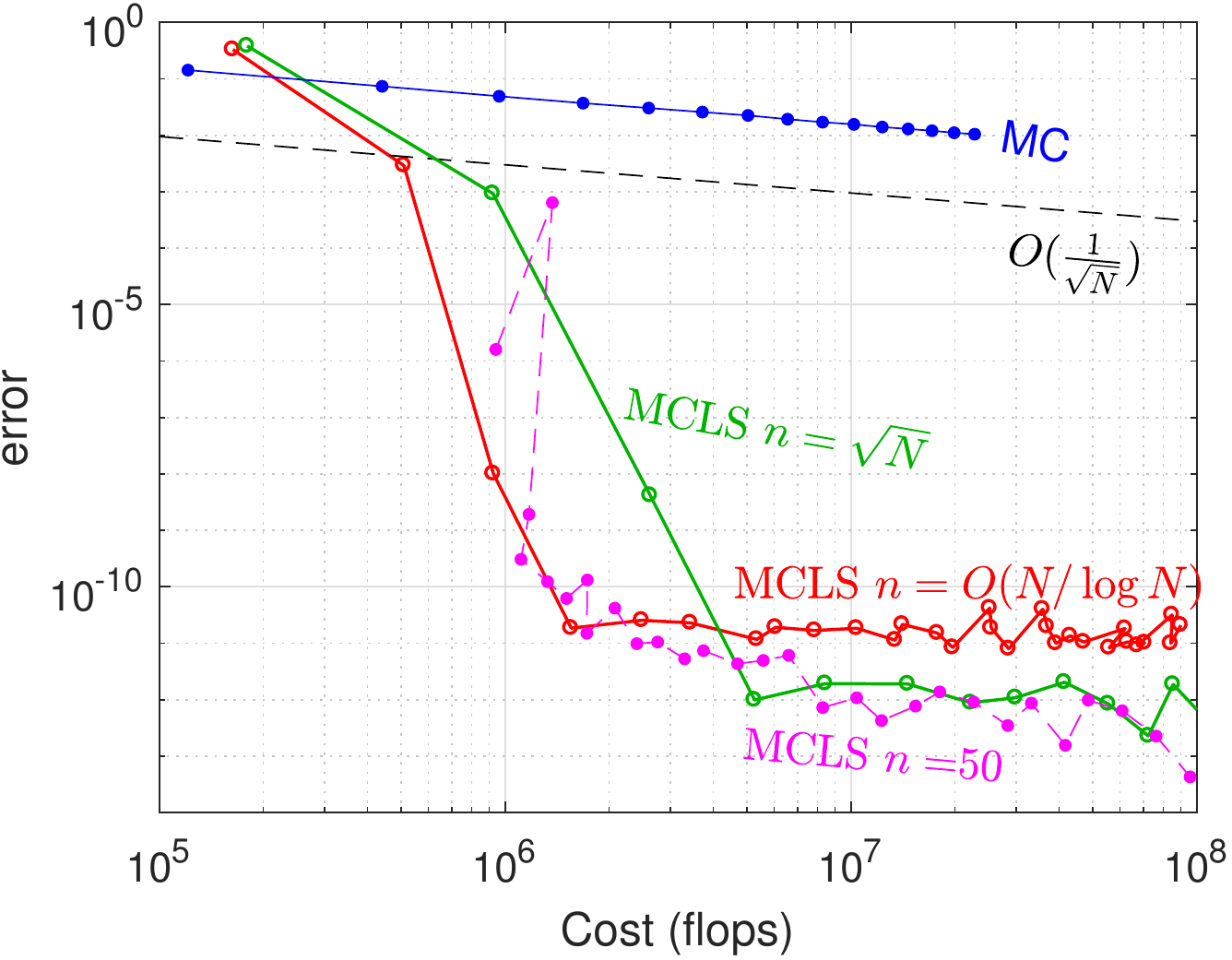}
  \end{minipage}   

\caption{Cost vs Convergence plots for MC and MCLS with varying $n$: $n=50$, $n=\sqrt{N}$ and $n= N/\log N$ , for $d=1$. Cost is computed as $2N(n+1)k$, the flop counts in the CG iteration, where $k$ is the number of CG steps required.
Left: Non-smooth function $f(x)=|x-\frac{1}{2}|$.  Right: analytic function $f(x) = \sin(30x)$. 
}
\label{fig:costd=1}
\end{figure}

\begin{figure}[ht]
\centering
  \begin{minipage}[t]{0.49\hsize}
\includegraphics[width=1.0\textwidth]{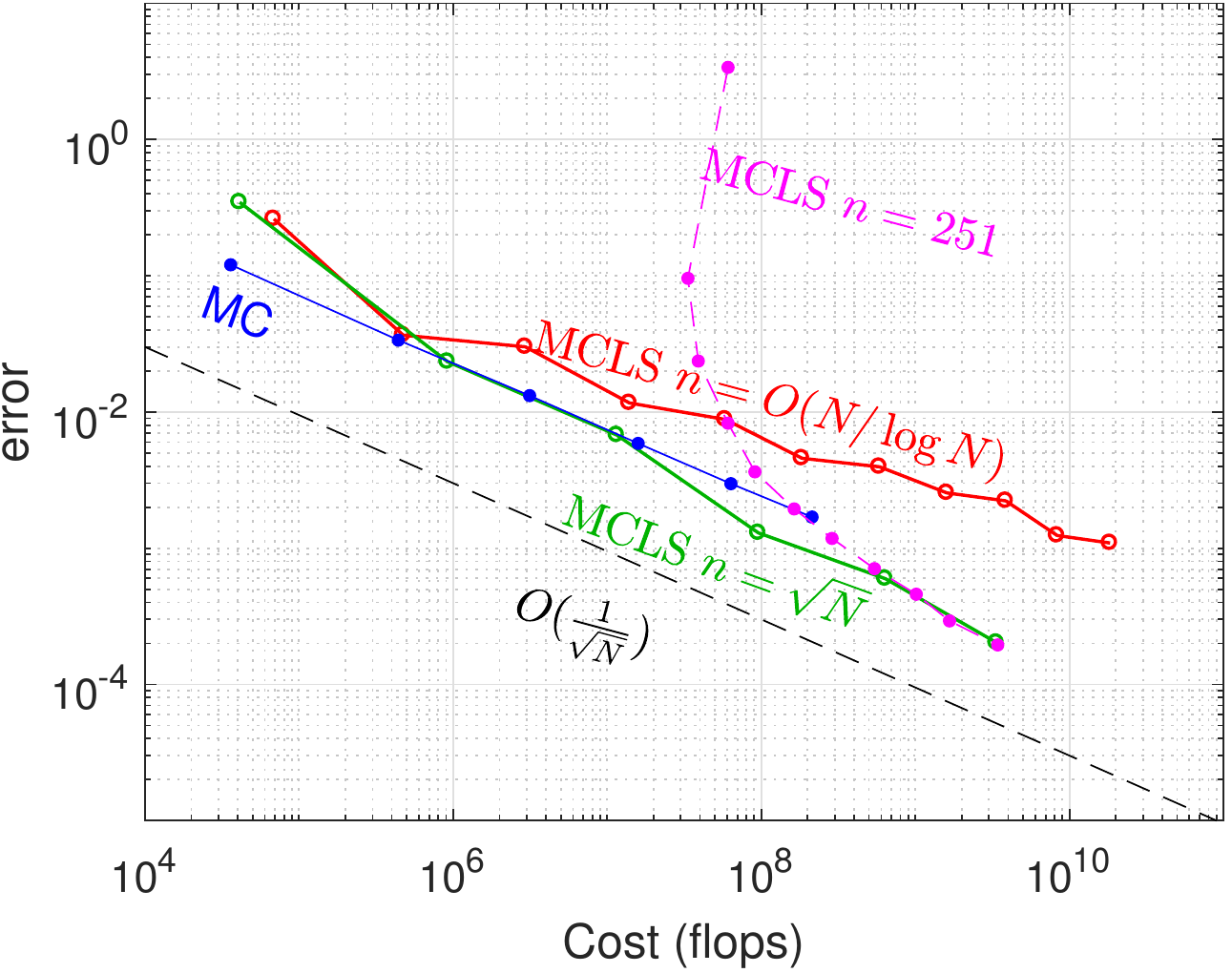}
  \end{minipage}
  \begin{minipage}[t]{0.5\hsize}
\includegraphics[width=1.06\textwidth]{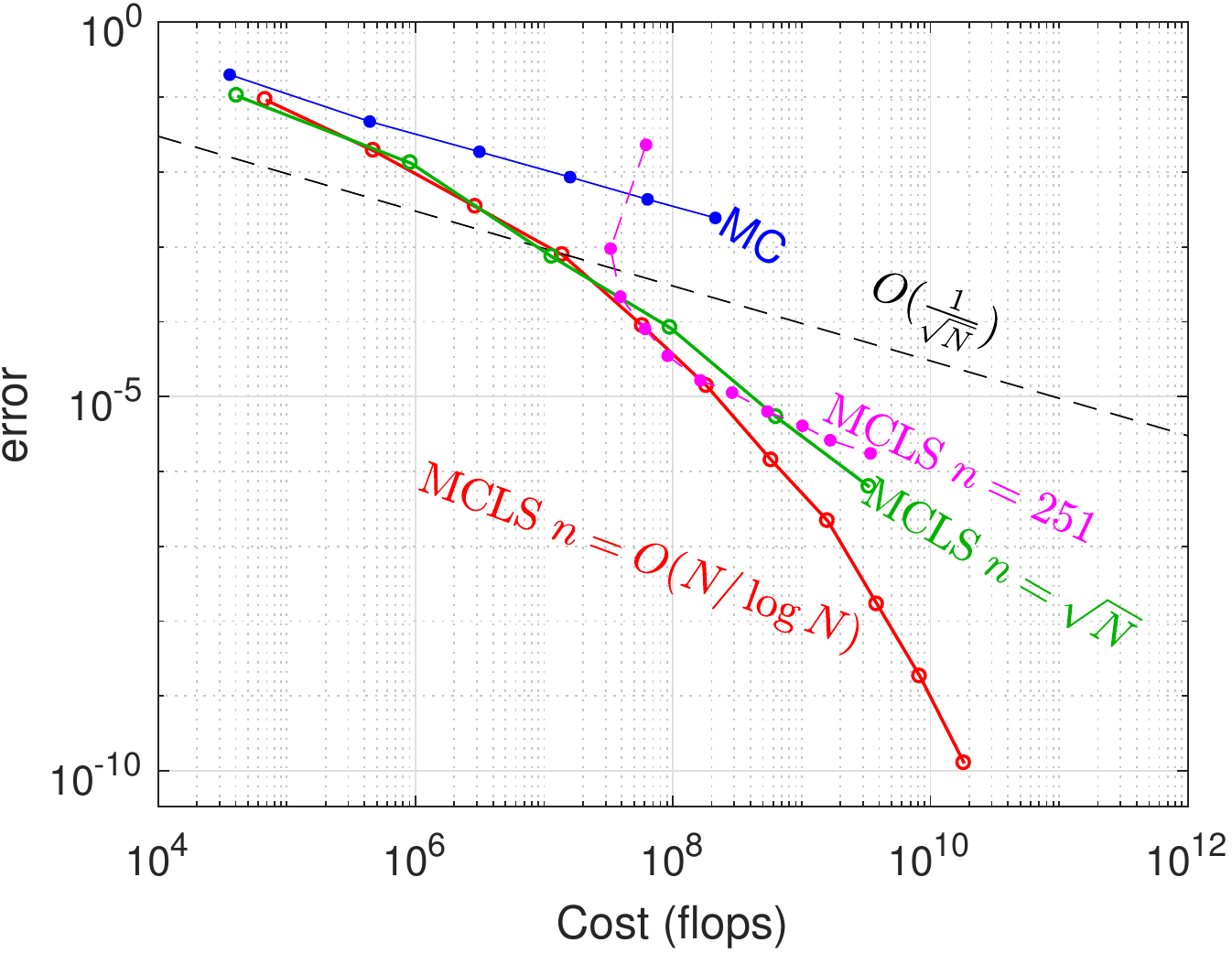}
  \end{minipage}   
\caption{Same as in~\Cref{fig:costd=1}, but with $d=5$. The fixed value $n=251$ comes from 
$n={d+k \choose k}-1$ 
for degree $k=5$. 
Left: non-smooth function $f(x)=\sum_{i=1}^d\mbox{exp}(-|x-\frac{1}{2}|)$. Right: analytic function $f(x) = \sin(\sum_{i=1}^d x_i)$. 
}
\label{fig:costd=5}
\end{figure}

To show the practical implication of this asymptotic analysis, in Figures~\ref{fig:costd=1} and~\ref{fig:costd=5} we examine the convergence of MC and MCLS. We consider the problem of integrating smooth and non-smooth functions for several dimensions $d$, on the unit cube $[0,1]^d$ and with respect to the Lebesgue measure. Even though the result of Proposition \ref{prop:asymptotic} holds for a fixed value of $n$, in practice the convergence rate can be improved by varying $n$ together with $N$, as illustrated in \cite{nakatsukasa2018approximate}, where such an adaptive strategy is denoted by MCLSA. For this reason, we show numerical results where we let the cost increase (represented on the x-axis) and for different choices of $n$ ($n$ fixed and $n$ varying)~\footnote{These figures differ from those in~\cite{nakatsukasa2018approximate} in that the $x$-axis is the cost rather than the number of sample points $N$.}.

As expected, the numerical results reflect our analysis presented above. 
For all dimensions and chosen functions, we achieve an efficiency gain by an appropriate choice of $n$ and $N$, asymptotically. Note that the erratic convergence with fixed $n$ is a consequence of ill-conditioning; an effect described also in~\cite{nakatsukasa2018approximate}. Namely, when the number of sample points $N$ is not enough, $\V$ tends to be ill-conditioned and 
the least-squares problem $\min_\cc \|\V\cc -\f\|_2$ requires many CG iterations, resulting in higher cost than with a larger $N$.
Therefore, the function ``N $\mapsto$ Cost(N)'' is not necessarily monotonically increasing in $N$. We observe that some of the curves in Figures~\ref{fig:costd=1} and~\ref{fig:costd=5}, for instance for $n$ fixed, are not functions of Cost(N), as they are not always single-valued. This shows that indeed the mapping ``N $\mapsto$ Cost(N)'' is not always monotone.

\section{Application: European option pricing}\label{sec-MCLS option pricing}

The option pricing problem is one of the main tasks in financial mathematics and can be summarized as follows. 
First, we fix a filtered probability space $(\Omega, \mathcal{F}, \mathcal{F}_t, \mathbb{Q})$, where $\mathbb{Q}$ denotes a risk neutral pricing measure. In this framework a stochastic process $(X_t)_{0 \leq t \leq T}$ defined on a time horizon $[0,T]$ for $T>0$ and taking values in a state space $E \subseteq \R^d$ is used to model the price of the financial assets. Then, the price at time $t=0$ of a European option with payoff function $f:E \to \R$ and maturing at time $T$ is given by
\begin{equation}\label{price}
e^{-rT}\mathbb{E}[f(X_T)]=e^{-rT}\int_E f(\x) d\mu(\x),
\end{equation}
where $r$ is a risk-free interest rate and $\mu$ denotes the distribution of $X_T$ whose support is assumed to be $E$ and $f \in L^1(\mu)$.
\subsection{MCLS for European option pricing}
In this section we explain how to apply MCLS to compute European option prices.
When applying MCLS for computing \eqref{price} we observe  two potential issues. First, the distribution $\mu$ often is not known explicitly. Therefore, we can not directly perform the sampling part, namely the first step of MCLS, as described in Algorithm \ref{Algo1}. Second, it is crucial that the basis functions $\{\phi_j \}_{j=0}^n$ are easily integrable with respect to $\mu$. Therefore we need to find an appropriate selection of the basis functions.

Concerning the sampling part, if $\mu$ is explicitly known, as for example in the Black and Scholes framework (see Section \ref{BS example} and Section \ref{BS example2} for two examples), we can just generate sample points according to $\mu$. If $\mu$ is not explicitly known, typically the process $(X_t)_{0 \leq t \leq T}$ can still be expressed as the solution of a stochastic differential equation (SDE). In this case, we propose to simulate $N$ paths of $X_t$ by discretizing its governing SDE and collect the realizations of $X_T$. More details follow below and an example can be found in Section \ref{JH example}.

To obtain an appropriate choice of the basis functions $\{\phi_j\}_{j=0}^n$ we need $\E[\phi_j(X_T)]$ to be easy to evaluate. To do so we exploit the structure of the underlying asset model.
If $X_t$ belongs to the wide class of affine processes, which is true for a large set of popular models including Black and Scholes, Heston and Levy models, then the characteristic function of $X_t$ can be easily computed, as explained e.g. in \cite{duffie2003affine}. Therefore, the natural choice of basis functions is to choose exponentials. If $X_t$ is a polynomial diffusion \cite{filipovic2016polynomial} (as in our numerical examples in Section \ref{JH example}) or a  polynomial jump-diffusion \cite{filipovic2017polynomial}, then its conditional moments are given in closed form. Therefore, polynomials are an excellent choice of basis functions.

To summarize, the main steps of MCLS for option pricing are as follows (if $\mu$ is not known explicitly):
\begin{enumerate}
\item Simulate $N$ paths of the process $X_t$, from $t=0$ to $t=T$ (time to maturity), by discretization of the governing SDE.
\item Let $\x_i$ for $i=1, \ldots, N$ be the realizations of  $X_T$ for each simulated path. Then, we evaluate $f(\x_i)$ and $\phi_j(\x_i)$, for $i=1,\ldots,N$ and $j=1,\ldots,n$.
\item Solve the least-squares problem \eqref{LS} to obtain the approximation of $f$. The solver can be chosen according to the scheme represented in Figure~\ref{scheme-algo}.
\item Finally, the option price is approximated by (we omit the discounting factor)
\begin{equation*}
\E[f(X_T)] = \int_{E} f(x) d \mu(x) \approx \hat I_{\mu,N}:=\sum_{j=0}^n c_j \int_{E} \phi_j(x) d \mu(x) = \sum_{j=0}^n c_j \E[\phi_j(X_T)].
\end{equation*}
Note that we selected the basis functions in such away that   the quantities $\E[\phi_j(X_T)]$ can be easily evaluated. In particular, no Monte Carlo simulation is required.
\end{enumerate}
Algorithm \ref{mainAlgo} summarizes this procedure.

\begin{algorithm}[t]
\caption{Generalized MCLS for European option pricing}
\begin{algorithmic}[1]\label{mainAlgo}
\REQUIRE{Payoff function $f$, basis functions $\{\phi_j\}_{j=0}^n$, $\phi_0\equiv 1$, integer $N (>n)$, governing SDE of $X_T$.}
\ENSURE{Approximate option price $\hat I_{\mu,N}\approx \int_E f(\x)d\mu(x)$}
\STATE Simulate $N$ paths of the process $X_t$ from $t=0$ to $t=T$, and collect the realizations of $X_T$ in $\x_i$, $i=1,\ldots,N$.
\STATE Evaluate $f(\x_i)$ and $\phi_j(\x_i)$, for $i=1,\ldots,N$ and $j=1,\ldots,n$.
\STATE Solve the least-squares problem \eqref{LS} for $\mathbf{c}=[c_0,c_1,\ldots,c_n]^T$.
\STATE Compute $\hat I_{\mu,N}=\sum_{j=0}^n c_j\int_{E} \phi_j(\x)d\mu(x)$.
\end{algorithmic}
\end{algorithm}
In the case that $\mu$ is explicitly known, the error resulting from MCLS is analysed in Proposition \ref{thm_convW} and Proposition \ref{thm_CM}. In case we discretize the governing SDE of $X_t$, we introduce a second source of error, which we address in the following.

Assume that $X_t$ is the solution of an SDE of the form 
\begin{align}
\label{SDE}
\begin{split}
dX_t&=b(X_t)dt+\Sigma(X_t)dW_t,\\
X_0&=x_0,
\end{split}
\end{align}
where $W_t$ denotes a $d$-dimensional Brownian motion, $b : \mathbb{R}^d \mapsto \mathbb{R}^{d}$, $\Sigma : \mathbb{R}^d \mapsto \mathbb{R}^{d \times d}$. and $x_0\in \R^d$. An approximation of the solution $X_t$ of \eqref{SDE} can be computed via a uniform Euler-Maruyama scheme, defined in the following.
\begin{definition}\label{euler-maruyama}
Consider an equidistant partition of $[0,T]$ in $N_s$ intervals, i.e.  
\begin{equation*}
\Delta t= T/N_s, \quad t_i=i \Delta t \quad \text{for } i=0,\cdots, N_s,
\end{equation*}
together with
\begin{equation*}
\Delta \widetilde{W}_i = W_{t_{i+1}}-W_{t_i} \quad \text{for } i=0,\cdots, N_s.
\end{equation*}
Then, the Euler-Maruyama discretization scheme of \eqref{SDE} is given by
\begin{align}\label{dSDE}
\begin{split}
\bar{X}_{i+1}&=\bar{X_i}+b(\bar{X_i})\Delta t+\Sigma(\bar{X_i}) \Delta \widetilde{W}_i, \quad \text{for } i=0,\cdots, N_s-1,\\
\bar{X_0}&=x_0,
\end{split}
\end{align}
and the Euler-Maruyama approximation of $X_T$ is given by $\bar{X}_{N_s}$. 
\end{definition}
Assume that we sample $N$ independent copies of $\bar{X}_{N_s}$ (first step of Algorithm \ref{mainAlgo}) and we apply MCLS to approximate \eqref{price}. Then the error naturally splits into two components as 
\begin{align*}
| \E[f(X_T)]-\bar{I}_{\mu,N}| \leq |\E[f(X_T)]-\E[f(\bar{X}_{N_s})]|+|\E[f(\bar{X}_{N_s})]-\bar{I}_{\mu,N}|.
\end{align*}
The second summand can then be approximated as in \eqref{eq:error_asympt}. We collect the result in the following proposition. Note that for simplicity we assume a vanishing interest rate.
\begin{proposition}\label{prop:disc error}
Let $\bar{I}_{\mu,N}$ be the MCLS estimator obtained by sampling according to the Euler-Maruyama scheme as in Definition \ref{euler-maruyama}. Then, the MCLS error asymptotically (for $n$ fixed and $N \to \infty$) satisfies
\begin{align}\label{eq:asymptotic_error_Euler}
| \E[f(X_T)]-\bar{I}_{\mu,N}|  \lessapprox   |\E[f(X_T)]-\E[f(\bar{X}_{N_s})]|+\frac{ \min_{\textbf{c}} \| f-\sum_{j=0}^n c_j \phi_j\|_{\bar{\mu}}}{\sqrt{N}},
\end{align}
where $\bar{\mu}$ is the distribution of $\bar{X}_{N_s}$.
\end{proposition}
The first term in the right-hand-side of \eqref{eq:asymptotic_error_Euler} is usually referred to as \textsl{time-discretization error}, while the second summand denotes the so-called \textsl{statistical error}.
The time-discretization error and, more generally, the Euler-Maruyama scheme together with its properties, are well studied in the literature, see e.g.~\cite{kloeden1992numerical}. Depending on the regularity properties\footnote{For example, if $b$ and $\Sigma$ are four times continuously differentiable and $f$ is continuous and bounded, then the scheme converges weakly with order $1$. See \cite{kloeden1992numerical} for details.} of $f, b$ and $\Sigma$, one can conclude, for example, that the time-discretization error is bounded from above by $C |\Delta t|$, for a constant $C>0$. In this case, we say that the Euler-Maruyama scheme converges \textsl{weakly} with order~$1$.
Finally, note that the statistical error can be further approximated as in \eqref{sigmaLS} using
\begin{equation*}
\min_c \| f-\sum_{j=0}^n c_j \phi_j\|_{\bar{\mu}}\approx \frac{1}{N-n-1}\sum_{i=1}^N(f(\x_i)-p(\x_i))^2=\frac{1}{N-n-1}\|\V \mathbf{c}-\f\|^2_2,
\end{equation*}
where the $\x_i$'s are sampled according to $\bar{\mu}$.
\subsection{Numerical examples for option pricing in polynomial models}
Next, we apply MCLS to numerically compute European option prices \eqref{price} for several types of payoff functions $f$ and in different models. In particular, the considered models belong to the class of polynomial diffusion models, introduced in \cite{filipovic2016polynomial}. All algorithms have been implemented in {\sc Matlab} version 2017a and run on a standard laptop (Intel Core i7, 2 cores, 256kB/4MB L2/L3 cache).

In all of our numerical experiments the solver for numerical solution of the least-squares problem \eqref{LS} is chosen according to the scheme in Figure \ref{scheme-algo}. 
The choice of the examples lead us to test all of the three choices in the scheme. For the univariate pricing examples in Heston's and the Jacobi model, Section \ref{JH example} the CG algorithm is appropriate. In Section \ref{BS example}, a basket option price of medium dimensionality in the multivariate Black-Scholes model, a QR based method is employed, because the condition number of $\V$ was usually larger than $O(1)$. In these both cases we directly sample from the distribution of the underlying random variable $X_T$, where in the univariate case we solve an SDE. Finally, we consider pricing a rainbow option in a high dimensional multivariate Black-Scholes model in Section \ref{BS example2}, where the randomized extended Kazcmarz algorithm combined with the weighted sampling strategy yields a good performance.

\subsubsection{Call option in stochastic volatility models}\label{JH example}
We consider the Heston model as in \cite{heston1993closed}. The log asset price $X_t$ (meaning that the asset price $S_t$ is of the form $S_t=e^{X_t}$) and the squared volatility process $V_t$ are defined via the SDE
\begin{align*}
&dV_t=\kappa(\theta -V_t)dt+\sigma \sqrt{V_t}dW^1_{t},\\
&dX_t=(r-V_t/2)d_t + \rho \sqrt{V_t} dW^1_{t}+\sqrt{V_t}\sqrt{1-\rho^2}dW^2_{t},
\end{align*}
where $W^1_{t}$ and $W^2_{t}$ are independent standard Brownian motions and the model parameters satisfy the conditions $\kappa \geq 0$, $\theta \geq 0$, $\sigma >0$, $r \geq 0$, $\rho \in [-1,1]$. The state space is $E = \R_+ \times \mathbb{R}$.
The log-asset process in the Heston model is a polynomial diffusion and its moments can be computed according to the moment formula introduced in \cite[Theorem 3.1]{filipovic2016polynomial}. In this case the formula is given by
\begin{equation}\label{moment-formula}
\E[p(X_T,V_T)|\mathcal{F}_t]=H_n(X_t,V_t)e^{G_n(T-t)}\vec{p},
\end{equation}
where $p$ is an arbitrary multivariate polynomial belonging to the space $\text{Pol}_n(\mathbb{R}^2)$ of bivariate polynomials of total maximal degree smaller than $n$, $H_n$ is a basis vector of $\text{Pol}_n(\mathbb{R}^2)$ and $\vec{p}$ is the coordinate vector of $p$ with respect to $H_n$. Finally, $G_n$ is the matrix representation of the action of the generator of $(V_t,X_t)$ restricted to the space $\text{Pol}_n(\mathbb{R}^2)$. 
Note that the matrix $G_n$ can be constructed as explained in \cite{kressner2017incremental}, with respect to the monomial basis.

In the following we apply MCLS in the Heston model in order to price single-asset European call options with payoff function given by 
\begin{equation*}
f(x)=(e^x-e^k)^+,
\end{equation*}
for a log-strike value $k$. We compare MC and MCLS to the Fourier pricing method introduced in \cite{heston1993closed}.

In this experiment we use an ONB (with respect to the corresponding $L^2_\mu$ space, where $\mu$ is the distribution of $X_T$) of polynomials as basis functions $\phi_j$. Conveniently the ONB can be obtained by applying the Gram-Schmidt orthogonalization process to the monomial basis.  Note that, even if the distribution $\mu$ is not known explicitly, we still can apply the Gram-Schmidt orthogonalization procedure since the corresponding scalar product and the induced norm can be computed via the moment formula \eqref{moment-formula}. 

Since the distribution of $X_T$ is not known a priori, we apply the Euler-Maruyama scheme as defined in \eqref{dSDE} and obtain
\begin{align}\label{EM Heston}
\begin{split}
V_{0}&=v_0,\\
X_{0}&=x_0,\\
V_{t_i}&= V_{t_{i-1}}+\kappa(\theta -V_{t_{i-1}})\Delta t+\sigma \sqrt{V_{t_{i-1}}} \sqrt{\Delta t} Z^1_i, \\
X_{t_i}&= X_{t_{i-1}}+(r -V_{t_{i-1}}/2)\Delta t+\rho \sqrt{V_{t_{i-1}}} \sqrt{\Delta t} Z^1_i + 
\sqrt{V_{t_{i-1}}}\sqrt{1-\rho^2V_{t_{i-1}}} \sqrt{\Delta t} Z^2_i,
\end{split}
\end{align}
for all $\quad i=1,\cdots, N_s$ and where $Z^1_i$ and $Z^2_i$ are independent standard normal distributed random variables.
For the following numerical experiments we consider the set of model parameters 
\begin{align*}
 \sigma=0.15,\enskip  v_0=0.04,\enskip x_0=0, \enskip \kappa = 0.5, \enskip \theta=0.01, \enskip \rho = -0.5, \enskip r = 0.01.
\end{align*}
As long as the square roots in \eqref{EM Heston} are positive the Euler-Maruyama scheme is well-defined. In our numerical experiments this was the case. 
To guarantee well-definedness, the scheme can be modified by taking the absolute value or the positive part of the arguments of the square roots. Such a modification is discussed, e.g., in \cite{kloeden2013converence}. The same remark holds for the forthcoming numerical examples.

\begin{figure}[ht]
\centering
\includegraphics[width=0.49\textwidth]{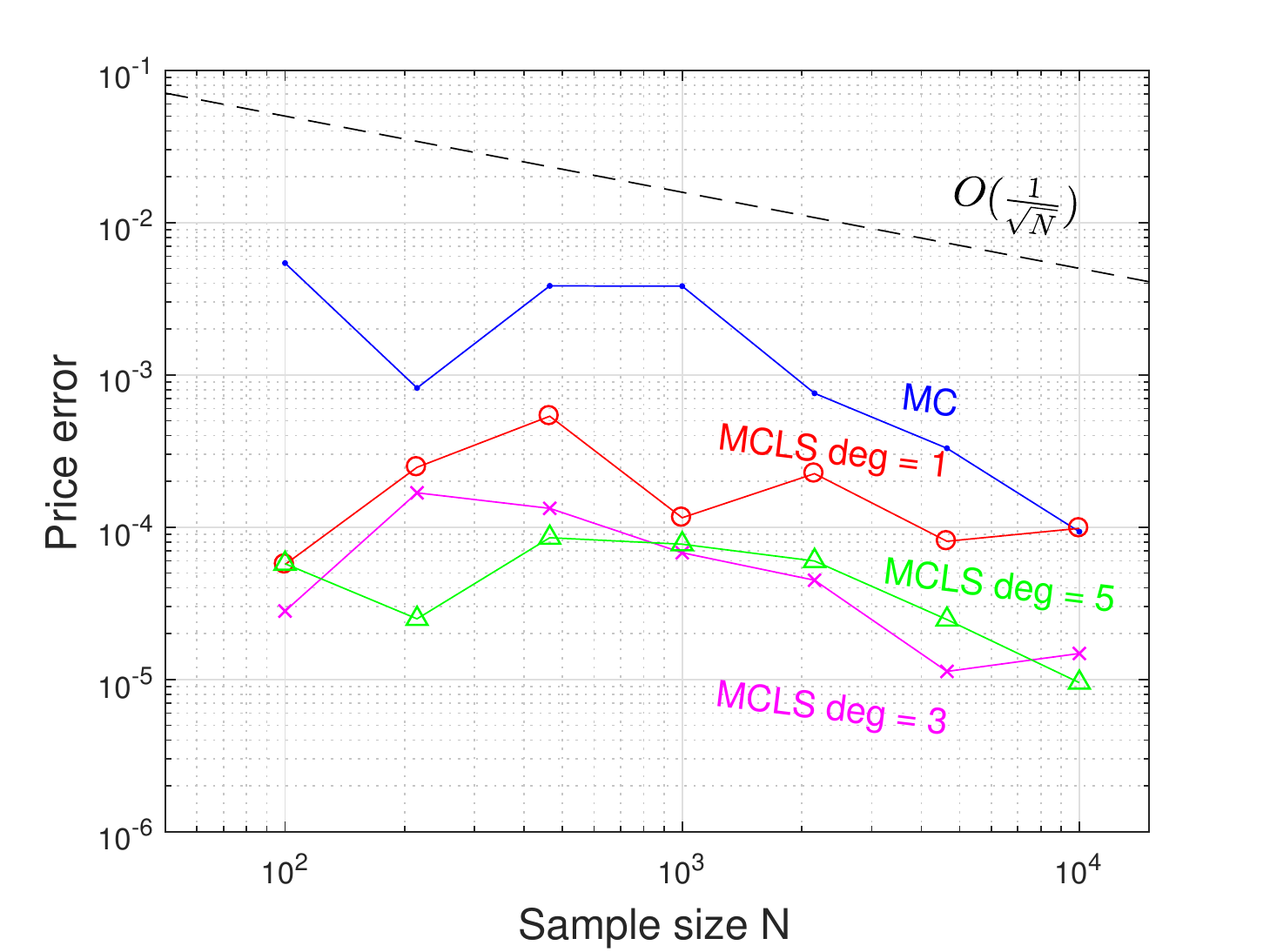}
\includegraphics[width=0.49\textwidth]{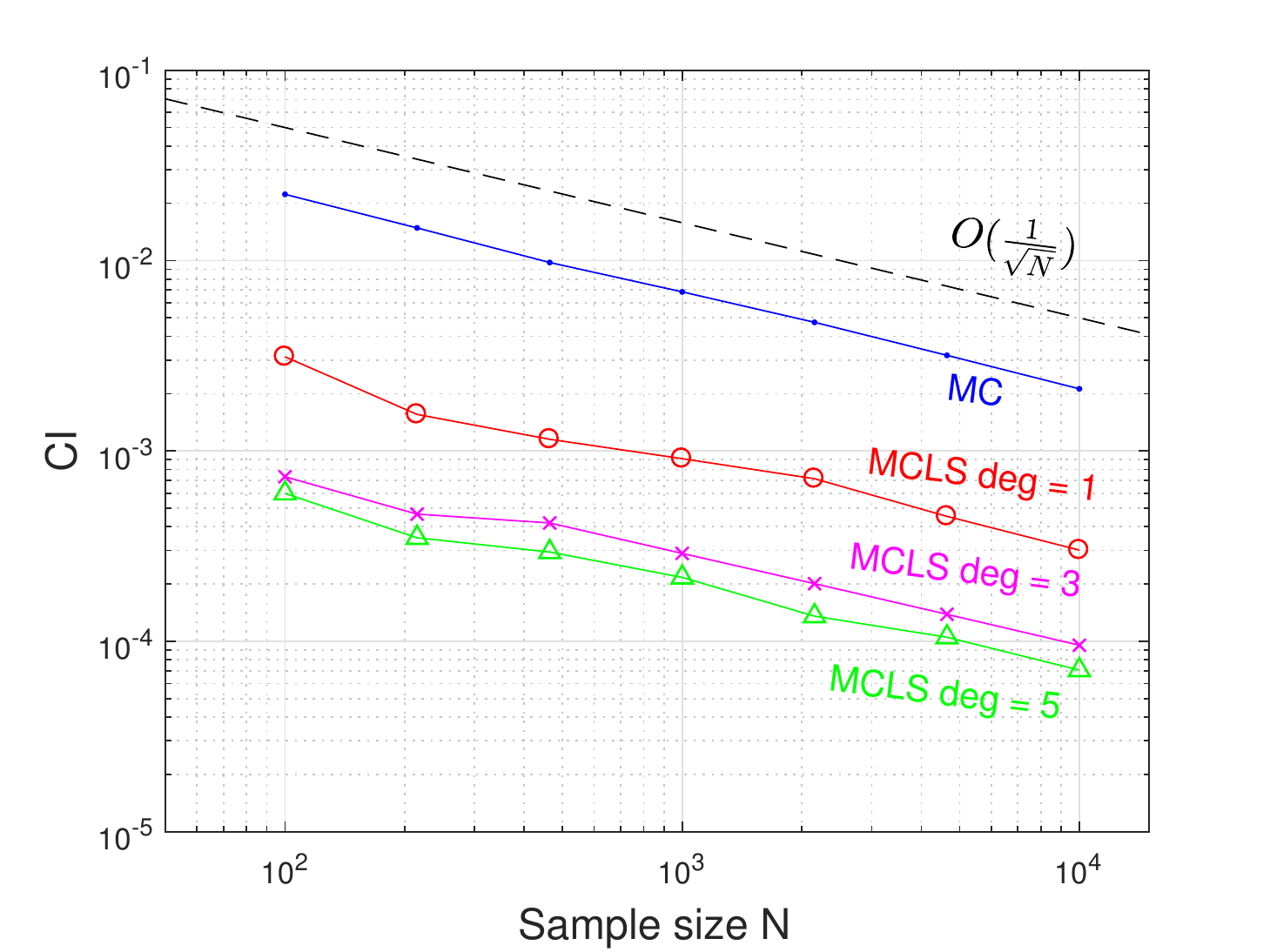}
\caption{MCLS for ITM call option in Heston model for different polynomial degrees. Left: Absolute price error. Right: Width of $95\%$ confidence interval. \label{ITM_Heston}}
\end{figure}

First, we apply MCLS to an in-the-money example, with payoff parameters
\[ k=-0.1, \quad T=1/12, \]
and we use $N_s=100$ time steps for the discretization of the SDE. 
We use an ONB consisting of polynomials of maximal degrees $0$ (standard MC), $1,3$ and $5$ and we obtain the results shown in Figure \ref{ITM_Heston}. In particular, we plot the absolute error of the prices and the width of the obtained $95\%$ confidence interval computed as in \eqref{sigmaLS} and \eqref{Cinterval}, against the number of simulated points N.

\begin{figure}[ht]
\centering
\includegraphics[width=0.49\textwidth]{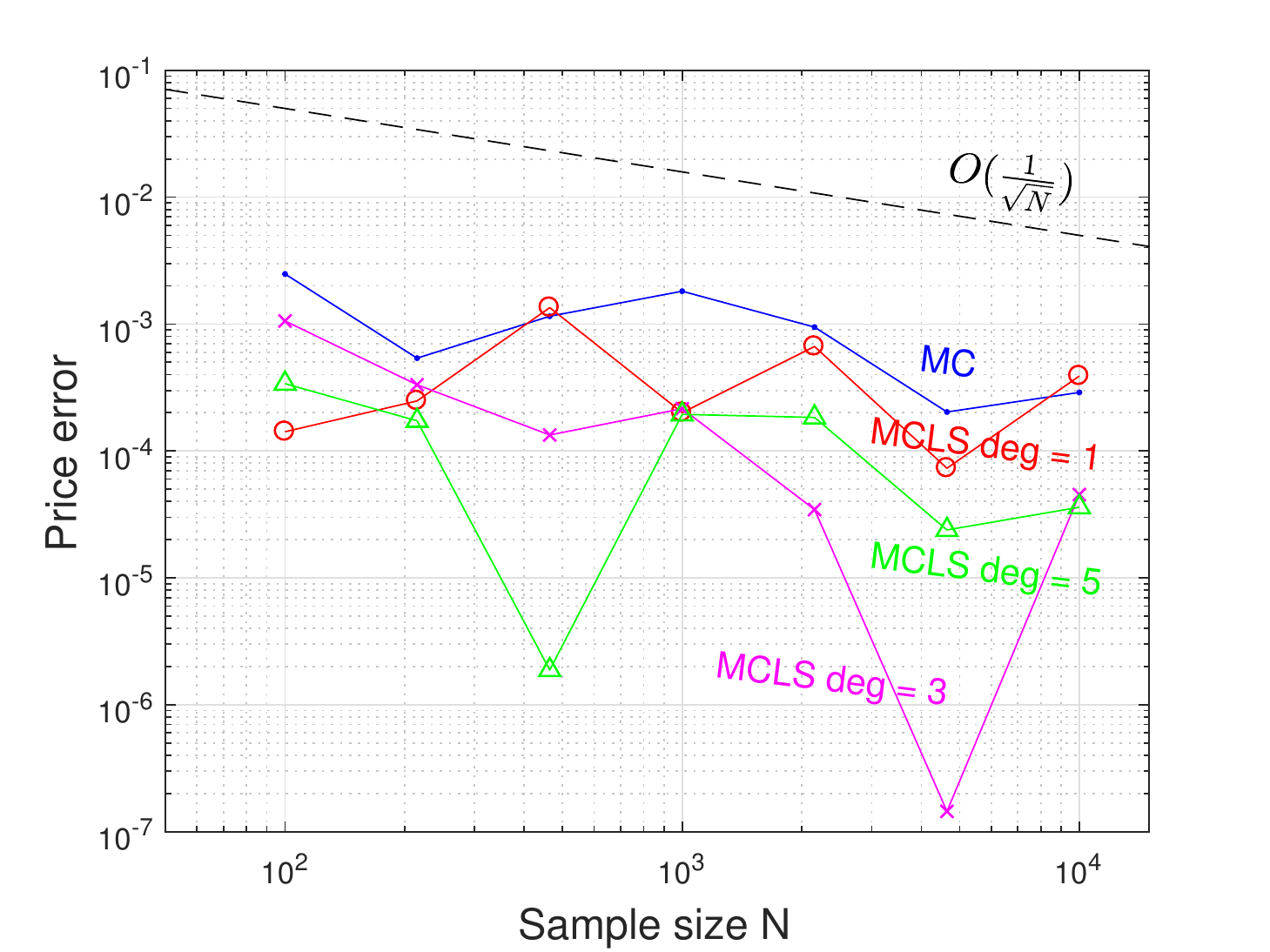}
\includegraphics[width=0.49\textwidth]{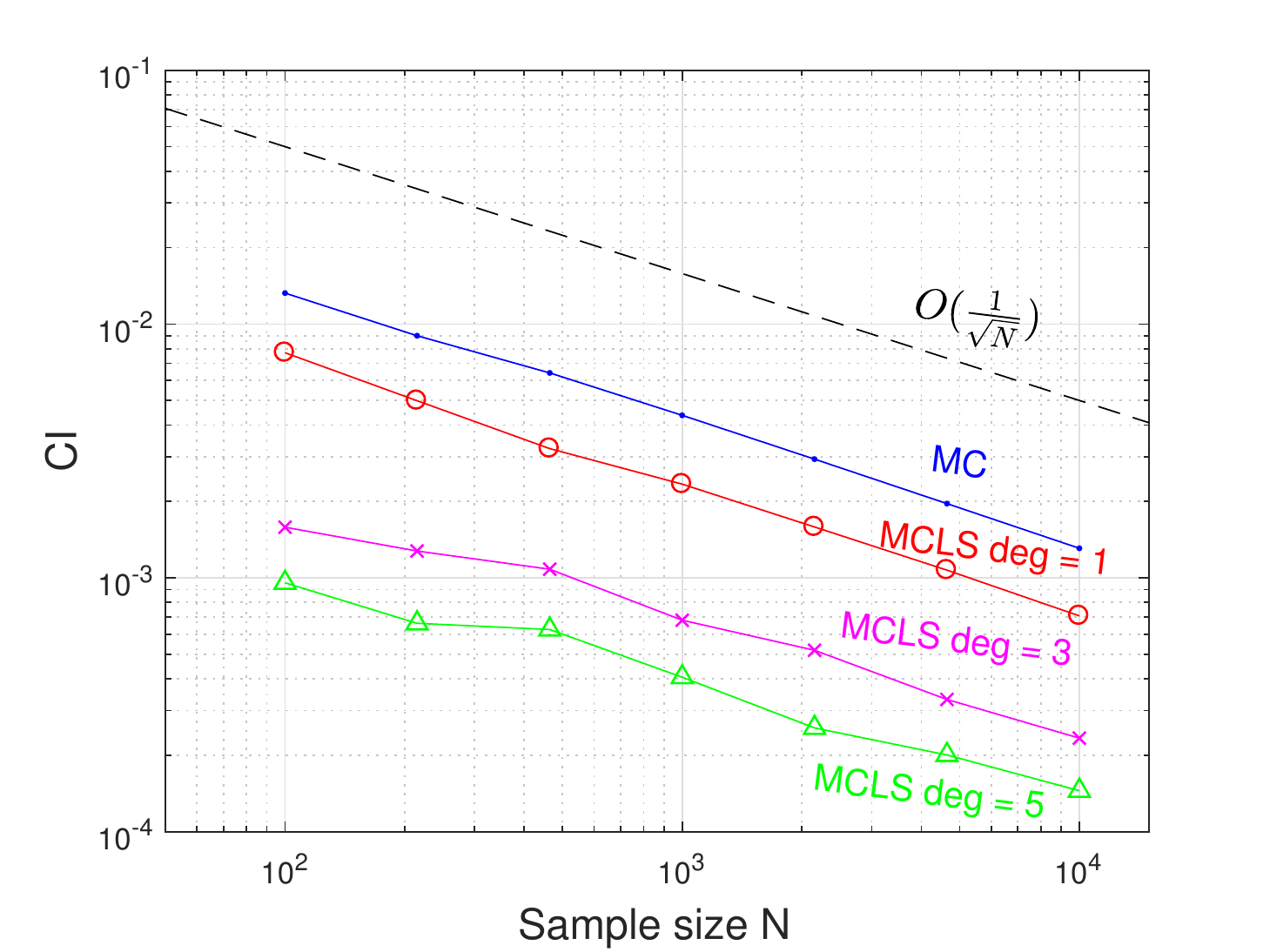}
\caption{MCLS for ATM call option in Heston model for different polynomial degrees. Left: Absolute price error. Right: Width of $95\%$ confidence interval. \label{ATM_Heston}}
\end{figure}

\begin{figure}[ht]
\centering
\includegraphics[width=0.49\textwidth]{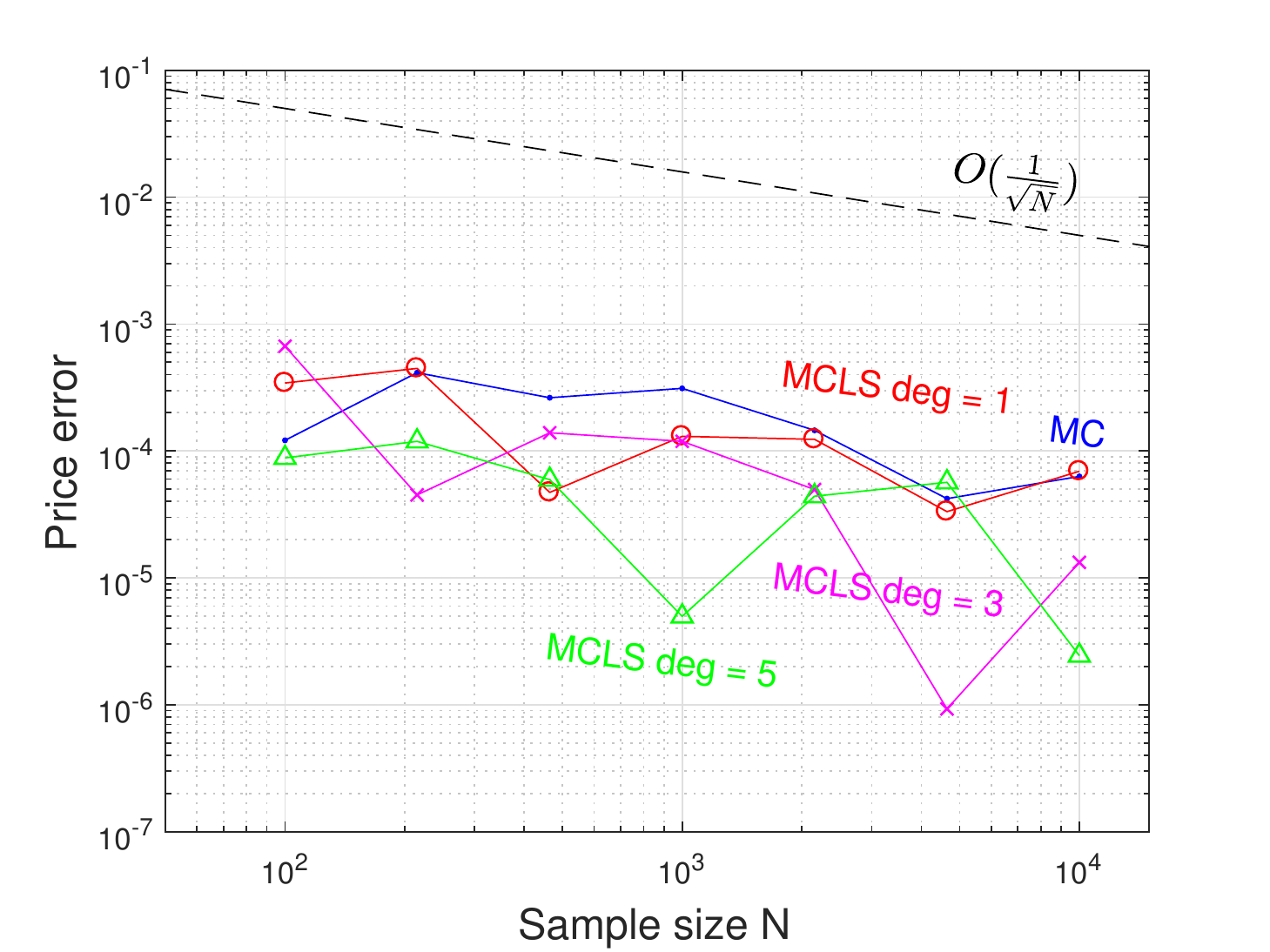}
\includegraphics[width=0.49\textwidth]{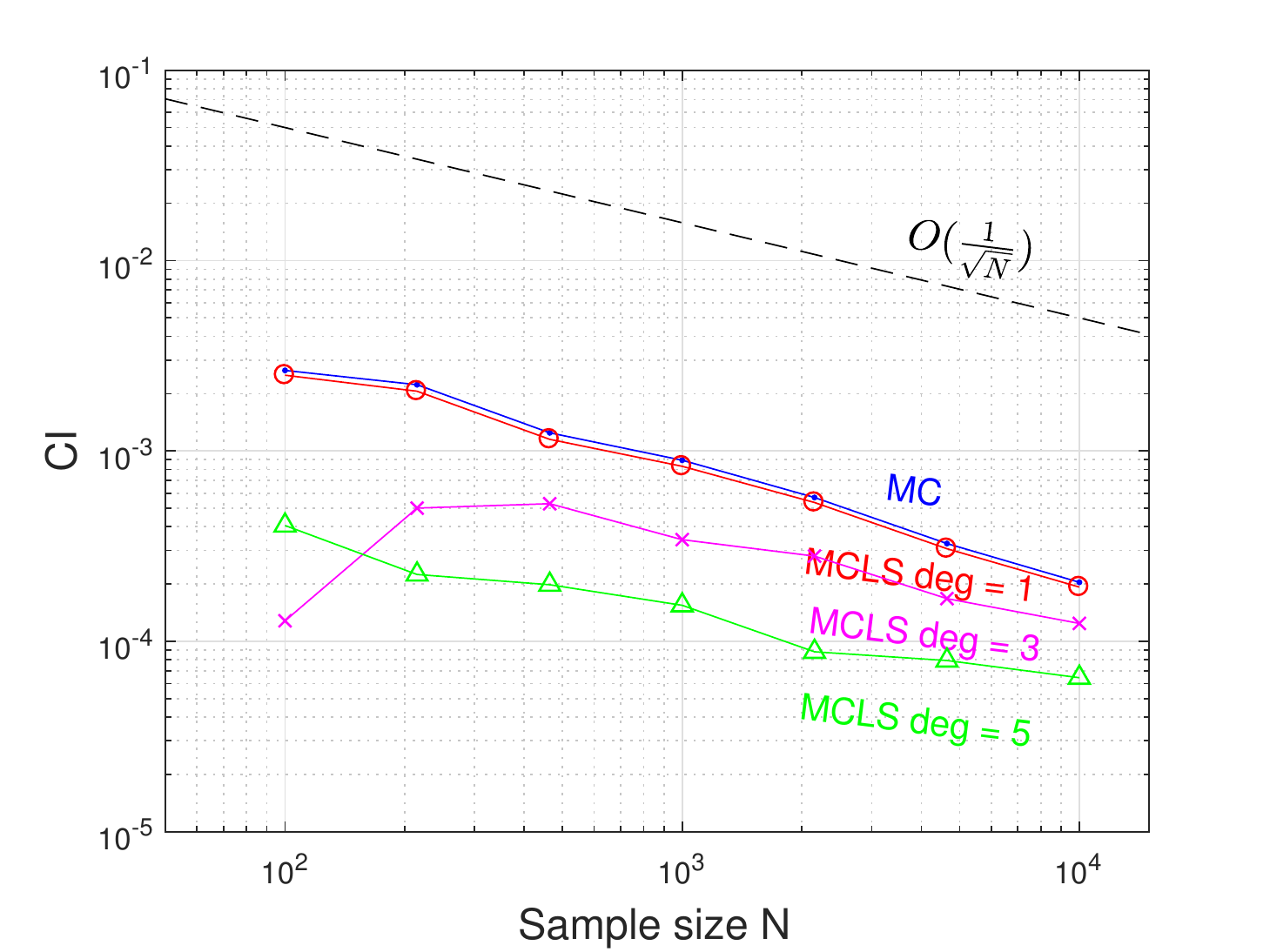}
\caption{MCLS for OTM call option in Heston model for different polynomial degrees. Left: Absolute price error. Right: Width of $95\%$ confidence interval. \label{OTM_Heston}}
\end{figure}

Second, we apply again MCLS but this time to an at-the-money call option with parameters 
\[ k=0, \quad T=1/12 \]
and to an out-of-the-money call option with parameters
\[ k=0.1, \quad T=1/12. \]
The results are shown in Figure \ref{ATM_Heston} and in Figure \ref{OTM_Heston}, respectively. 

In this setting, for all different choices of payoff parameters, we show in Table \ref{imp_vols_Heston} the implied volatility\footnote{For a given call option price $C$ the implied volatility is defined as the volatility parameter that renders the corresponding Black Scholes price equal to $C$.} absolute errors for the MC and MCLS prices computed with a basis of polynomials of maximal degree $5$. The implied volatility error is measured against the implied  volatility of the reference method.

\begin{table}[t!]
\centering
\begin{tabular}{cccc}
\multicolumn{4}{c}{\textbf{Implied vol absolute errors}} \\[0.5ex]
\hline\\[-2ex]
 & $k=-0.1$ & $k=0$ & $k=0.1$  \\
N & \hspace{0.15cm} MC \hspace{0.15cm} MCLS& \hspace{0.15cm} MC \hspace{0.15cm} MCLS &\hspace{0.15cm} MC \hspace{0.15cm} MCLS  \\[0.5ex]
\hline\\[-2ex]
100 &-- \hspace{0.55cm}  0.21		&2.16 \enskip 0.29 	&0.50 \enskip 0.37\\
215 &4.35 \enskip  0.09	 		&0.47 \enskip 0.15 	&1.56 \enskip 0.49\\
464 &9.16 \enskip  0.31 			&1.00 \enskip 0.00	&1.03 \enskip 0.26\\
1000 &9.13 \enskip 0.28			&1.58 \enskip 0.17	&1.21 \enskip 0.02\\
2154 &2.44 \enskip 0.22 			&0.82 \enskip 0.16	&0.59 \enskip 0.19\\
4642 &1.15 \enskip  0.09    		&0.18 \enskip 0.02	&0.18 \enskip 0.24\\
10000 &0.34 \enskip 0.04	 		&0.25 \enskip 0.03	&0.28 \enskip 0.01\\
\end{tabular}
\captionof{table}{Implied volatility errors (in $\%$) for MC and MCLS with basis of polynomials of maximal degree $5$ in the Heston model, for different sizes $N$ of the sample set. \label{imp_vols_Heston}}
\end{table}

Before commenting on the numerical results, we apply MCLS to a second stochastic volatility model, the Jacobi model as in \cite{ackerer2016jacobi}.
Here, the log asset price $X_t$ and the squared volatility process $V_t$ are defined through the SDE
\begin{align*}
&dV_t=\kappa(\theta -V_t)dt+\sigma \sqrt{Q(V_t)}dW^1_{t},\\
&dX_t=(r-V_t/2)dt + \rho \sqrt{Q(V_t)} dW^2_{t}+\sqrt{V_t-\rho^2Q(V_t)}dW^2_{t},
\end{align*}
where 
\begin{equation*}
Q(v)=\frac{(v-v_{min})(v_{max}-v)}{(\sqrt{v_{max}}-\sqrt{v_{min}})^2},
\end{equation*}
for some $0 \leq v_{\min} < v_{\max}$. Here, $W^1_{t}$ and $W^2_{t}$ are independent standard Brownian motions and the model parameters satisfy the conditions $\kappa \geq 0$, $\theta \in [v_{\min},v_{\max}]$, $\sigma >0$, $r \geq 0$, $\rho \in [-1,1]$.
The state space is in this case $E = [v_{\min}, v_{\max}] \times \mathbb{R}$. The matrix $G_n$ in \eqref{moment-formula} can be constructed as explained in the original paper \cite{ackerer2016jacobi} (with respect to a Hermite polynomial basis) or as in \cite{kressner2017incremental} (with respect to the monomial basis). For the numerical experiments we consider the set of model parameters 
\begin{align*}
 &\sigma=0.15,\enskip  v_0=0.04,\enskip x_0=0, \enskip \kappa = 0.5, \enskip \theta=0.04, \\ 
  &v_{\min}=10^{-4}, \enskip v_{\max}=0.08, \enskip \rho = -0.5, \enskip r = 0.01.
\end{align*}
We again consider single-asset European call options with payoff parameters
\[ k=\{-0.1,0,0.1\}, \quad T=1/12. \]
As reference pricing method we choose the polynomial expansion technique introduced in \cite{ackerer2016jacobi}, where we truncate the polynomial expansion of the price after $50$ terms.

We simulate the whole path of $X_t$ from $0$ to $T$ in order to get the sample points $x_i$, $i=1, \cdots, n$. The discretization scheme of the SDE is given by
\begin{align*}
V_{0}&=v_0,\\
X_{0}&=x_0,\\
V_{t_i}&= V_{t_{i-1}}+\kappa(\theta -V_{t_{i-1}})\Delta t+\sigma \sqrt{Q(V_{t_{i-1}})} \sqrt{\Delta t} Z^1_i,\\
X_{t_i}&= X_{t_{i-1}}+(r -V_{t_{i-1}}/2)\Delta t+\rho \sqrt{Q(V_{t_{i-1}})} \sqrt{\Delta t} Z^1_i + 
\sqrt{V_{t_{i-1}}-\rho^2Q(V_{t_{i-1}})} \sqrt{\Delta t} Z^2_i
\end{align*}
for all $i=1,\cdots, N_s$, where $Z^1_i$ and $Z^2_i$ are independent standard normal distributed random variables and the rest of the parameters are as specified in the example for the Heston model.

\begin{figure}[ht]
\centering
\includegraphics[width=0.49\textwidth]{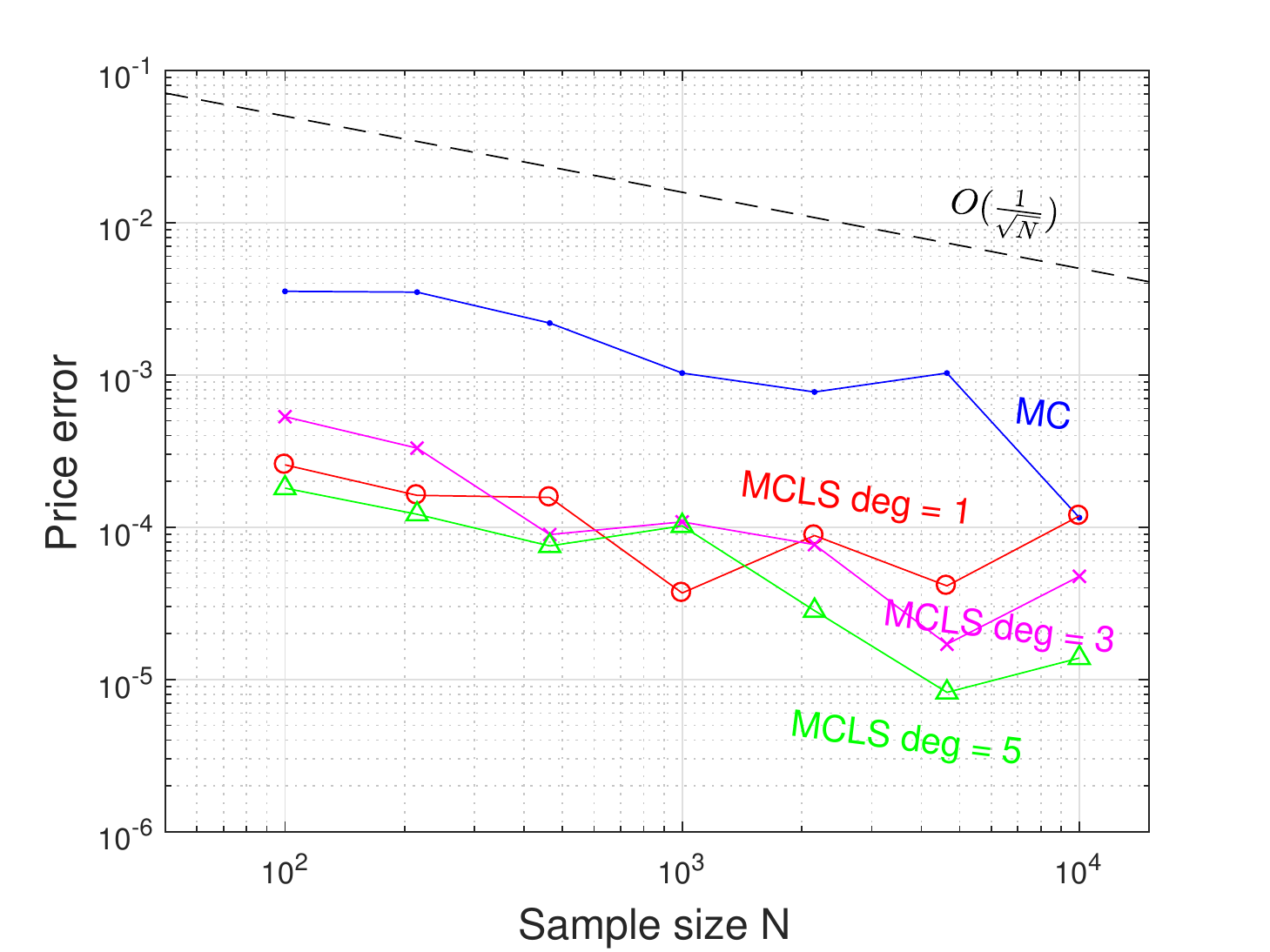}
\includegraphics[width=0.49\textwidth]{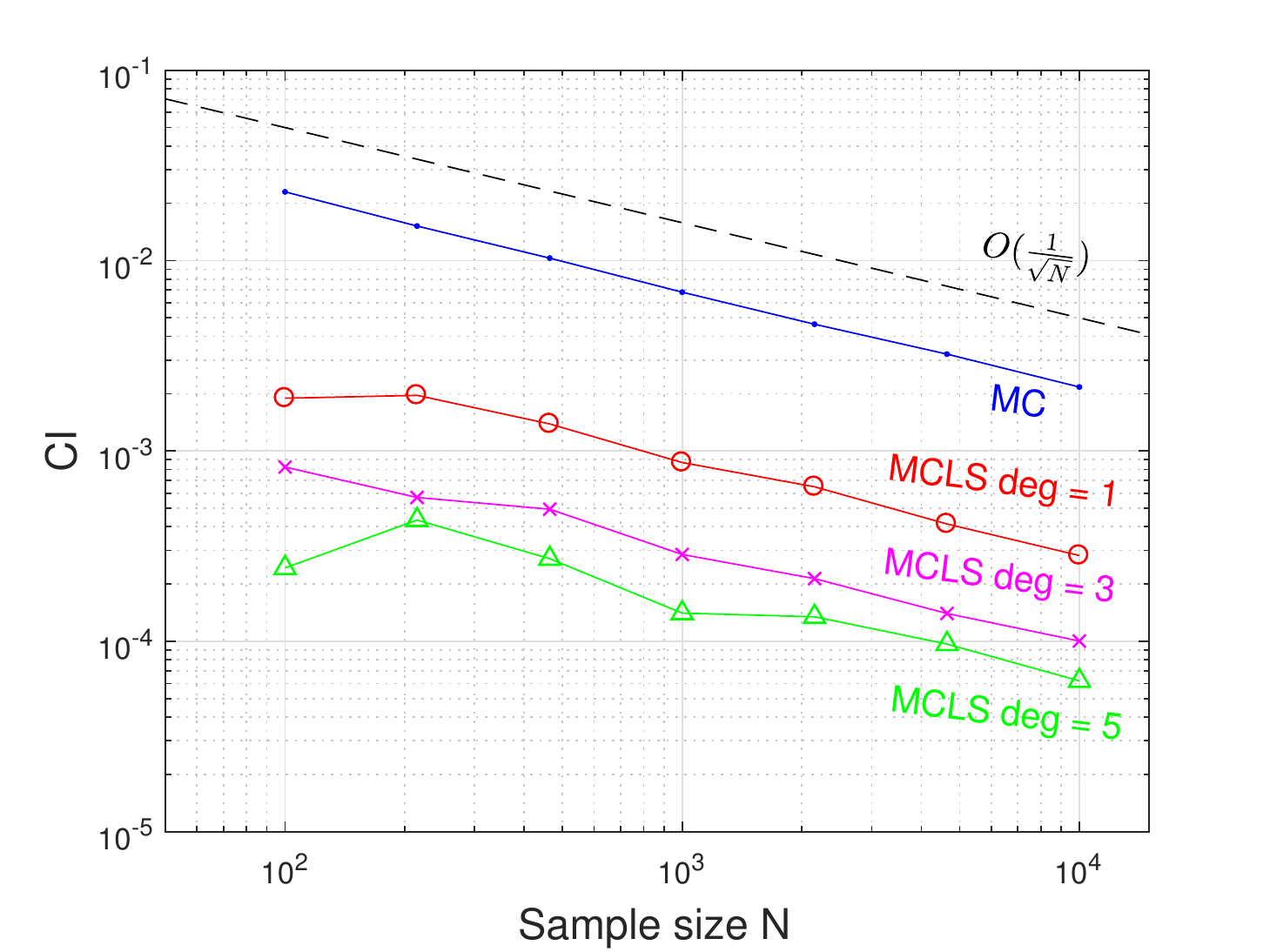}
\caption{MCLS for ITM call option in Jacobi model for different polynomial degrees. Left: Absolute price error. Right: Width of $95\%$ confidence interval. \label{ITM_Jacobi}}
\end{figure}

\begin{figure}[ht]
\centering
\includegraphics[width=0.49\textwidth]{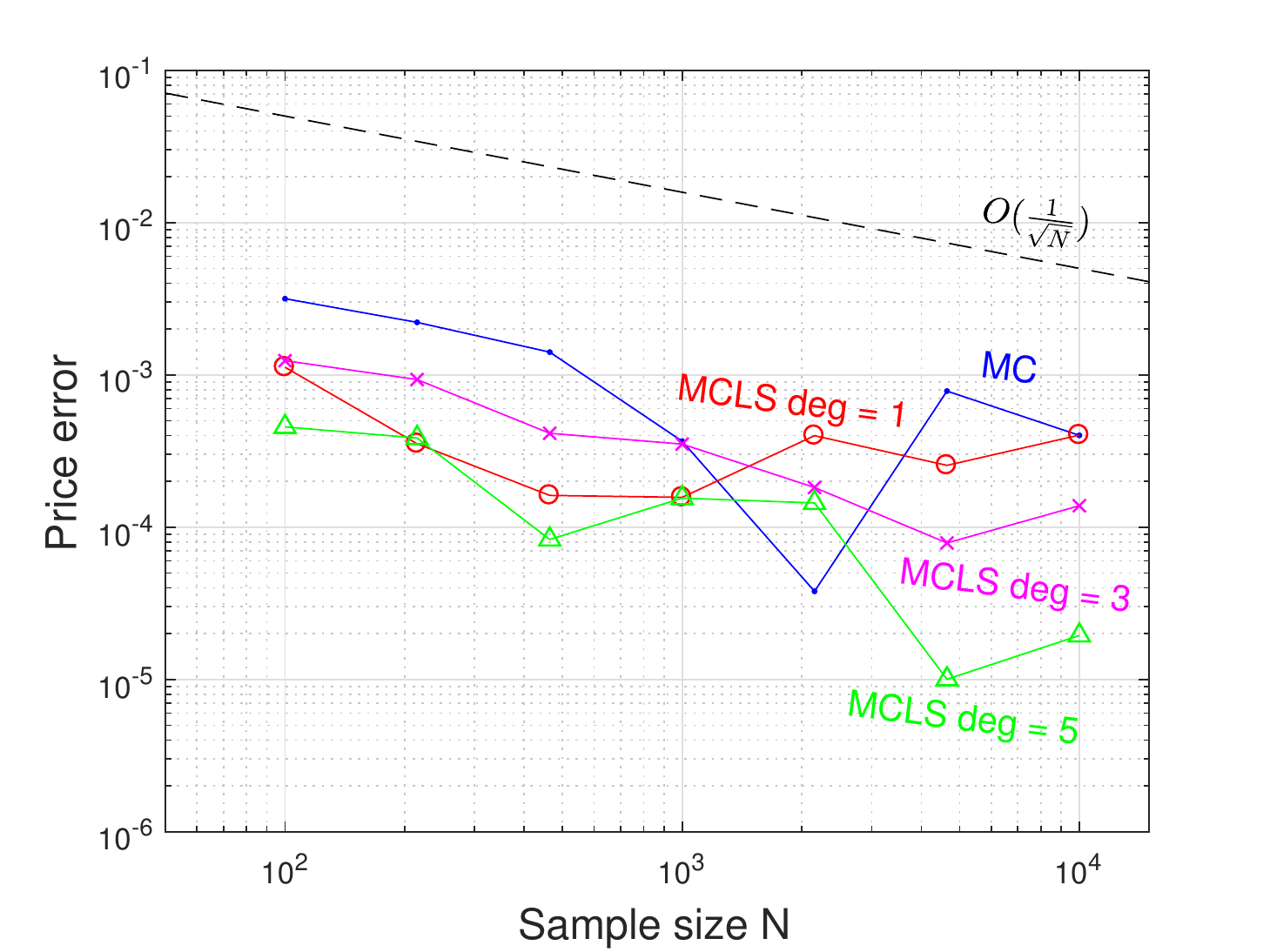}
\includegraphics[width=0.49\textwidth]{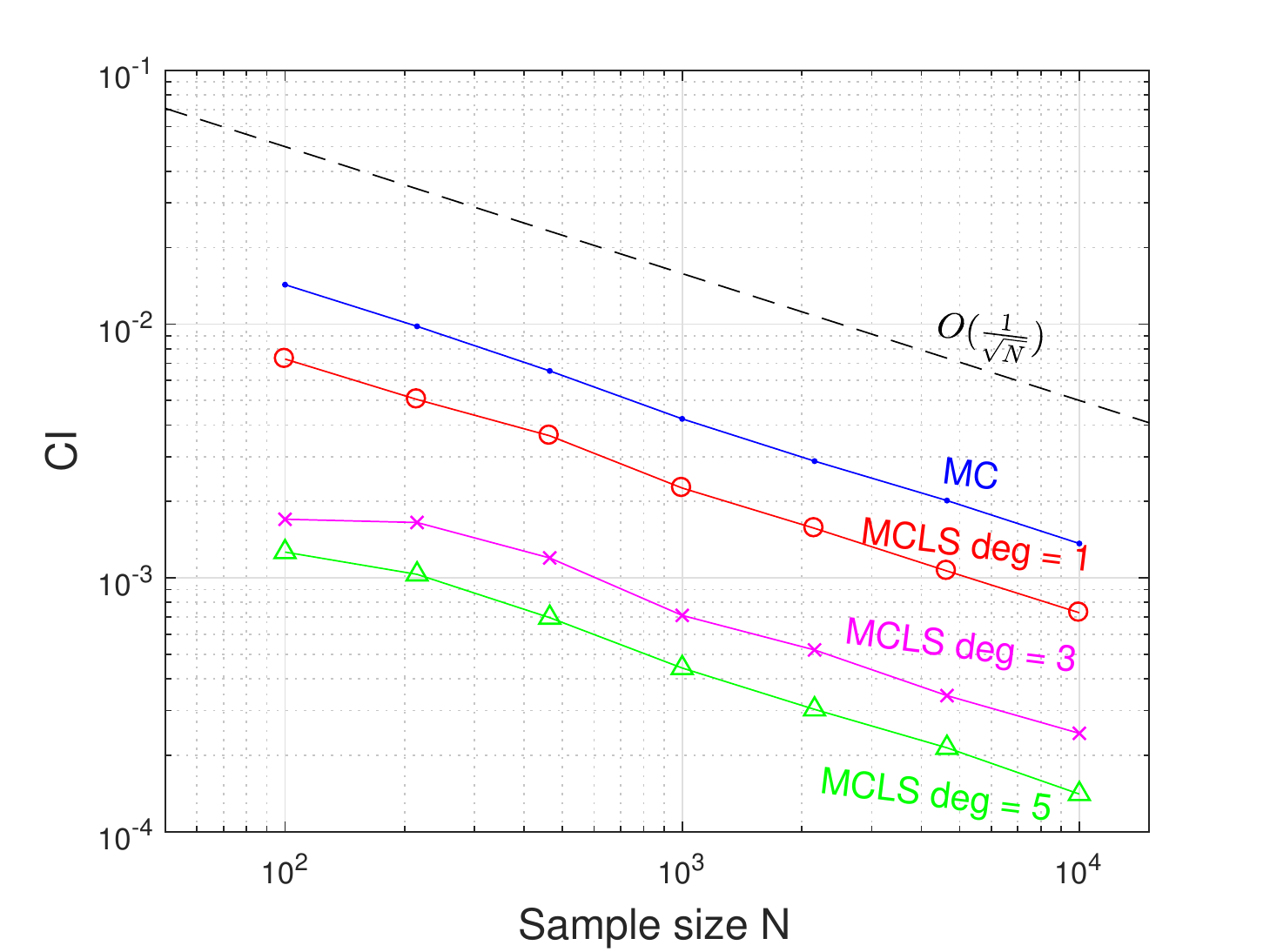}
\caption{MCLS for ATM call option in Jacobi model for different polynomial degrees. Left: Absolute price error. Right: Width of $95\%$ confidence interval. \label{ATM_Jacobi}}
\end{figure}

\begin{figure}[ht]
\centering
\includegraphics[width=0.49\textwidth]{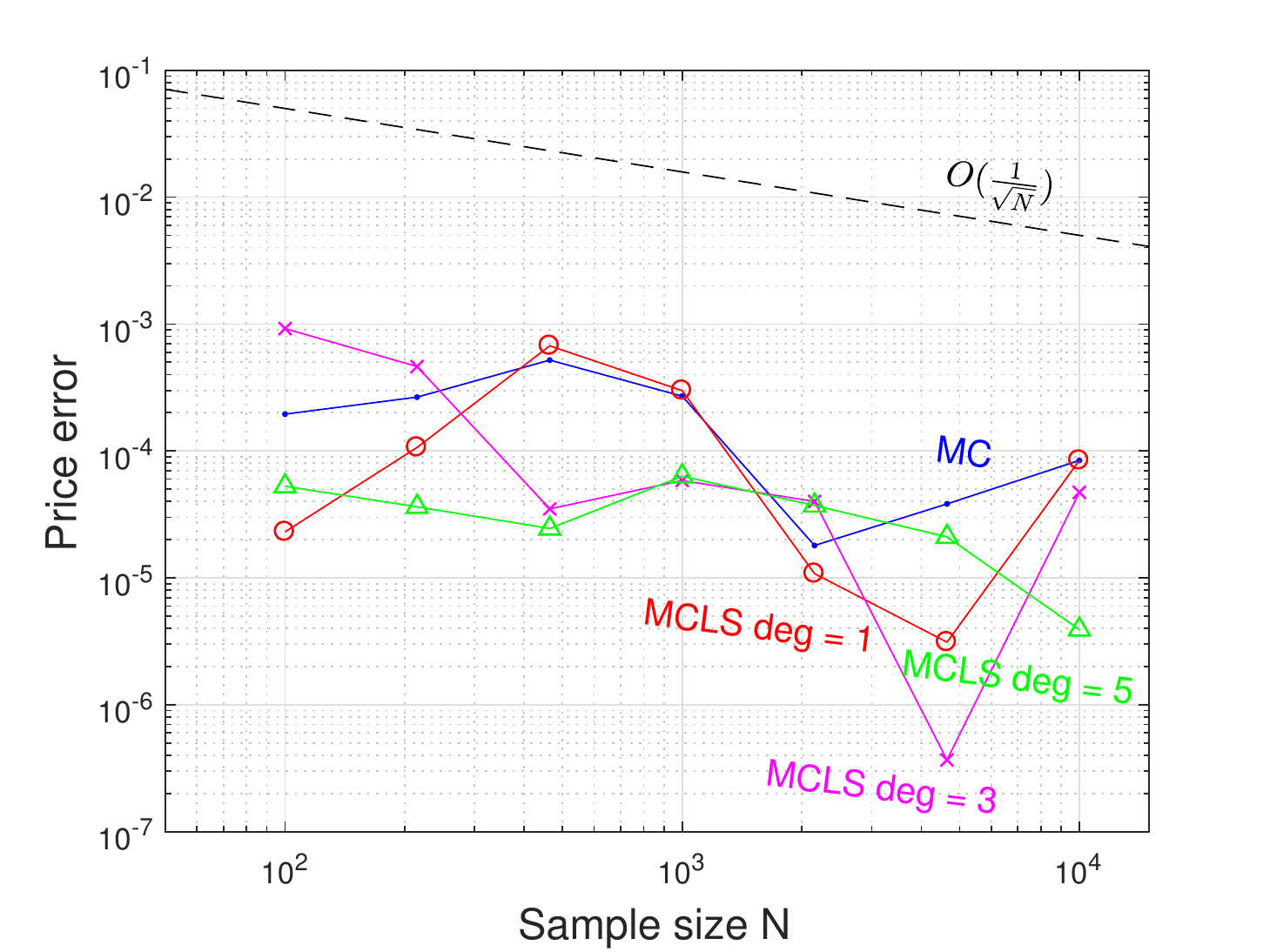}
\includegraphics[width=0.49\textwidth]{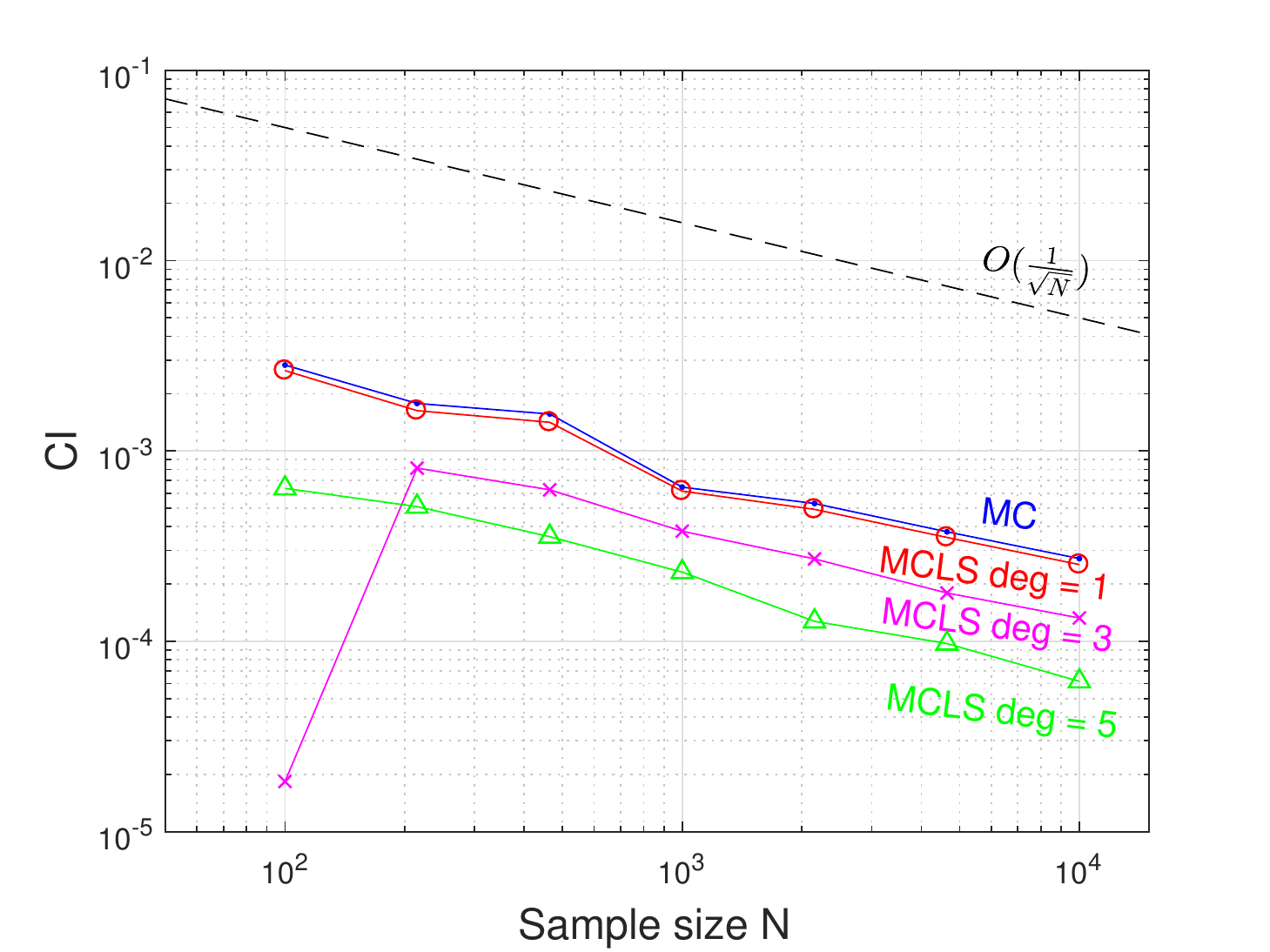}
\caption{MCLS for OTM call option in Jacobi model for different polynomial degrees. Left: Absolute price error. Right: Width of $95\%$ confidence interval. \label{OTM_Jacobi}}
\end{figure}

We use again an ONB consisting of polynomials of maximal degrees $0$ (standard MC), $1,3$ and $5$ and we obtain the results shown in Figures \ref{ITM_Jacobi}, \ref{ATM_Jacobi} and \ref{OTM_Jacobi}, for ITM, ATM and OTM call options, respectively. Lastly, we show in Table \ref{imp_vols_Jacobi} the implied volatility absolute errors for the MC and MCLS prices computed with a basis of polynomials with maximal degree $5$. 

\begin{table}[t]
\centering
\begin{tabular}{cccc}
\multicolumn{4}{c}{\textbf{Implied vol absolute errors}} \\[0.5ex]
\hline\\[-2ex]
 & $k=-0.1$ & $k=0$ & $k=0.1$  \\
N & \hspace{0.15cm} MC \hspace{0.15cm} MCLS& \hspace{0.15cm} MC \hspace{0.15cm} MCLS &\hspace{0.15cm} MC \hspace{0.15cm} MCLS  \\[0.5ex]
\hline\\[-2ex]
100 &8.75 \enskip 0.67 		&2.75 \enskip 0.40 	&0.72 \enskip 0.20\\
215 &8.67 \enskip 0.46	 	&1.92 \enskip 0.33 	&0.96 \enskip 0.14\\
464 &-- \hspace{0.55cm}  0.30 	&1.23 \enskip 0.07	&1.77 \enskip 0.10\\
1000 &3.27 \enskip 0.39	        &0.32 \enskip 0.13	&1.16 \enskip 0.24\\
2154 &2.55 \enskip 0.11 		&0.03 \enskip 0.13	&0.07 \enskip 0.14\\
4642 & 3.26 \enskip 0.03    	&0.68 \enskip 0.01	&0.15 \enskip 0.08\\
10000 &0.47 \enskip 0.05	 	&0.35 \enskip 0.02	&0.32 \enskip 0.02\\
\end{tabular}
\captionof{table}{Implied volatility errors (in $\%$)  for MC and MCLS with basis of polynomials with maximal degree $5$ in the Jacobi model, for different sizes $N$ of the sample set. \label{imp_vols_Jacobi}}
\end{table}
We can observe that MCLS strongly outperforms the standard MC in terms of price errors, confidence interval width and implied volatility errors, for every type of moneyness, in both chosen stochastic volatility models.

The last remark concerns the condition number of the Vandermonde matrix $\V$. Thanks to the choice of the ONB, in both models, its condition number is at most of order $10$. Therefore, the CG algorithm has been selected. As another consequence of the low condition number we did not implement weighted sampling.

\subsubsection{Basket options in Black Scholes models - medium size problems}\label{BS example}
In this section we address multi-dimensional option pricing problems of medium size, meaning with number of assets $d\leq 10$ and $N\leq 10^5$. The asset prices $S^1_t,\ldots,S^d_t$ follow a $d$-dimensional Black Scholes model, i.e.  
\begin{equation}\label{sde-BS}
dS_t^i=rS_t^i+\sigma_{i}S_t^i dW_{t}^i,\quad i=1,\cdots, d,
\end{equation}
for some volatility values $\sigma_i$, $i=1,\ldots,d$, a risk-free interest rate $r$ and $d$ correlated Brownian motions $W_{t}^i$ with correlation parameters $\rho_{ij} \in [-1,1]$ for $i\neq j$. The state space is $E = \mathbb{R}^d_+ $ and the explicit solution of \eqref{sde-BS} is given by 
\begin{equation*}
S_t^i = S_0^i \exp \big ( (r-\frac{\sigma_i^2}{2})t+\sigma_i W_t^i \big ).
\end{equation*}
The process $(S_t^1,\dots, S_t^d)$ is a polynomial diffusion and the moment formula is given by 
\begin{equation*}
\E[p(S_T^1,\dots, S_T^d)|\mathcal{F}_t]=H_n(S_t^1,\dots, S_t^d)e^{G_n(T-t)}\vec{p},
\end{equation*}
where the involved quantities are defined along the lines following \eqref{moment-formula}.
The matrix $G_n$ can be computed with respect to the monomial basis as in the following lemma, and turns out to be diagonal, making Step 4 of Algorithm \ref{mainAlgo} even more efficient.
\begin{lemma}
Let $\mathcal{H}_n$ be the monomial basis of $\text{Pol}_n(\mathbb{R}_+^d)$. Let
\begin{align*}
\pi:\mathcal E\rightarrow \Bigg \{1,\ldots, \binom{n+d}{n}\Bigg\}
\end{align*}
be an enumeration of the set of tuples $\mathcal E=\{\mathbf{k}\in \mathbb{N}_0^d:|\mathbf{k}|\le n\}$.
Then, the matrix representation $G_n$ of the infinitesimal generator of the process $(S^1_t,\cdots, S^d_t)$ with respect to $\mathcal{H}_n$ and restricted to $\text{Pol}_n(\mathbb{R}_+^d)$ is diagonal with diagonal entries
\begin{align*}
G_{\pi(\mathbf{k}),\pi(\mathbf{k})}= \frac{1}{2} \sum_{i=1}^d \sum_{j=1}^d  \sigma_i \sigma_j \rho_{ij} (k_ik_j \mathbf{1}_{i \neq j}+k_i (k_i-1)\mathbf{1}_{i=j})+r\sum_{i=1}^d k_i.
\end{align*}
\begin{proof}
The infinitesimal generator $\mathcal{G}$ of $(S^1_t,\cdots,S^d_t)$ is given by
\begin{equation*}
\mathcal{G}f = \frac{1}{2} \sum_{i=1}^d \sum_{j=1}^d  \sigma_i \sigma_j \rho_{ij} s_i s_j \partial_{s_i s_j}f+
r\sum_{i=1}^d s_i \partial_{s_i} f,
\end{equation*}
which implies that for any monomial of the form $s_1^{k_1}\cdots s_d^{k_d}$ one has
\begin{equation*}
\mathcal{G} s_1^{k_1}\cdots s_d^{k_d} =s_1^{k_1}\cdots s_d^{k_d} \Big (\frac{1}{2} \sum_{i=1}^d \sum_{j=1}^d  \sigma_i \sigma_j \rho_{ij} (k_ik_j \mathbf{1}_{i \neq j}+k_i (k_i-1)\mathbf{1}_{i=j})+r\sum_{i=1}^d k_i \Big ).
\end{equation*}
It follows that $G_n$ is diagonal as stated above.
\end{proof}
\end{lemma}
For the following numerical experiments we consider basket options with payoff function
\begin{equation}\label{payoff basket}
f(s_1,\cdots, s_d) = \Big(\sum_{i=1}^d w_i s_i - K\Big)^+
\end{equation}
for different moneyness with payoff parameters
\begin{equation*}
K=\{0.9,1,1.1\}, \quad T=1,\quad w_i=\frac{1}{d} \enskip \forall i.
\end{equation*}
Model parameters are chosen to be
\begin{equation*}
S_0^i = 1 \enskip \forall i, \quad \sigma_i = \text{rand}(0,0.5) \enskip \forall i, \quad \{\rho_{ij}\}_{i,j=1}^d=R_d, \quad r=0.01,
\end{equation*}
where $R_d$ denotes a random correlation matrix of size $d \times d$, where we choose $d=5$ and $d=10$.

We compare MCLS to a reference price computed via a standard Monte Carlo algorithm with $10^6$ simulations.  
We plot again the absolute price errors and the width of the $95\%$ confidence intervals (computed as in \eqref{sigmaLS} and \eqref{Cinterval}) for different chosen polynomial degrees (maximally $1$ and maximally $3$). To be more precise, we used the monomial basis as functions $\{\phi_{j}\}_{j=0}^n$. Note that the distribution of the price process $(S_t^1, \cdots, S_t^d)$ is known to be the geometric Brownian distribution so that there is no need to simulate the whole path but only the process at final time $T$. 

\begin{figure}[ht]
\centering
\includegraphics[width=0.49\textwidth]{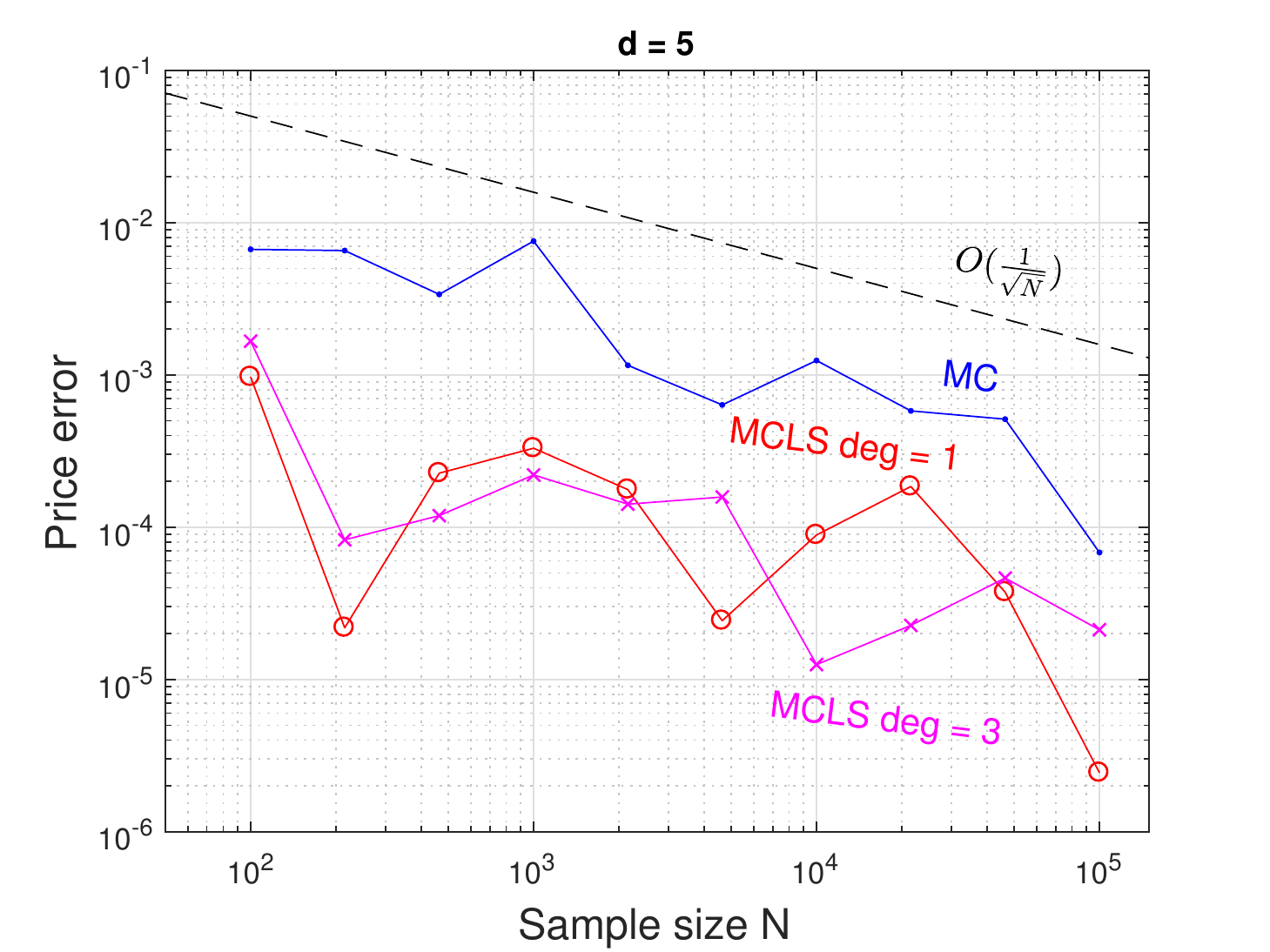}
\includegraphics[width=0.49\textwidth]{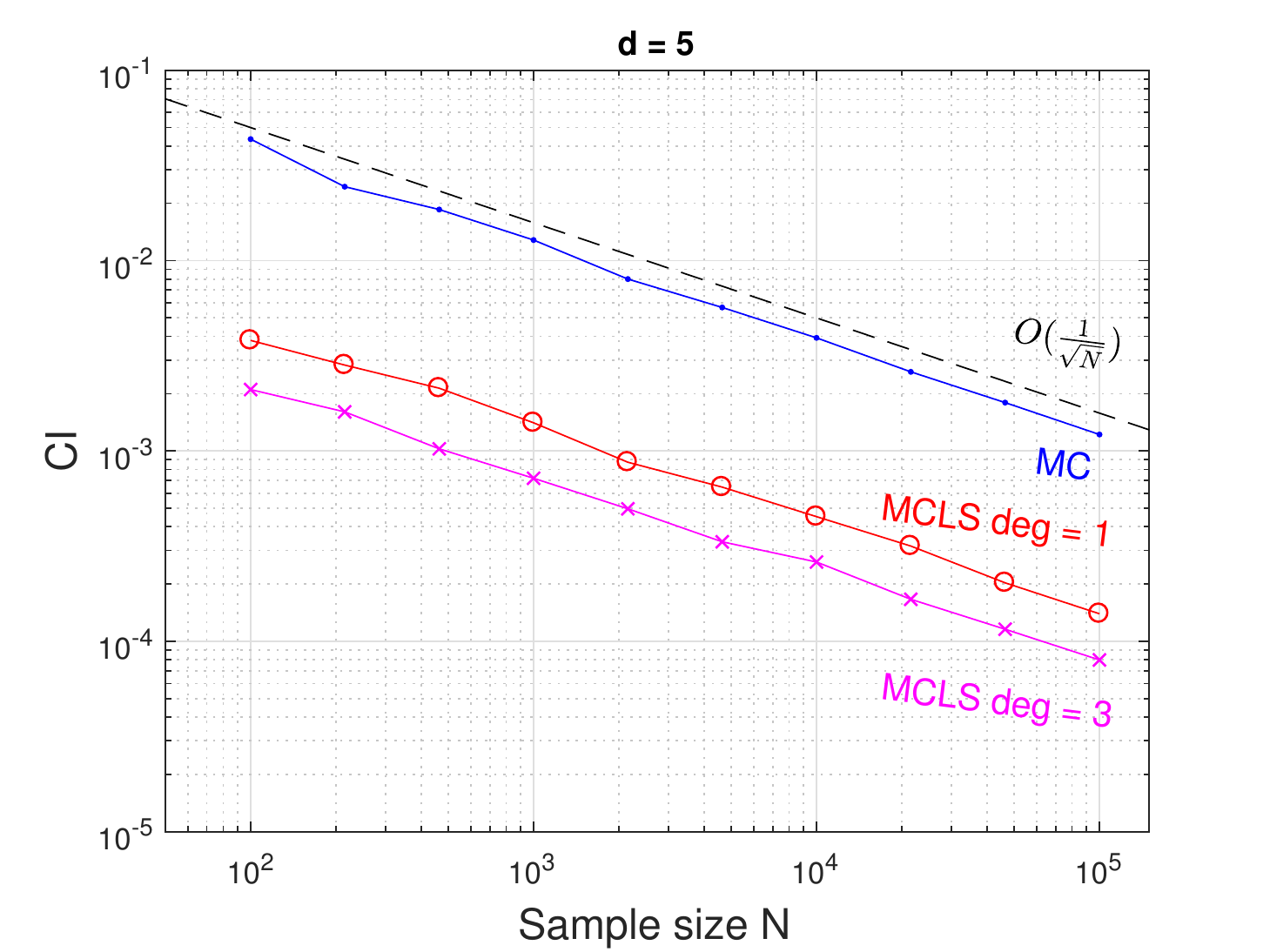}
\includegraphics[width=0.49\textwidth]{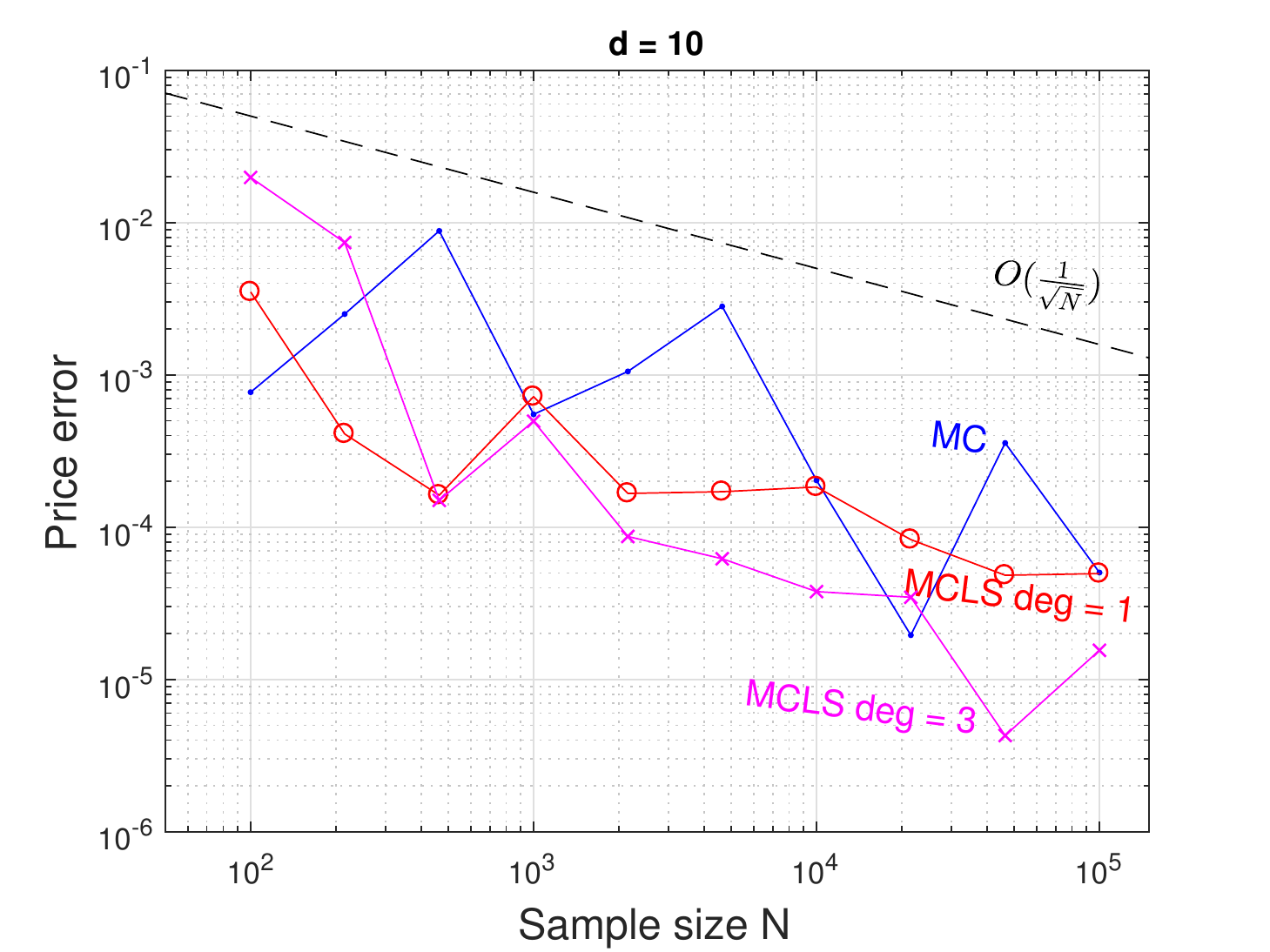}
\includegraphics[width=0.49\textwidth]{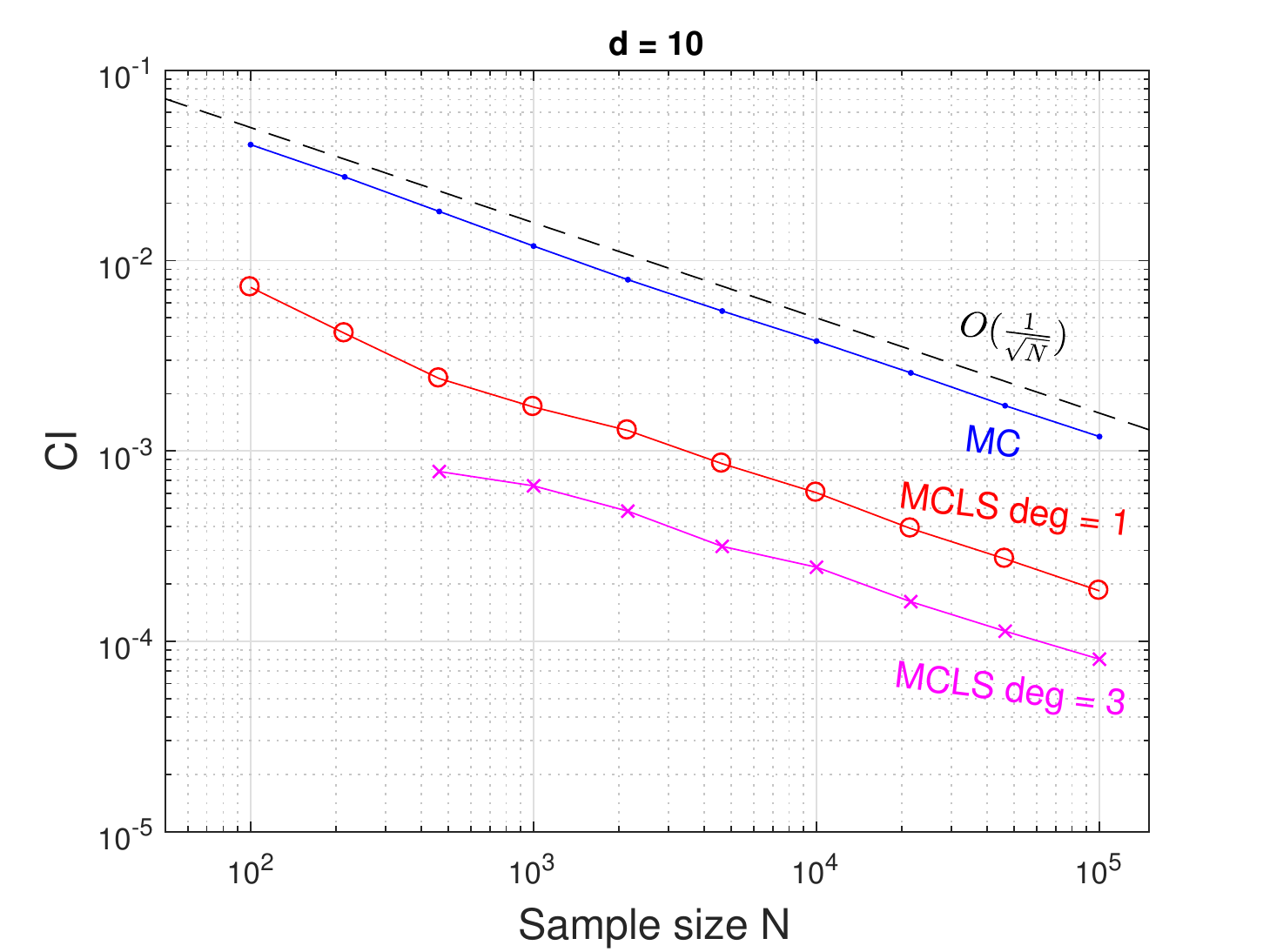}
\caption{MCLS for ITM basket option in Black Scholes model for different dimensions and polynomial degrees. Left: absolute price errors with respect to a reference price computed with $10^6$ simulations. Right: Width of $95\%$ confidence interval. \label{ITM_BS}}
\end{figure}

\begin{figure}[ht]
\centering
\includegraphics[width=0.49\textwidth]{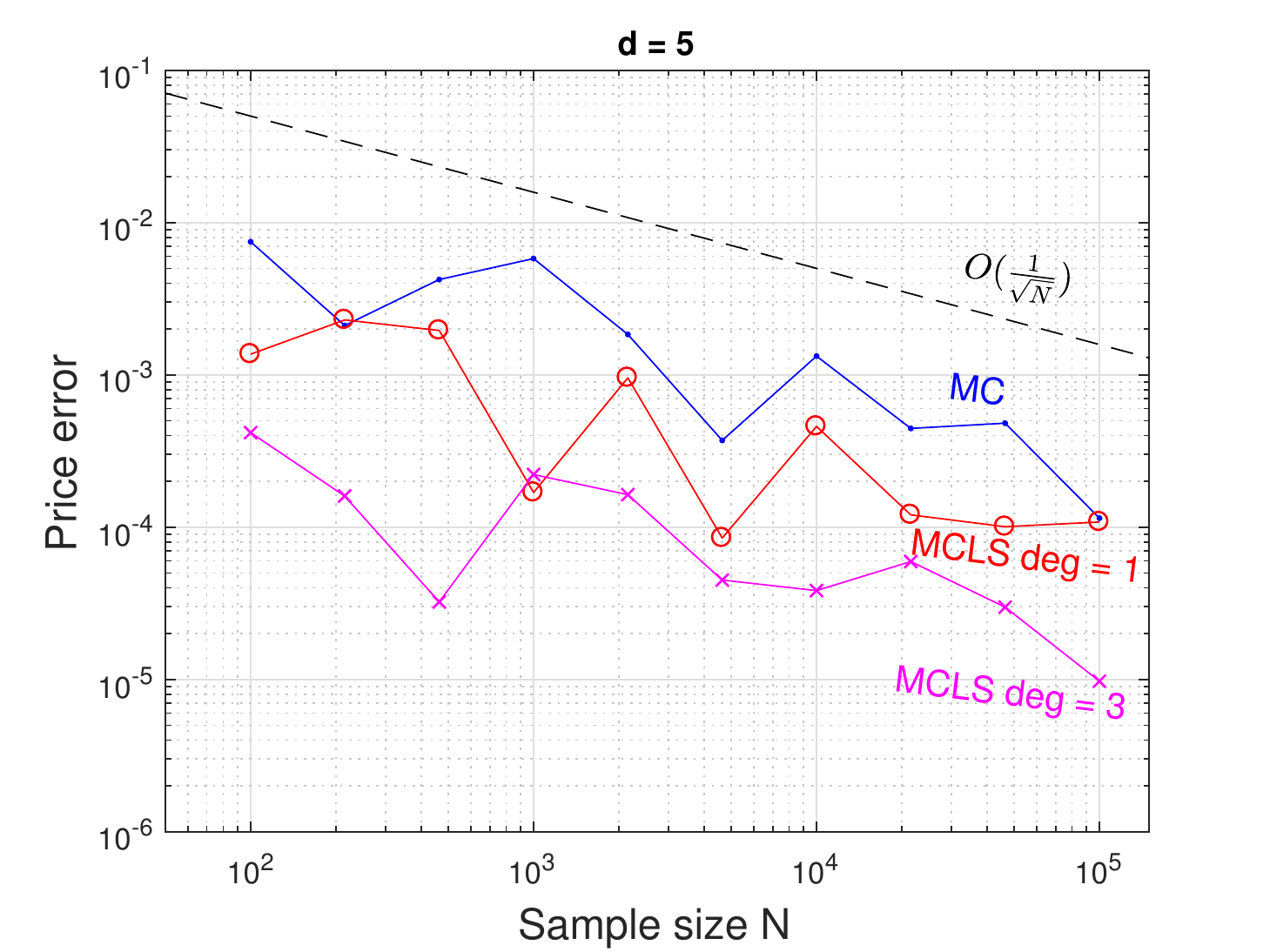}
\includegraphics[width=0.49\textwidth]{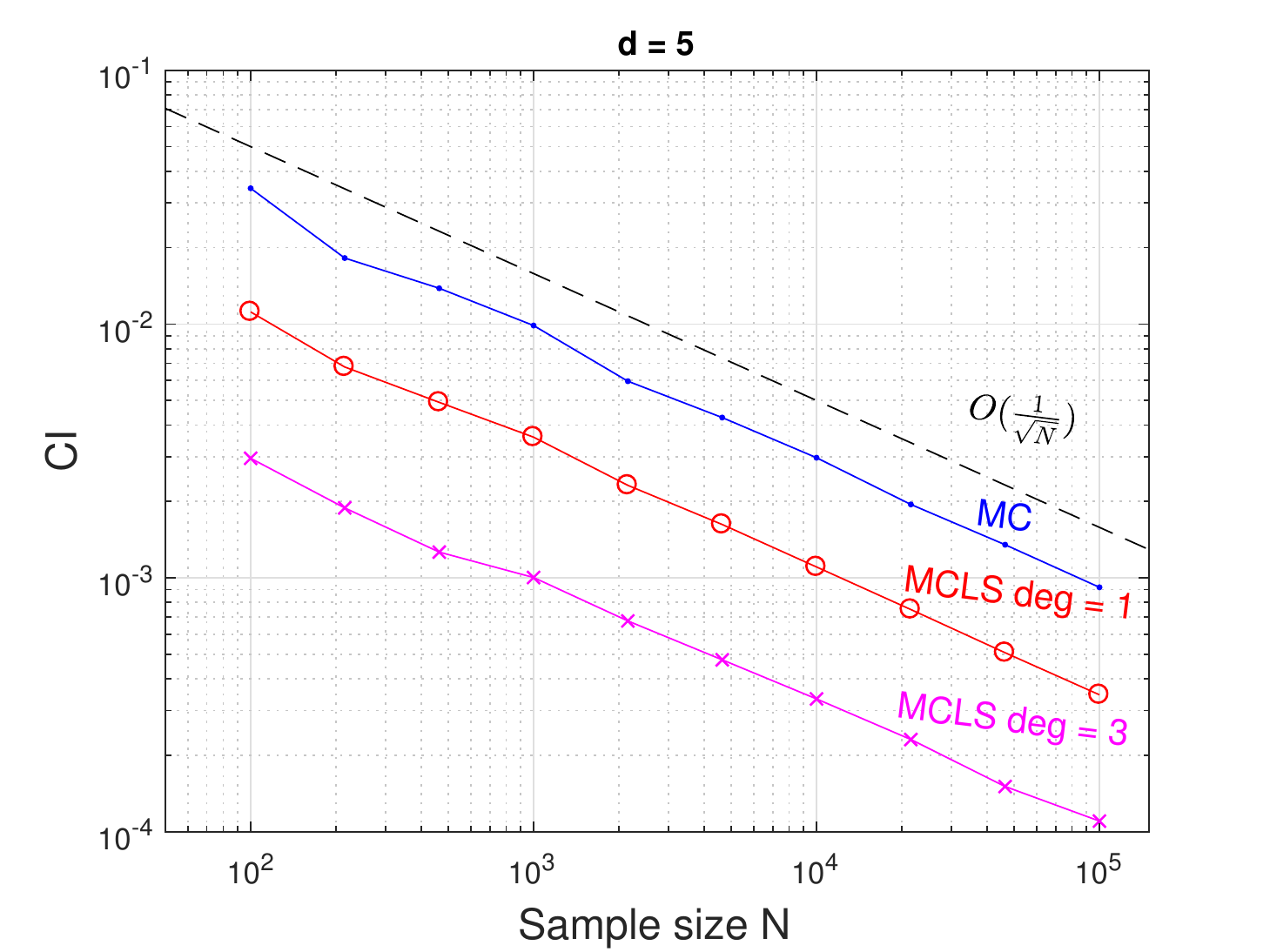}
\includegraphics[width=0.49\textwidth]{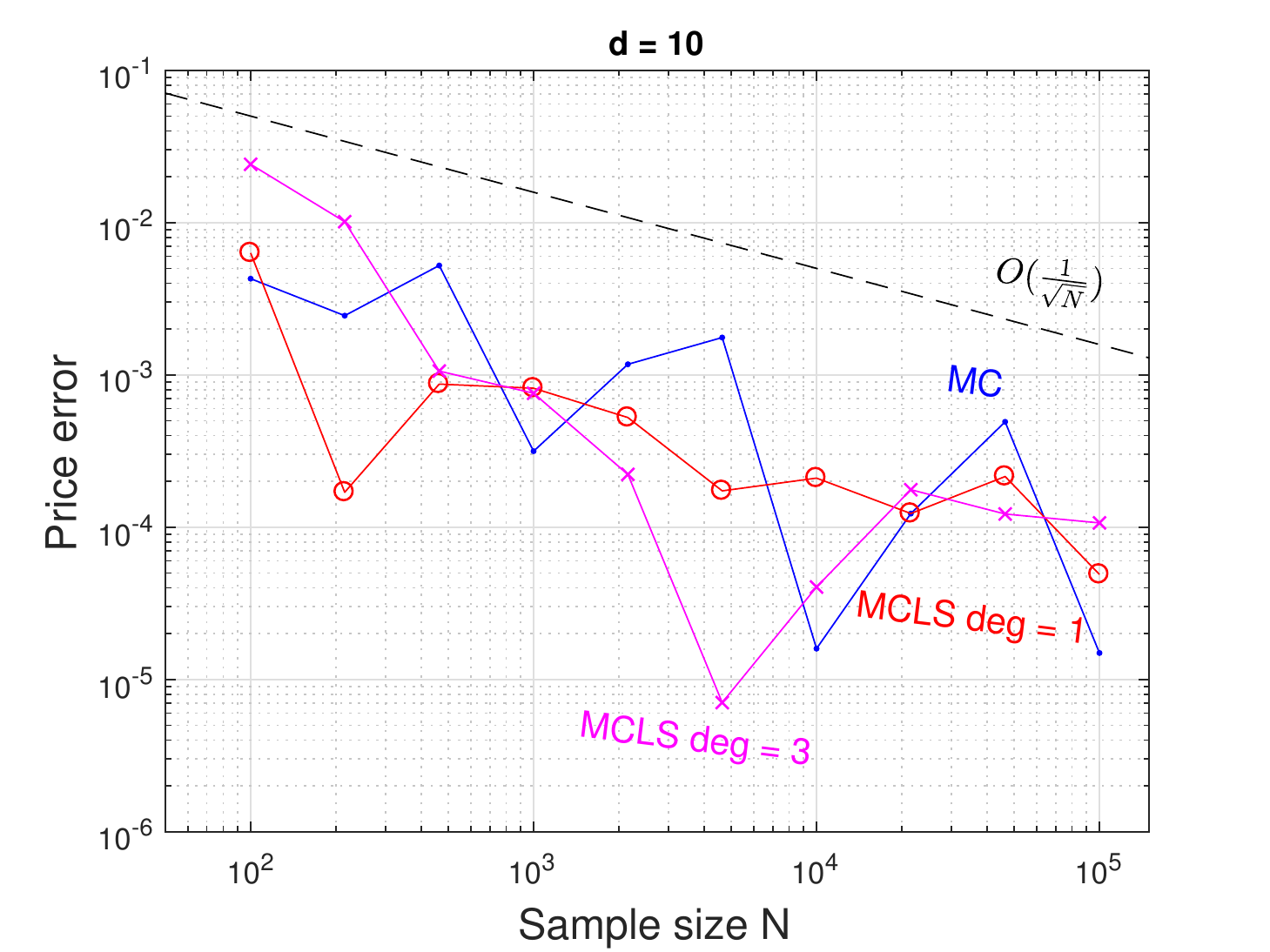}
\includegraphics[width=0.49\textwidth]{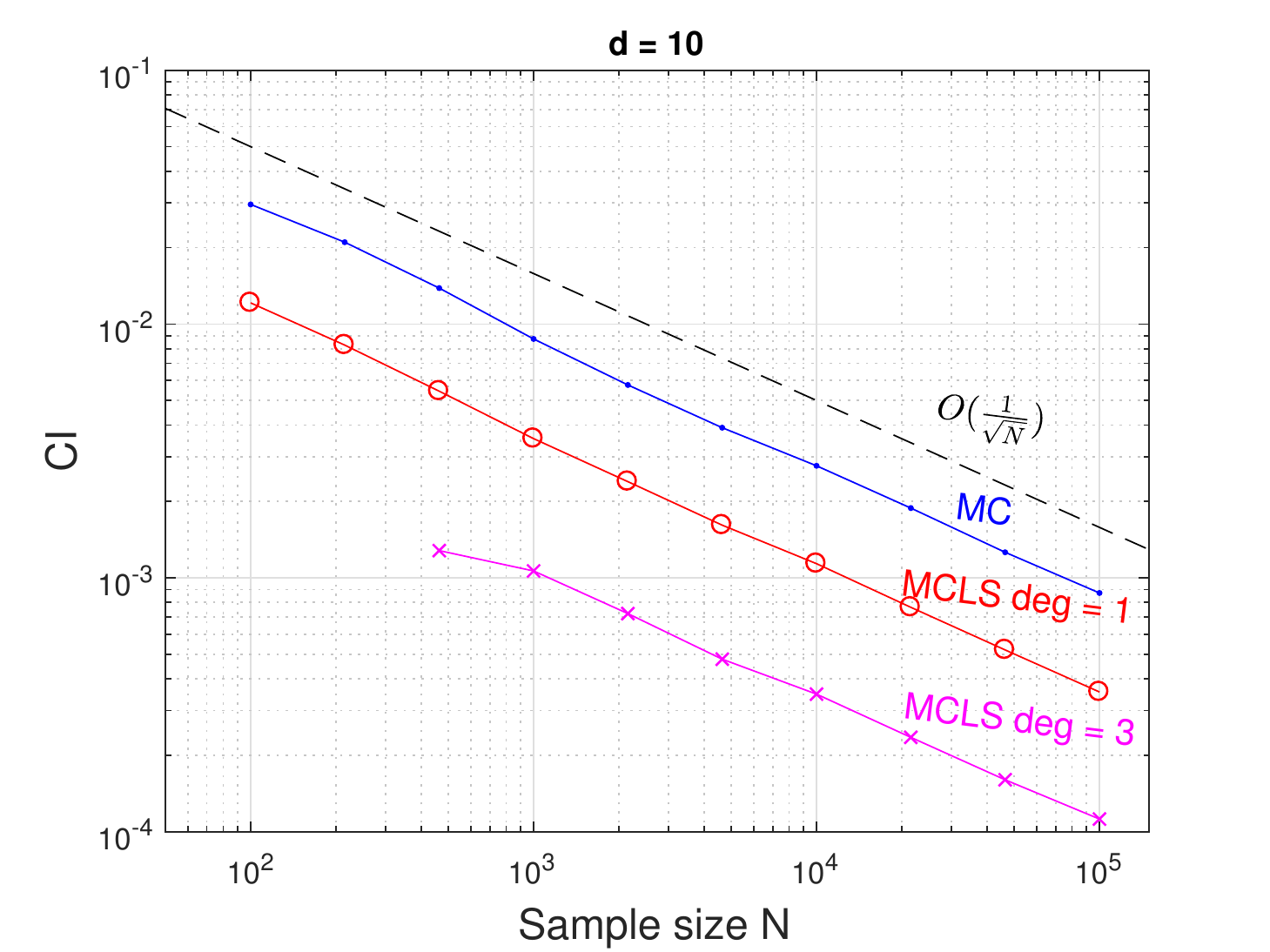}
\caption{MCLS for ATM basket option in Black Scholes model for different dimensions and polynomial degrees. Left: absolute price errors with respect to a reference price computed with $10^6$ simulations. Right: Width of $95\%$ confidence interval. \label{ATM_BS}}
\end{figure}

\begin{figure}[ht]
\centering
\includegraphics[width=0.49\textwidth]{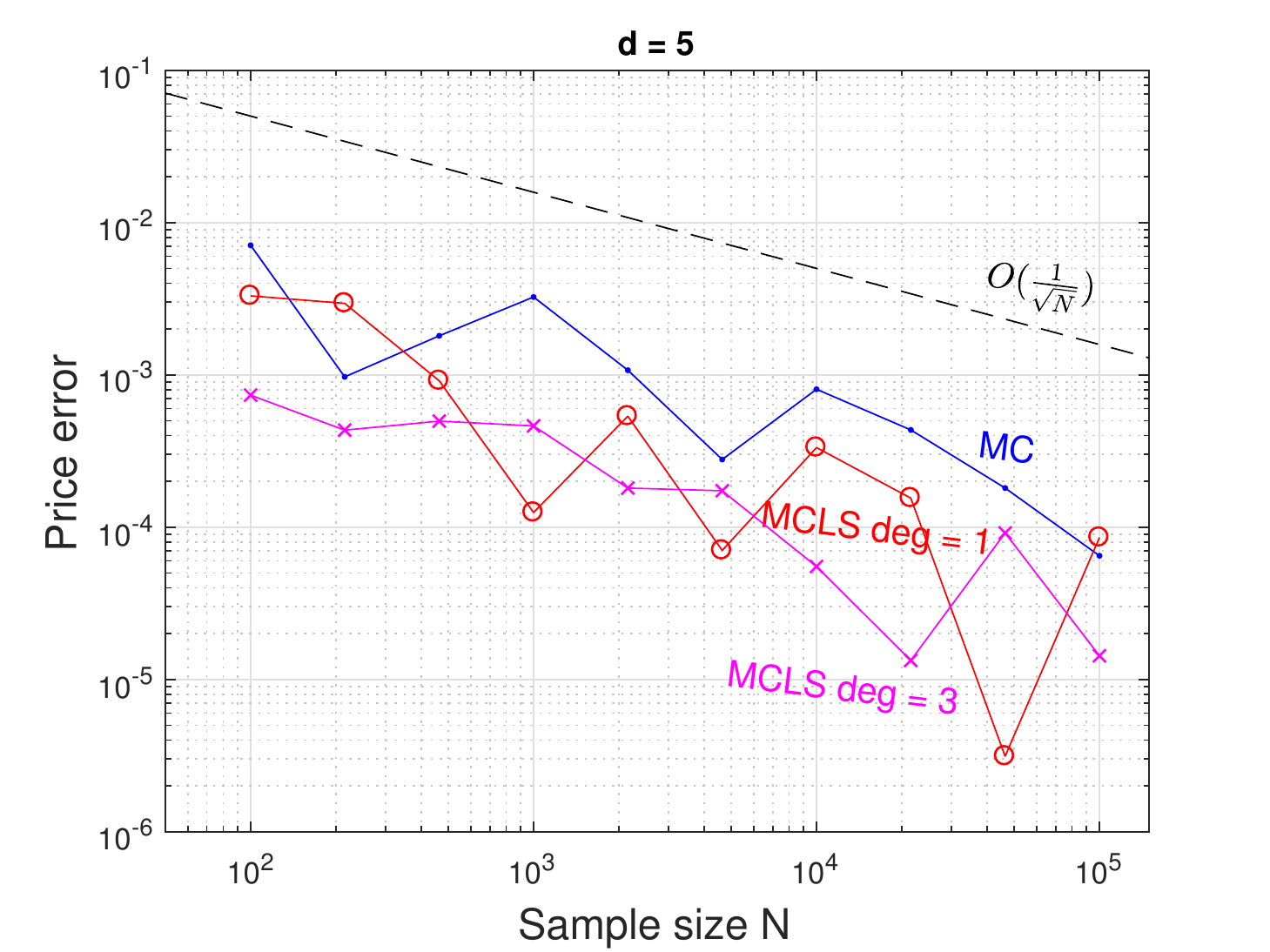}
\includegraphics[width=0.49\textwidth]{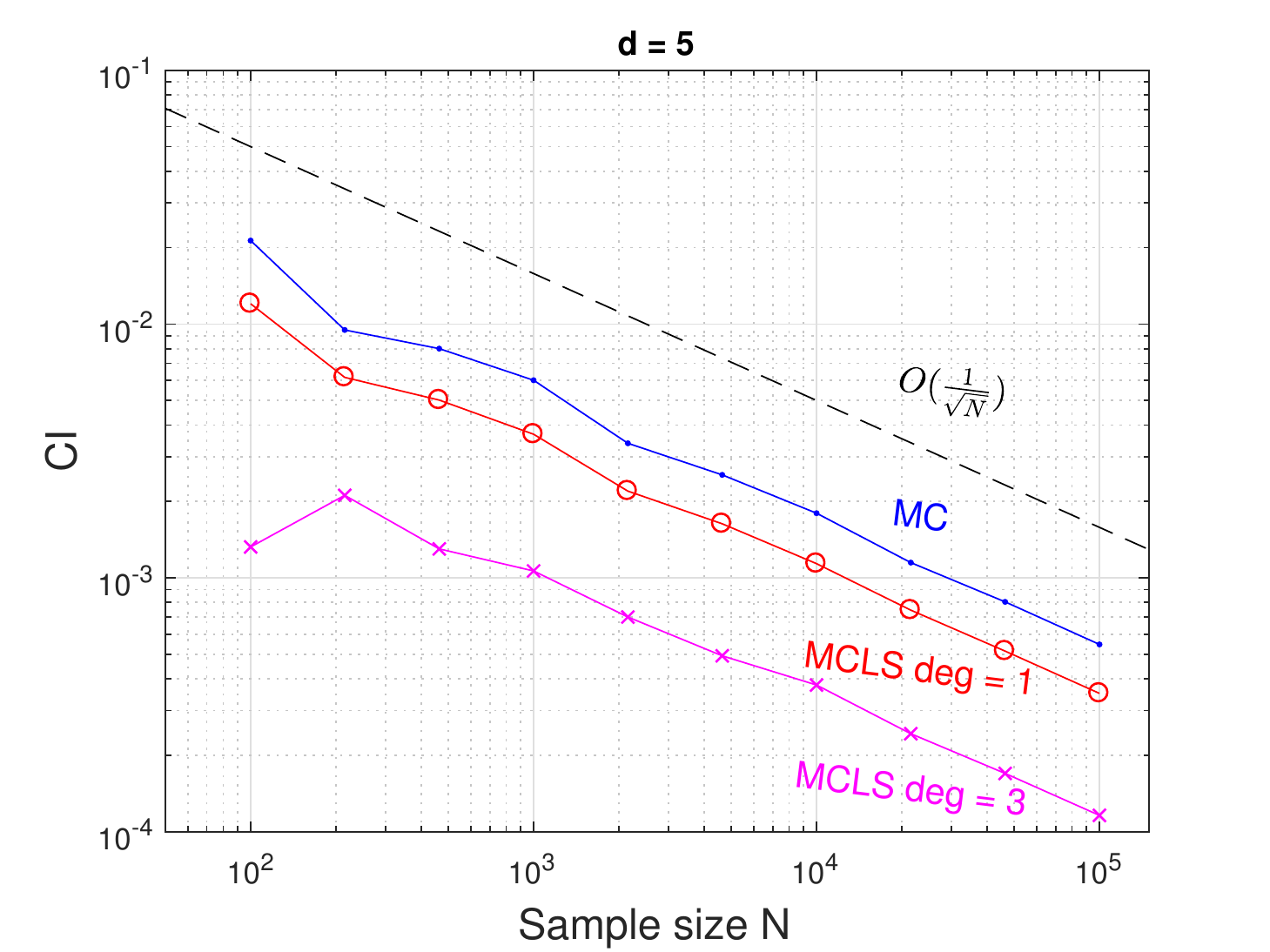}
\includegraphics[width=0.49\textwidth]{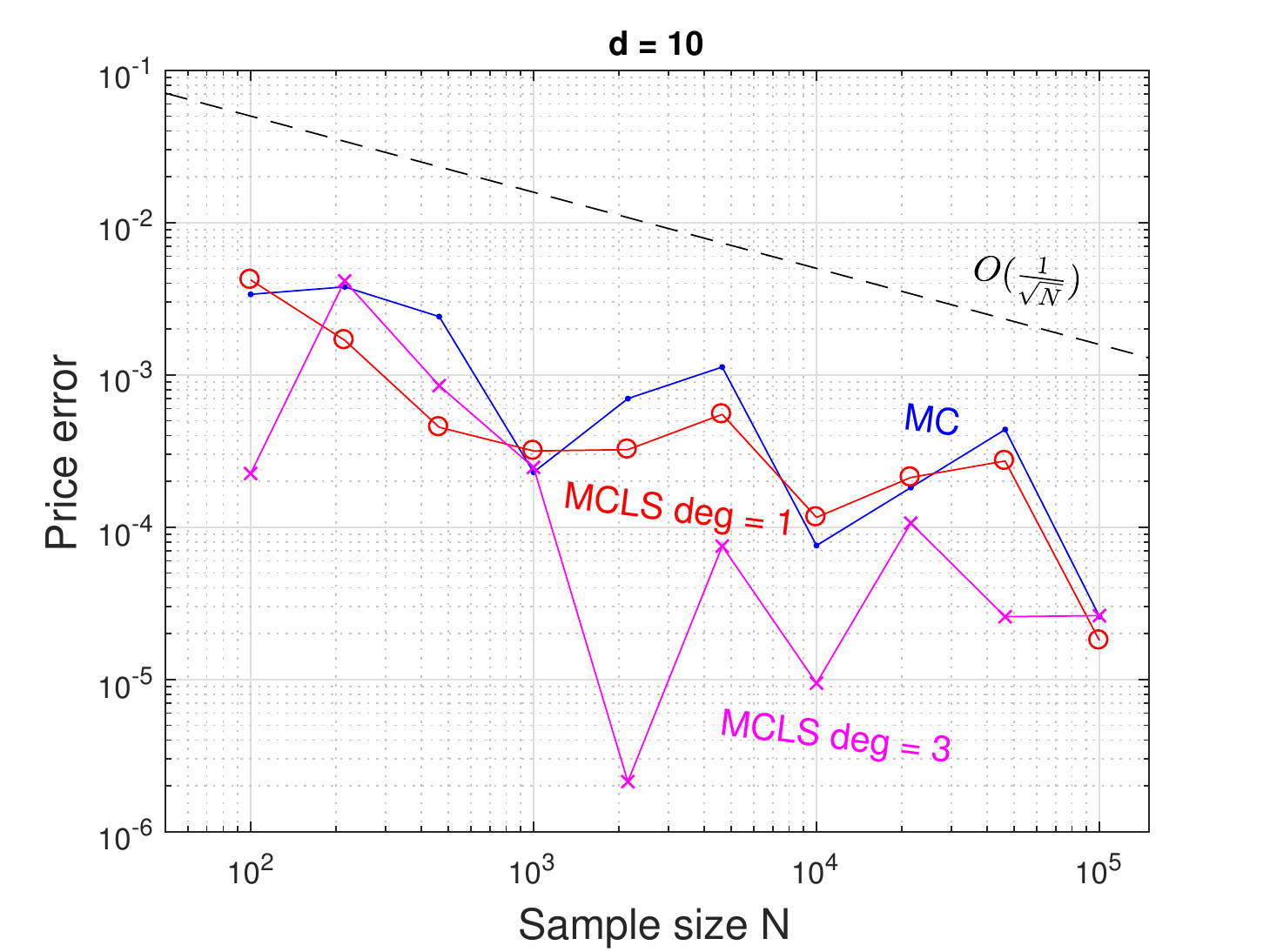}
\includegraphics[width=0.49\textwidth]{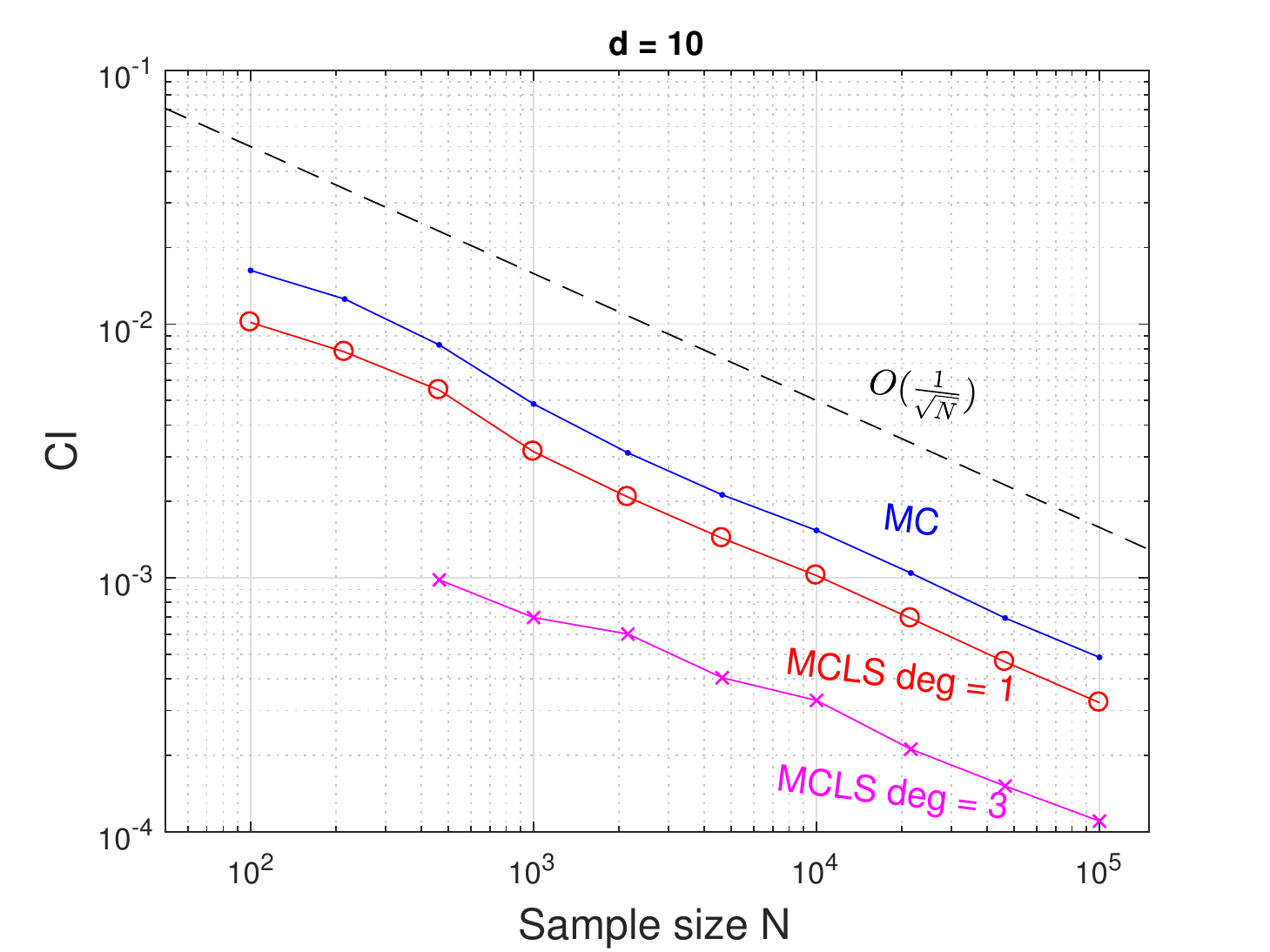}
\caption{MCLS for OTM basket option in Black Scholes model for different dimensions and polynomial degrees. Left: absolute price errors with respect to a reference price computed with $10^6$ simulations. Right: Width of $95\%$ confidence interval. \label{OTM_BS}}
\end{figure}

The results are shown in Figures \ref{ITM_BS}, \ref{ATM_BS} and \ref{OTM_BS}. In the legend the represented number indicates again the maximal total degree of the basis monomials. For instance, if $d=2$ and the maximal total degree is $\operatorname{deg}=3$, this means the the basis functions $\phi_j$ are chosen to be $\{1,s_1,s_2,s_1^2, s_1s_2, s_2^2, s_1^3, s_1^2 s_2, s_1 s_2^2, s_2^3\}$.


We observe that also in these multidimensional examples MCLS strongly outperforms the standard MC in terms of absolute price errors and width of the confidence intervals. Due to the use of the multivariate monomials as basis functions,  the condition number of $V$ is relatively high, reaching values up to order $10^5$. However, the QR based algorithm chosen according to the selection scheme \ref{scheme-algo} for the numerical solution of the least-squares problem \eqref{LS} still yields accurate results. The Vandermonde matrix $V$ is here still storable, being of size at most $10^5 \times 286$. 
In the next section we treat problems of higher dimensionality leading to a Vandermonde matrix of bigger size. There, its storage is not feasible any more and neither CG nor QR based solver can be used.

\subsubsection{Basket options in Black-Scholes models - large size problems}\label{BS example2}
In the multivariate Black-Scholes model we now consider rainbow options with payoff function
\begin{equation*}
f(s_1,\cdots, s_d) = (K-\min(s_1,\cdots, s_d))^+,
\end{equation*}
so that we apply MCLS in order to compute the quantity 
\begin{align*}
\e^{-rT}\mathbb{E}[ (K-\min(S_T^1,\cdots,S_T^d))^+] = \e^{-rT}\int_{\R^d_+}  (K-\min(s_1,\cdots,s_d))^+ d\mu(s_1, \cdots, s_d),
\end{align*}
where $\mu$ is the distribution of $(S_T^1, \cdots, S_T^d)$. In contrast to the payoff \eqref{payoff basket} which presents one type of irregularity that derives from taking the positive part $(\cdot)^+$, this payoff function presents two types of irregularities: one again due to $(\cdot)^+$, and the second one deriving from the $\min(\cdot)$ function. This example is therefore more challenging.

As in \cite{nakatsukasa2018approximate}, we rewrite the option price with respect to the Lebesgue measure
\begin{equation*}
\e^{-rT}\int_{[0,1]^d}  \bigg(K-\min_{i=1,\dots,d}\Big(S_0^i \exp \big ( (r-\frac{\sigma_i^2}{2})t+\sigma_i \sqrt{T} \mathbf{L} \Phi^{-1}(\x) \big ) \Big)\bigg)^+ d\x,
\end{equation*}
where $\mathbf{L}$ is the Cholesky decomposition of the correlation matrix and $\Phi^{-1}$ is the inverse map of the cumulative distribution function of the multivariate standard normal distribution.

The model and payoff parameters are chosen to be
\begin{equation*}
S_0^i=1,\quad K=1, \quad  \sigma_i=0.2\enskip \forall i,\quad  \Sigma= I_d \quad T=1,\quad r=0.01,
\end{equation*}
so that we consider a basket option of uncorrelated assets.

We apply MCLS for $d=\{5, 10, 20\}$ using different total degrees for the approximating polynomial space and compare it with a reference price computed using the standard MC algorithm with $10^7$ simulations.  Also, we consider different numbers of simulations that go up to $10^6$. 
We choose a basis of tensorized Legendre polynomials, which form an ONB with respect to the Lebesgue measure on the unit cube $[0,1]^d$, and we perform the sampling step of MCLS (step 1) according to the optimal distribution as introduced in \cite{cohen2017optimal} and reviewed in Section \ref{sec-optimally}. The solver for the least-squares problem is chosen according to the scheme shown in Figure~\ref{scheme-algo}, where we assume that the Vandermonde matrix $\V$ can be stored whenever the number of entries is less than $10^8$. This implies that also, for example, for the case $d=5$ with polynomial degree $5$ and $10^6$ simulations $\V$ can not be stored. Indeed, for $d=5$, $\text{deg}=5$ and $N=10^6$ the matrix $\V$ has $2.52 \cdot 10^8$ entries. For all of these cases, we therefore solve the least-squares problem by applying the randomized extended Kaczmarz algorithm.

\begin{figure}[ht]
\centering
\includegraphics[width=0.49\textwidth]{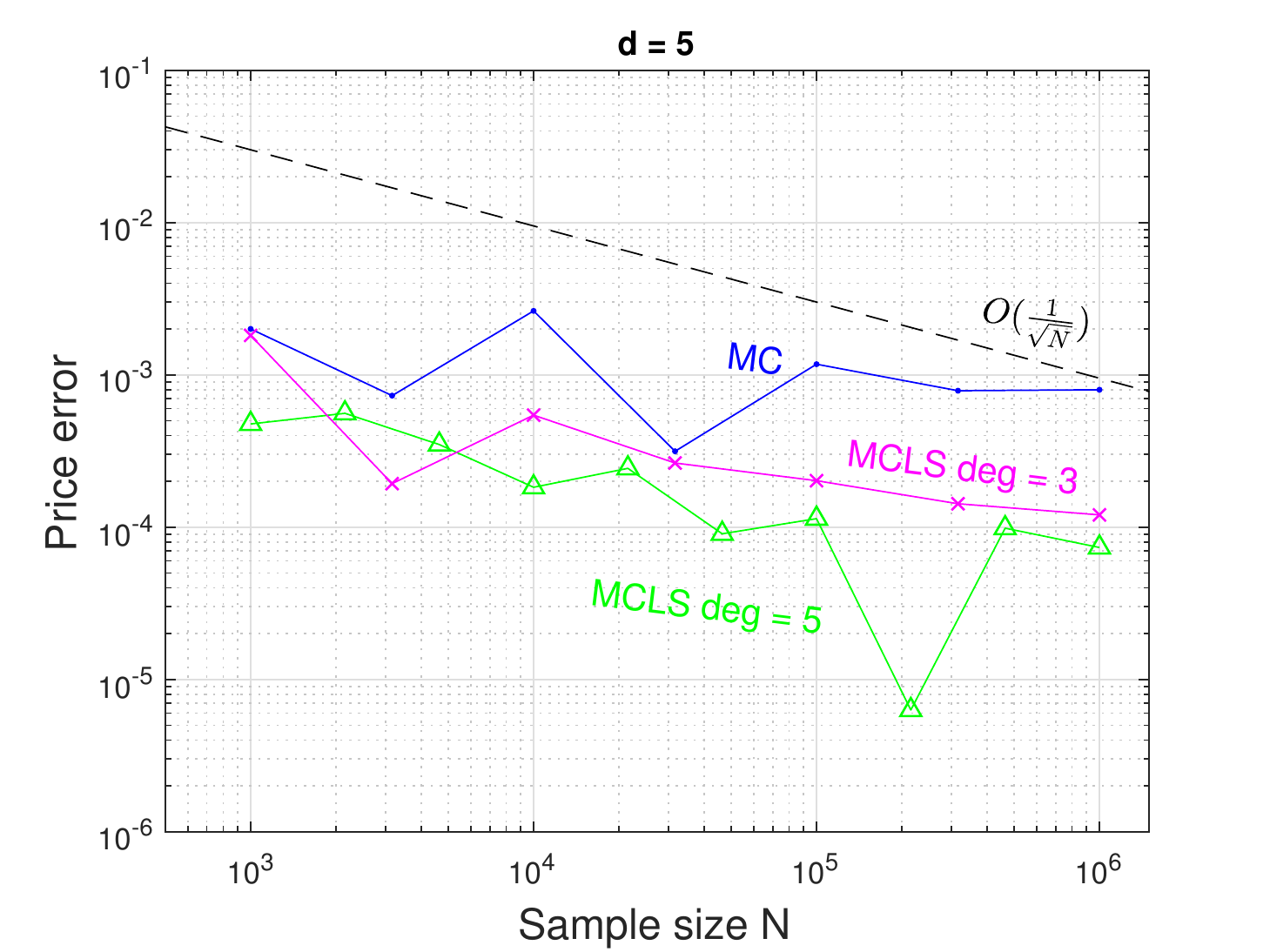}
\includegraphics[width=0.49\textwidth]{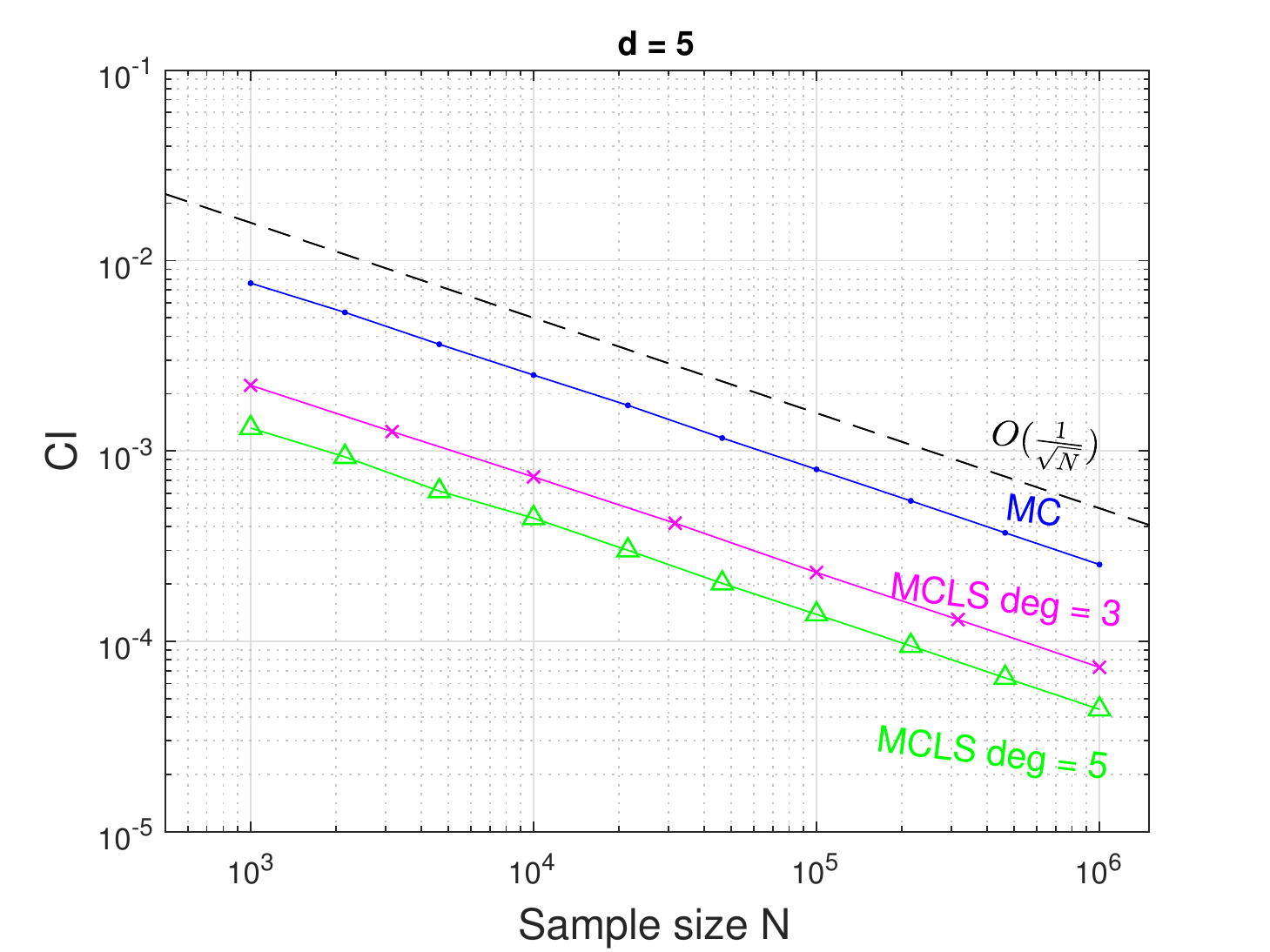}
\includegraphics[width=0.49\textwidth]{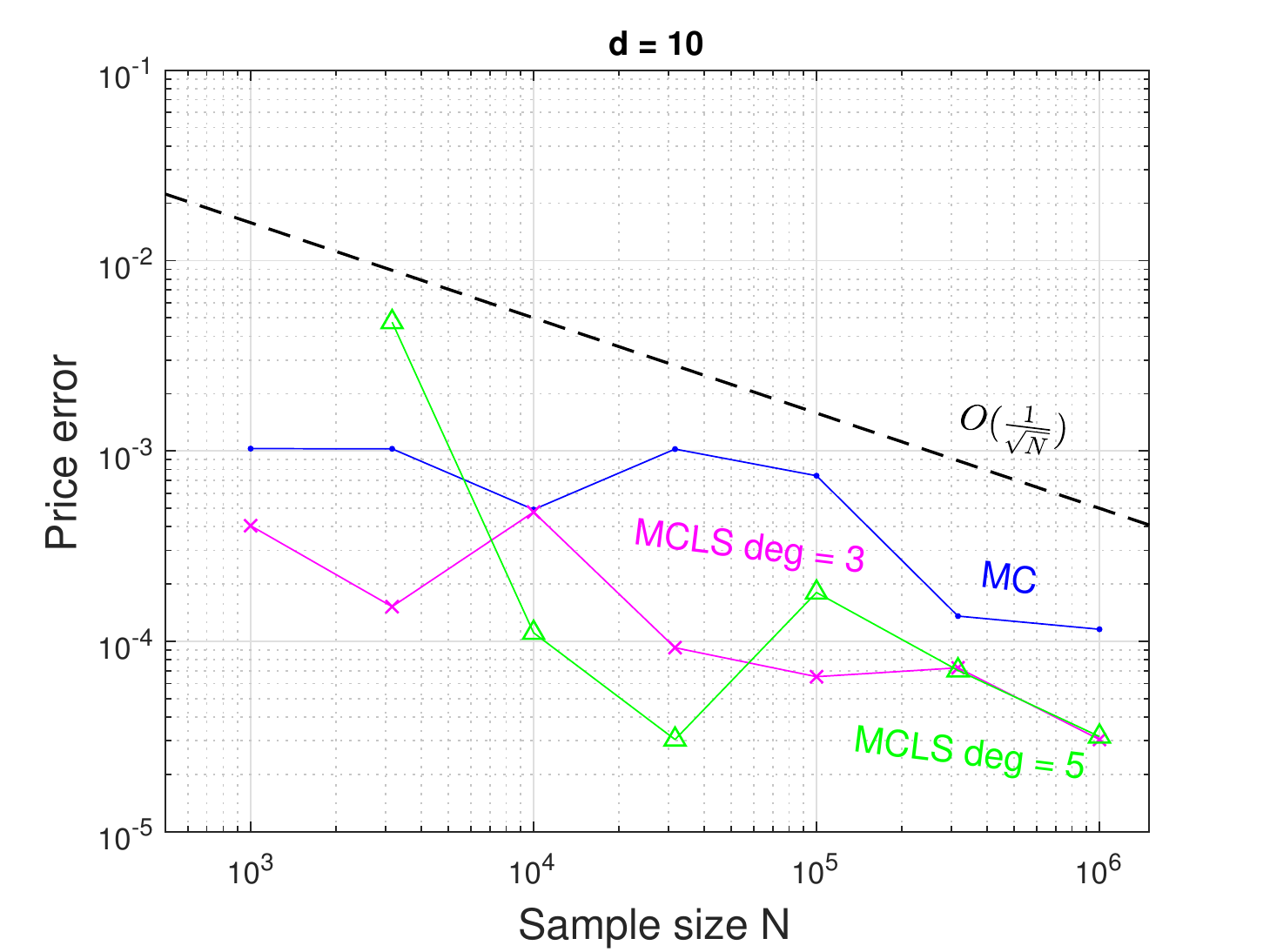}
\includegraphics[width=0.49\textwidth]{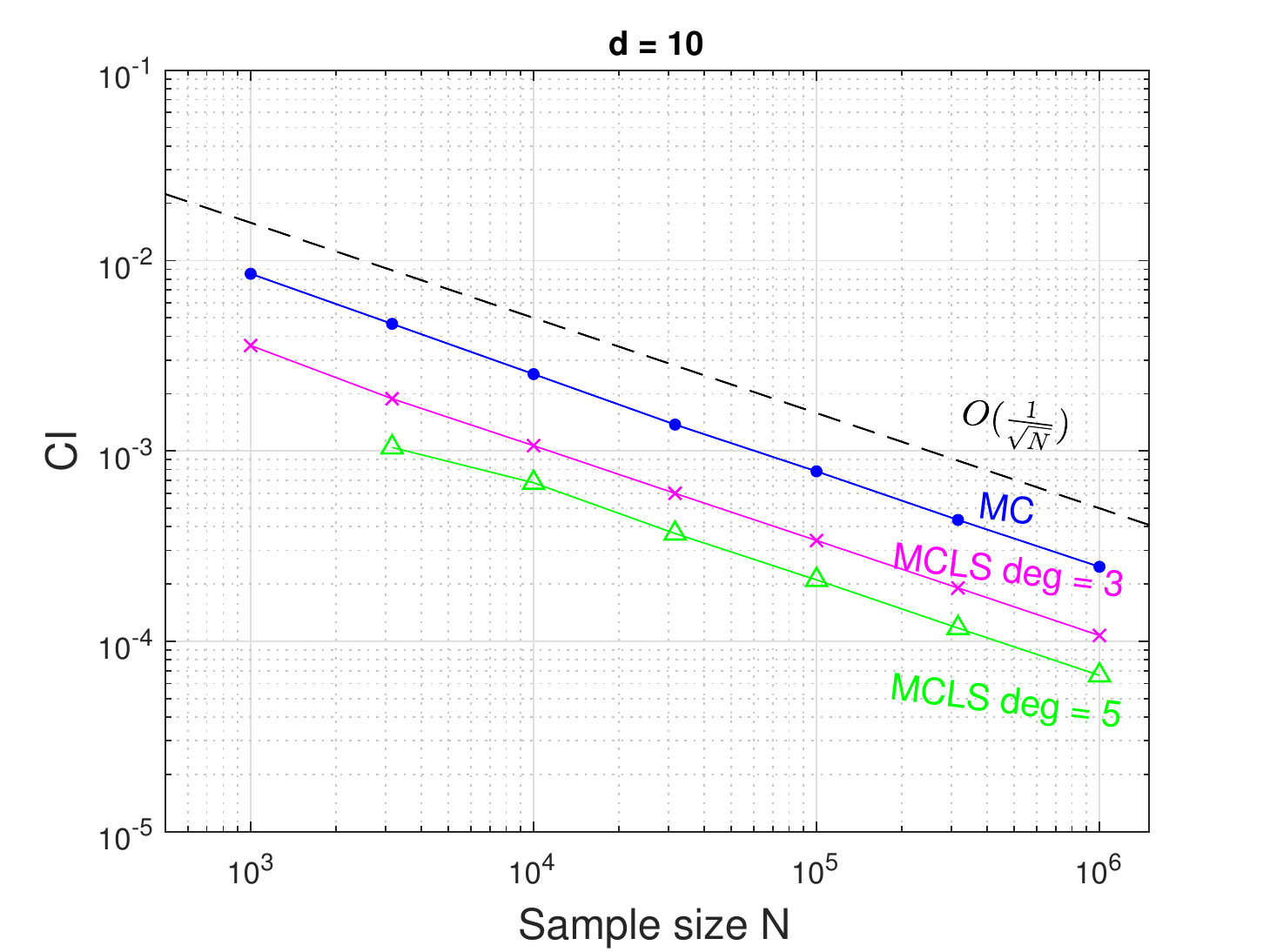}
\includegraphics[width=0.49\textwidth]{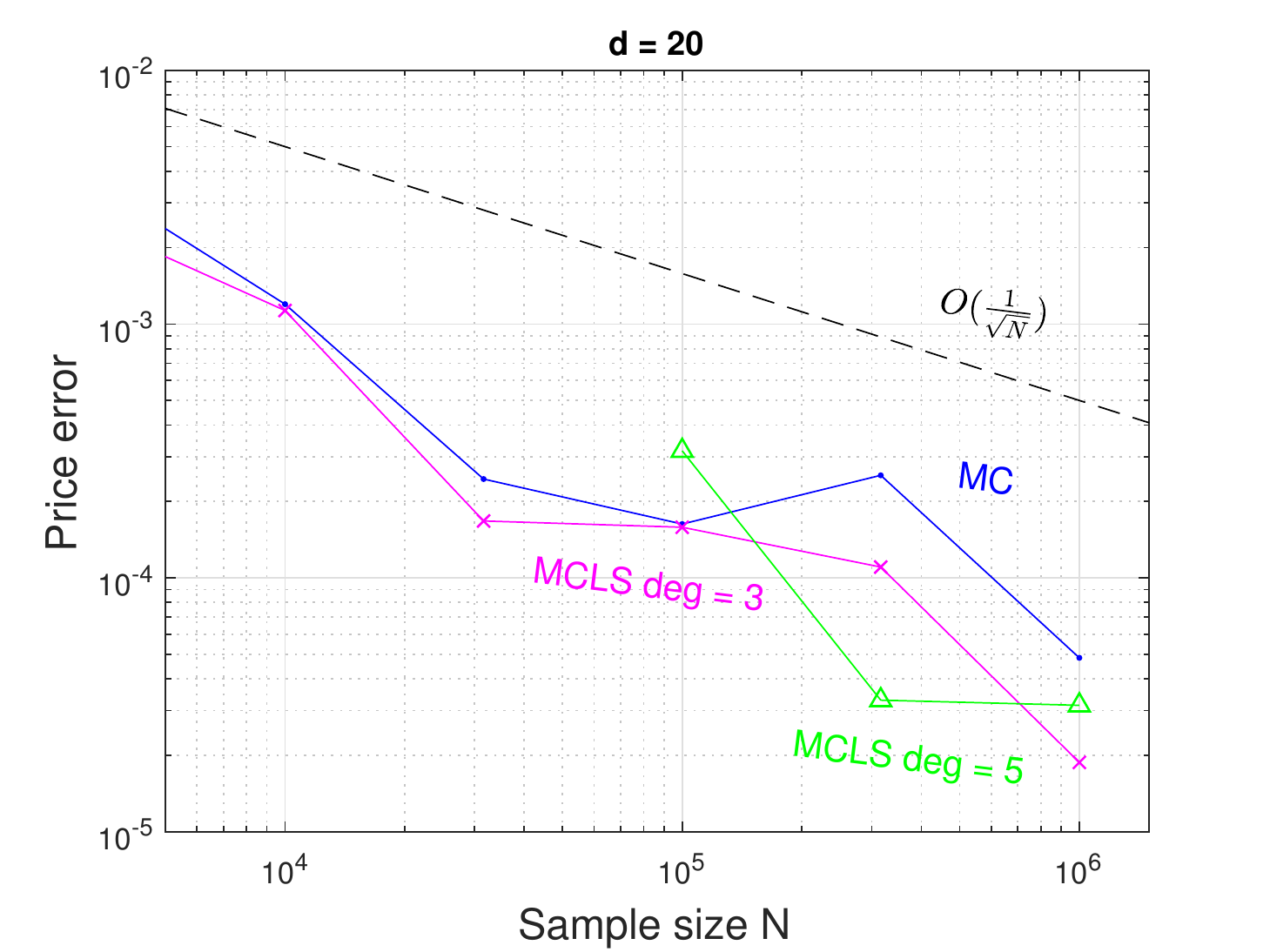}
\includegraphics[width=0.49\textwidth]{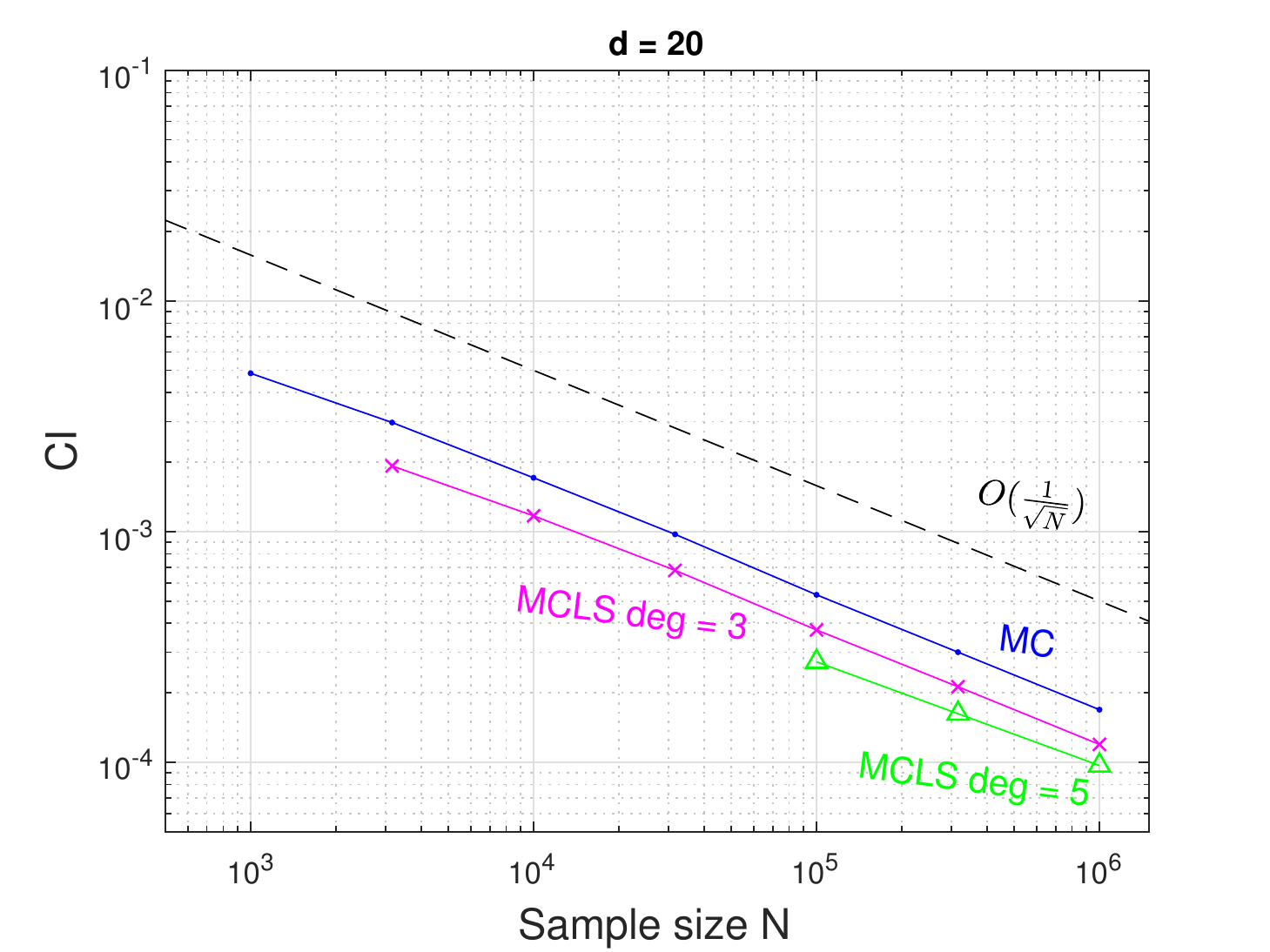}
\caption{MCLS for rainbow options in Black-Scholes model for different dimensions and polynomial degrees. Left: absolute price errors with respect to a reference price computed with $10^7$ simulations. Right: width of the $95\%$ confidence interval. \label{ATM_Kaczmarz}}
\end{figure}

In Figure~\ref{ATM_Kaczmarz} we plot the obtained price absolute errors and the width of the $95\%$ confidence intervals for all considered problems. We notice that MCLS outperforms again MC in terms of confidence interval width and price errors, as observed for medium dimensions. The choice of the weighted sampling strategy combined with the ONB allowed us to obtain a well conditioned matrix $\V$, according to the theory presented in the previous sections.

These examples and the obtained numerical results show therefore that our extension of MCLS is effective and allows us to efficiently price single and multi-asset European options. In the next section we test our extended MCLS in a slightly different setting where the integrating function is smooth.

\section{Application to high-dimensional integration}\label{sec-high dimensional application}
In this section we apply the extended MCLS algorithm to compute the definite integral 
\begin{equation}\label{eq:sin integral}
\int_{[0,1]^d} \sin \Big (\sum_{j=1}^d x_j \Big ) d\x.
\end{equation}
This is a classical integration problem~\cite{genz1984testing}, which was  also considered in~\cite{nakatsukasa2018approximate}, where MCLS is applied to compute \eqref{eq:sin integral} for dimension at most $d=6$ and with at most $N=10^5$ simulations. Our goal is to show that, thanks to the use of REK, we can increase the dimension $d$ and the number of simulations $N$.

Here we apply MCLS for $d=10$ and $d=30$ using a basis of tensorized Legendre polynomials of total degree $5$ and $4$, respectively. We compare it to the reference result which for $d=2$ $ (\text{mod } 4)$ is explicitly given by 
\begin{equation*}
\int_{[0,1]^d} \sin \Big (\sum_{j=1}^d x_j \Big ) d\x = \sum_{j=0}^d (-1)^{j+1} \binom{d}{j} \sin(j).
\end{equation*}
Also, we consider different sample sizes that go up to $10^7$. 
We perform the sampling step of MCLS (step 1) again according to the optimal distribution. The choice of the solver for the least-squares problem is again taken according to the scheme in Figure~\ref{scheme-algo} and we assume that the Vandermonde matrix $V$ can be stored whenever the number of entries is less than $10^8$.

\begin{figure}[ht]
\centering
\includegraphics[width=0.49\textwidth]{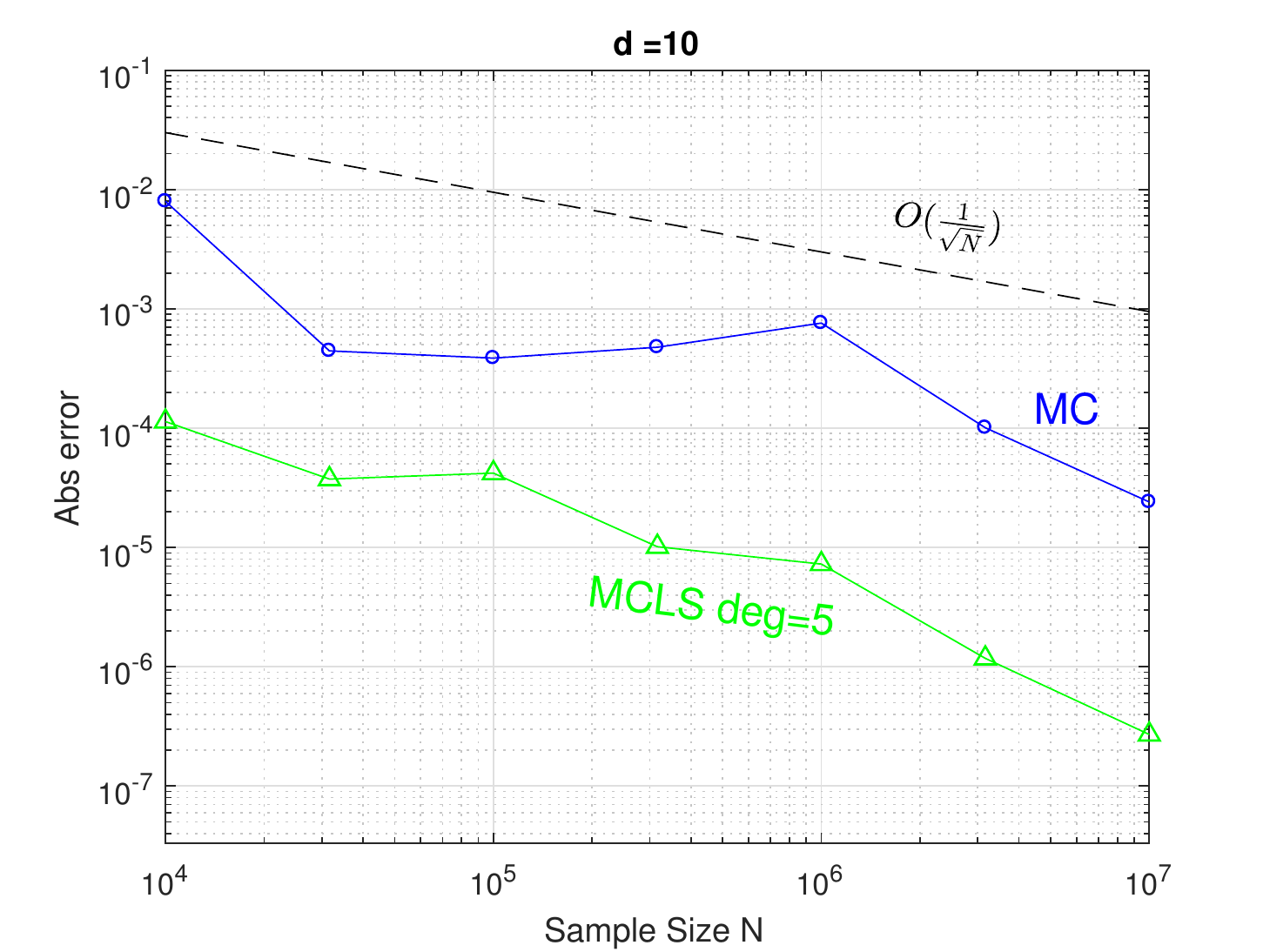}
\includegraphics[width=0.49\textwidth]{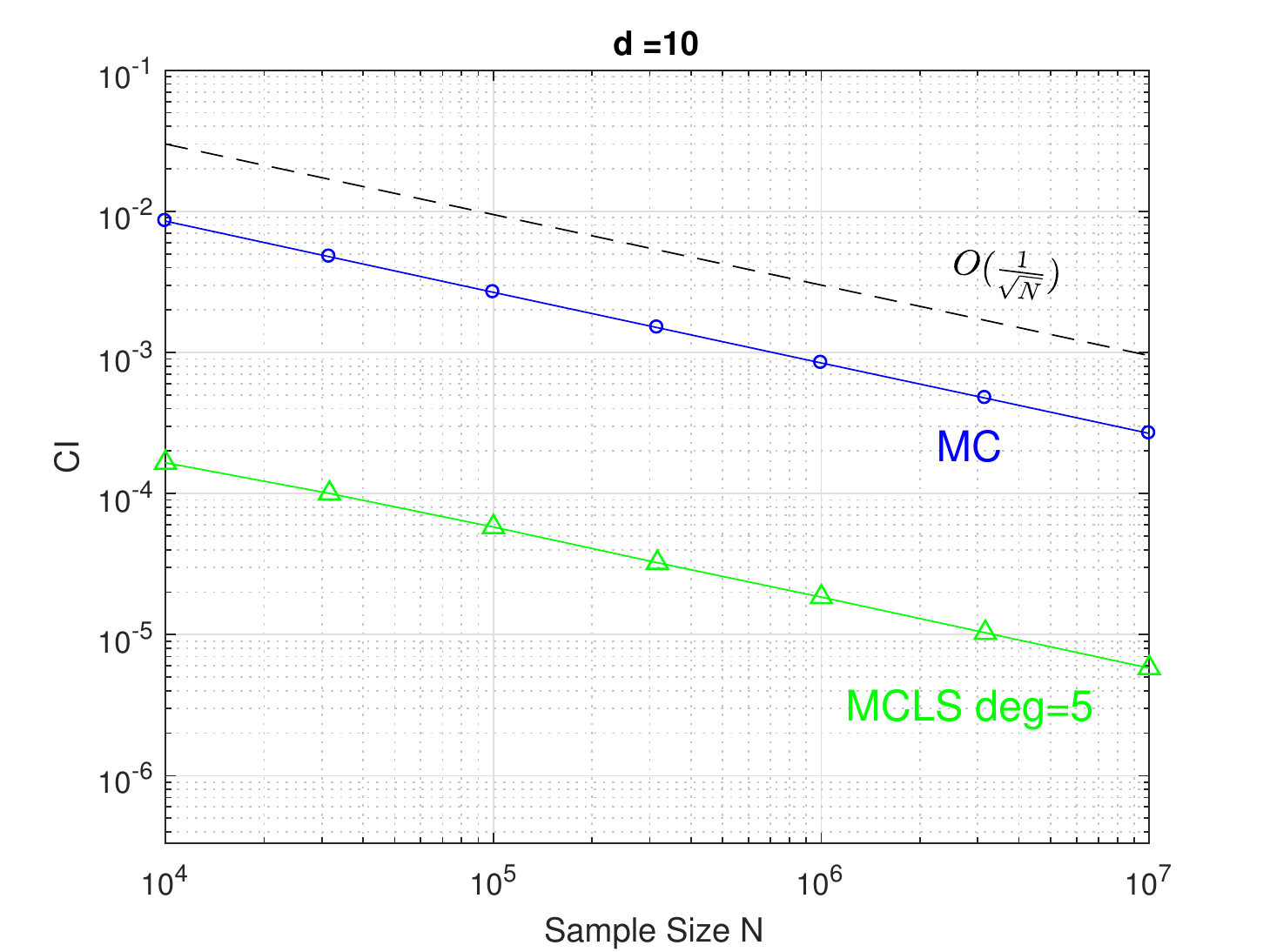}
\includegraphics[width=0.49\textwidth]{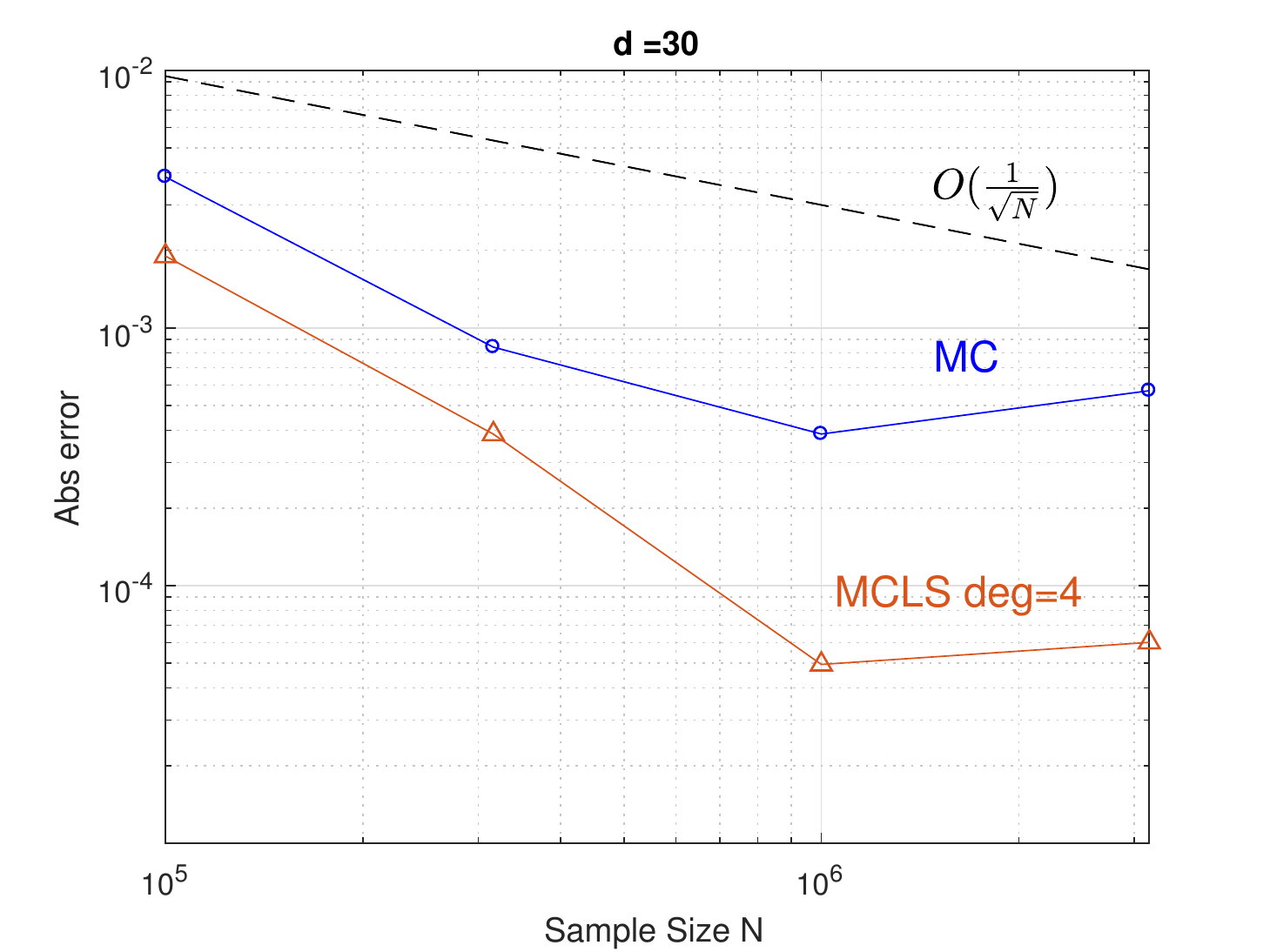}
\includegraphics[width=0.49\textwidth]{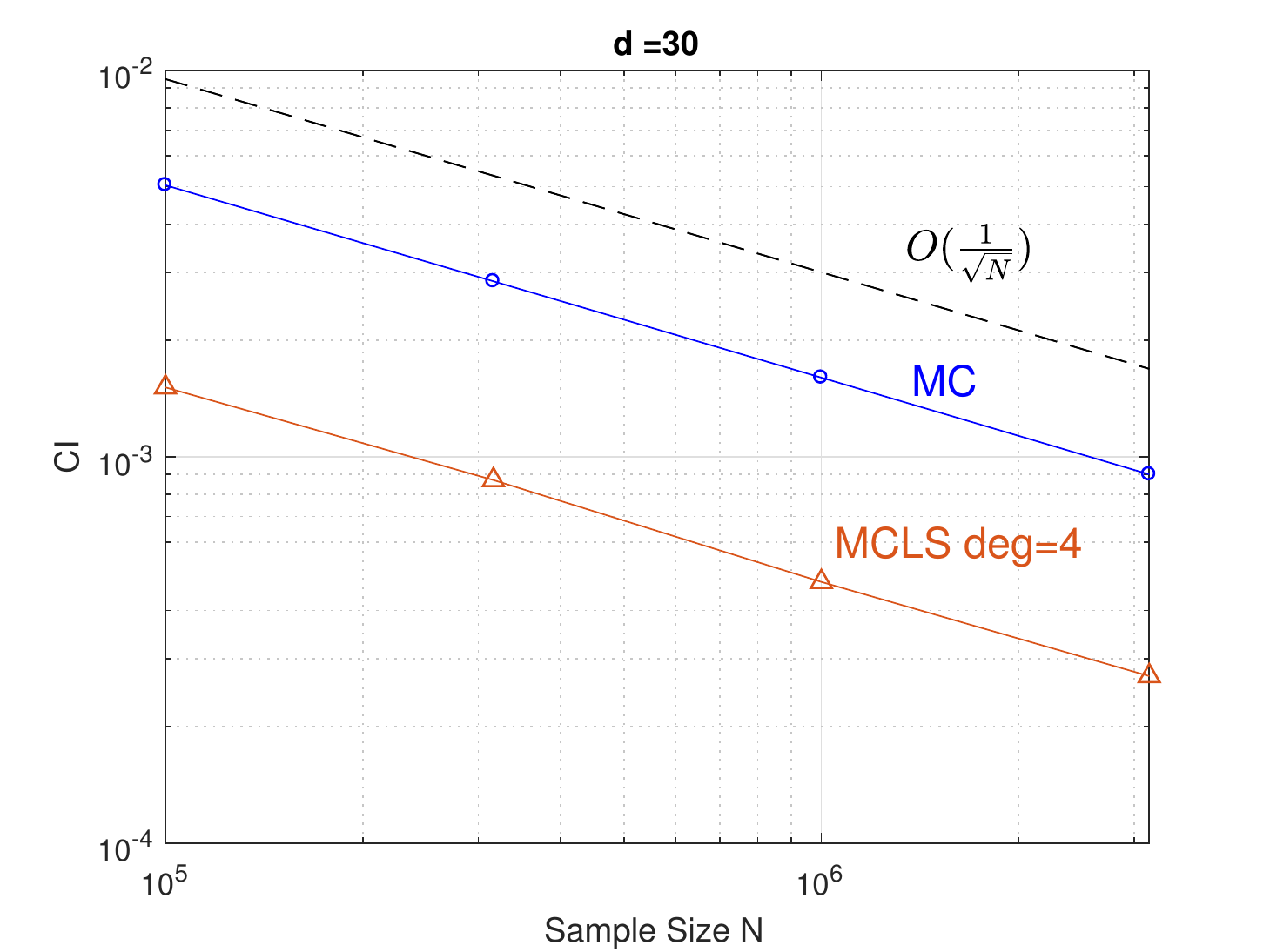}
\caption{MCLS for integrating the function $\sin(\sum_{j=1}^d x_j)$ for $d=\{10,30\}$ and different polynomial degrees. Left: absolute errors with respect to a reference result given in closed form. Right: width of the $95\%$ confidence interval. \label{SinFunction}}
\end{figure}

The results are shown in Figure \ref{SinFunction}. Again, we have plotted the obtained absolute error computed with respect to the reference result (left) and the width of the $95\%$ confidence interval (right). First, we note that MCLS performs much better than the standard MC, as in the previous examples. Furthermore, the results are considerably better than the ones obtained in the previous section, see Figure \ref{ATM_Kaczmarz}. This is due to the fact that the integrand is now smooth (it is analytic/entire), while in the multi-asset option example it was only continuous. Indeed, the function approximation error $\min_{\mathbf{c} \in \R^{n+1}} \|\sqrt{w} (f-\sum_{j=0}^n c_j \phi_j)\|_\mu$ is expected to be much smaller in this case, since polynomials provide a more suitable approximating space for smooth functions than for irregular functions. According to Proposition \ref{thm_convW}, this results in a stronger variance reduction and hence a better approximation of the integral. 

\section{Conclusion and future work}\label{sec-conclusion}
We have presented a numerical technique to price single and multi-asset European options and, more generally, to compute expectations of functions of multivariate random variables. The methodology consists of extending MCLS to arbitrary probability measures and combining it with weighted sampling strategies and the randomized extended Kaczmarz algorithm. The core concepts and algorithms have been presented in Section \ref{sec-MCLS}. 
In Section \ref{sec-cost analysis} we have proposed a new cost analysis. Here, we have shown that MCLS asymptotically outperforms the standard Monte Carlo method as the cost goes to infinity, provided that the integrand satisfies certain regularity conditions.

In Section \ref{sec-MCLS option pricing} we have applied the new method to price European options. First, we have adapted the generalization to compute multi-asset option prices, where we have proposed to modify the sampling step of MCLS by discretizing the governing SDE of the underlying price process, whenever needed. The modification of the first step introduces a new source of error, which has been analyzed in Proposition \ref{prop:disc error}. In the Sections \ref{JH example}, \ref{BS example} and \ref{BS example2} we have applied the algorithm to price multi-asset European options in the Heston model, in the Jacobi model and in the multidimensional Black Scholes model, where we have exploited the fact the they belong to the class of polynomial diffusions and the moments can be computed in closed form. For these examples, MCLS usually provides considerably high accuracy compared to the standard MC for the same sample sizes. This typically holds for different sample sizes and when accuracy is measured in terms of implied volatility, see Table \ref{imp_vols_Heston} and Table \ref{imp_vols_Jacobi}, and in terms of option price errors and confidence interval widths, see for instance Figures \ref{ITM_Heston}-\ref{OTM_Jacobi} and Figure \ref{ATM_Kaczmarz}. As expected, enlarging the number of basis functions $n$ for a given sample size $N$ leads to more accurate results. Moreover, in Section \ref{BS example2} employing REK allowed us to solve high dimensional problems with high accuracy. For instance also our experiments for basket options on $20$ assets shows that enlarging the number of basis functions yields higher accuracy in terms of confidence intervals. Finally, in Section \ref{sec-high dimensional application} we considered the approximation of a multidimensional integral of a smooth function. Storage requirements limit the feasibility of the basic MCLS for high dimension. Indeed, in \cite{nakatsukasa2018approximate} only cases with maximal dimension $d=6$ and $N=10^5$ could be treated. Thanks to the application of REK we were able to treat dimension $d=30$ and $N$ up to $10^7$. This illustrates the effectiveness of our extended approach.

To extend the approach further to even higher dimensions, other computational bottlenecks arising are to be addressed. Solving the storage issue in the least-squares problem with REK leaves us with a high number of function calls. We do not need to store the full Vandermode matrix, but instead rows and columns are required many times during the iteration. This leads to a high computational cost. One can reduce this cost by 1) reducing the number of function calls and by 2) making the function calls more efficient. To achieve 1), one can for instance store the rows and columns of the Vandermonde matrix which are called with highest probability. To achieve 2) one can exploit further insight of the functions, for instance using a low-rank approximation~\cite{grasedyck2013literature} or functional analogues of tensor decomposition approximation~\cite{gorodetsky2019continuous}.

\section*{Appendix}
Here, we present the proof of Proposition \ref{thm_convW}.

\begin{proof} 
Note that the approximate function $\sum_{j=0}^n\hat{c}_j\phi_j$ and thus  $\hat{I}_{\mu,N}$  only depends on the span of the basis functions $ \{\phi_j\}_{j=0}^n$ and not on the specific choice of the basis. Therefore, without loss of generality we can assume that the chosen basis functions $ \{\phi_j\}_{j=0}^n$ form an orthonormal basis (ONB) in $L^2_\mu$, i.e. $\int_E \phi_i(\x) \phi_j(\x) d\mu(\x)=\delta_{ij}$. 

We decompose the function $f$ into a sum of orthogonal terms
\begin{equation}\label{fdecomp}
f=\sum_{j=0}^n c^\ast_j \phi_j + g=: f_1+g,
\end{equation}
where $g$ satisfies $\int_E g(\x) \phi_j(\x) d\mu(\x)=0$ for all $j=0,\cdots, n$. Note that $\|g\|_\mu=\min_{{\mathbf c} \in \R^{n+1}}\|f-\sum_{j=0}^nc_j\phi_j\|_\mu$. 
Assume now that we sample according to $\frac{d\mu}{w}$ and obtain the points $\{\tilde{\x}_i\}_{i=1}^N$. Then, the vector of sample values in the weighted least-squares problem can be decomposed as
\[
\tilde{\f}= 
[\tilde{f}_1(\tilde{\x}_1)+\tilde{g}(\tilde{\x}_1),\ldots,\tilde{f}_1(\tilde{\x}_N)+\tilde{g}(\tilde{\x}_N)]^T
=\widetilde{\V} \mathbf{c}^\ast+\tilde{\g}, 
\]
where $\widetilde{\V}$ and $\tilde{f}$ are defined as in \eqref{eq:MCLSweight} and $\tilde{g}:=\sqrt{w}g$ and hence \[\tilde{\g}=[\sqrt{w(\tilde{\x}_1)}g(\tilde{\x}_1),\dots, \sqrt{w(\tilde{\x}_N)}g(\tilde{\x}_N)].\]
Let $\hat{\mathbf{c}}$ be again the least-squares solution to~\eqref{eq:MCLSweight}, 
then
\[\hat{\mathbf{c}}=\text{argmin}_{\mathbf{c} \in \R^{n+1}} \| \widetilde{\V} \mathbf{c} - (\widetilde{\V} \mathbf{c}^\ast+\tilde{\g})\|_2 =
 (\widetilde{\V}^T\widetilde{\V})^{-1}\widetilde{\V}^T(\widetilde{\V} \mathbf{c}^\ast+\tilde{\g})= \mathbf{c}^\ast+ (\widetilde{\V}^T\widetilde{\V})^{-1}\widetilde{\V}^T\tilde{\g},
 \]
 where
the second summand is exactly $\mathbf{c}_g:=\mbox{argmin}_{\mathbf{c}\in \R^{n+1}} \|\widetilde{\V}\mathbf{c} - \tilde{\g}\|_2 $. 
It thus follows that the integration error is $\hat I_{\mu,N}-I_\mu = c_{g,0}=[1,0,\ldots,0](\widetilde{\V}^T\widetilde{\V})^{-1}\widetilde{\V}^T\tilde{\g}$. 

Now  by the strong law of large numbers  we have 
\begin{align*}
\frac{1}{N}(\widetilde{\V}^T\widetilde{\V})_{i+1,j+1} = 
&\frac{1}{N}\sum_{l=1}^N w(\tilde{\x}_l)\phi_i(\tilde{\x}_l)\phi_j(\tilde{\x}_l)
\\
&\rightarrow \int_{E}w(\tilde{\x})\phi_i(\tilde{\x})\phi_j(\tilde{\x})\frac{d\mu(\tilde{\x})}{w(\tilde{\x})}=\int_{E}\phi_i(\tilde{\x})\phi_j(\tilde{\x})d\mu(\tilde{\x})
=\delta_{ij}
\end{align*}
almost surely and in probability as $N\rightarrow \infty$, by the orthonormality of $\{\phi_j\}_{j=0}^n$. Therefore we have 
$\frac{1}{N}\widetilde{\V}^T\widetilde{\V}\stackrel{p}{\rightarrow} \mathbf{Id}_{n+1}$  as $N\rightarrow \infty$, where $\mathbf{Id}_{n+1}$ denotes the identity matrix in $\R^{(n+1)\times(n+1)}$. Moreover, $\sqrt{N}\left(\frac{1}{N}\sum_{i=1}^N w(\tilde{\x}_i)g(\tilde{\x}_i)\right)
\stackrel{d}{\rightarrow} Z\sim \mathcal{N}(0,\|\sqrt{w}g\|_\mu^2)$ for $N\rightarrow \infty$ by the central limit theorem, where we used the fact $\int_E g(\x) d\mu(\x)=0$ for the mean and $\int_{E} (w(\tilde{x})g(\tilde{\x}))^2 \frac{d\mu(\tilde{\x})}{w(\tilde{\x})}=\|\sqrt{w}g\|_\mu^2$ for the variance. Thanks to Slutsky's theorem (see e.g. \cite[Chapter 5]{gut2013probability}) we finally obtain
 \begin{equation*}
\sqrt{N}(\hat I_{\mu,N}-I_\mu)=
\sqrt{N}[1,0,\ldots,0]^T
\frac{1}{N}(\frac{1}{N}\widetilde{\V}^T\widetilde{\V})^{-1}\widetilde{\V}^T\tilde{\g}
\xrightarrow{d} \mathcal{N}(0,\|\sqrt{w}g\|_\mu^2). 
 \end{equation*}
\end{proof}

\bibliographystyle{plain}
\bibliography{references}

\end{document}